\documentclass[aps,prx,nofootinbib,twocolumn,amsmath,amssymb,superscriptaddress,notitlepage]{revtex4-1}

\usepackage{latexsym}
\usepackage{graphicx}
\usepackage{times,psfrag,subfigure}
\usepackage{enumerate}
\usepackage{amsmath}
\usepackage{dsfont}
\usepackage{dcolumn}
\usepackage{bm,bbm}
\usepackage{color}
\usepackage{latexsym,amsmath,amssymb,bm,euscript}
\usepackage{dsfont}
\usepackage{textcomp}
\usepackage{tabularx}
\usepackage{setspace}
\usepackage{ctable}
\usepackage{sidecap}
\usepackage{placeins}
\usepackage{threeparttable}
\usepackage{multirow}
\usepackage{comment}
\usepackage{makecell}
\usepackage{pifont}
\newcommand{\cmark}{\ding{51}}%
\newcommand{\xmark}{\ding{55}}%

\let \oldbm \bm
\renewcommand{\vec}[1]{\oldbm{#1}}

\def\bk{{\vec k}}

\def\bs{{\vec s}}
\def\L{{\mathcal L}}

\def\bq{{\vec q}}

\def\bG{{\vec G}}

\def\bn{{\vec n}}
\def\bm{{\vec m}}

\def\br{{\vec r}}

\def\bfOne{{\bf 1}}
\def\bfTwo{{\bf 2}}

\def\tr{\mathop{\mathrm{tr}}}
\def\Tr{\mathop{\mathrm{Tr}}}

\def\T{\mathcal{T}}

\def\S{\mathcal{S}}

\def\G{\mathcal{G}}

\def\H{\mathcal{H}}

\def\U{{\rm U}}
\def\SU{{\rm SU}}

\def\hlinewd#1{%
\noalign{\ifnum0=`}\fi\hrule \@height #1 %
\futurelet\reserved@a\@xhline}

\newcommand{\beq}{\begin{equation}}
\newcommand{\eeq}{\end{equation}}
\newcommand{\beqarray}{\begin{eqnarray}}
\newcommand{\eeqarray}{\end{eqnarray}}

\allowdisplaybreaks

\graphicspath{{figures/}}

\begin{document}

\title{Soft modes in magic angle twisted bilayer graphene}

\date{\today}
\author{Eslam Khalaf}
\affiliation{Department of Physics, Harvard University, Cambridge, Massachusetts 02138, USA}

\author{Nick Bultinck}
\affiliation{Department of Physics, University of California, Berkeley, CA 94720, USA}
\affiliation{Department of Physics, Ghent university, 9000 Gent, Belgium}

\author{Ashvin Vishwanath}
\affiliation{Department of Physics, Harvard University, Cambridge, Massachusetts 02138, USA}

\author{Michael P. Zaletel}
\affiliation{Department of Physics, University of California, Berkeley, CA 94720, USA}

\begin{abstract}
    We present a systematic study of the low-energy collective modes for different insulating states at integer fillings in magic angle twisted bilayer graphene. In particular, we provide a simple counting rule for the total number of soft modes, and analyze their energies and symmetry quantum numbers in detail. To study the soft mode spectra, we employ time dependent Hartree-Fock whose results are  reproduced analytically via an effective low-energy sigma model description. We find two different types of low-energy modes - (i) approximate Goldstone modes associated with breaking an  enlarged  $\U(4)\times\U(4)$ symmetry and, surprisingly, a second branch (ii) of nematic modes with non-zero angular momentum under three-fold rotation. The modes of type (i) include true gapless Goldstone modes associated with exact continuous symmetries in addition to gapped "pseudo-Goldstone" modes associated with approximate symmetries. While the modes of type (ii) are always gapped, we show that their gap depends sensitively on the distribution of Berry curvature, decreasing as the Berry curvature grows more concentrated. For realistic parameter values, the gapped soft modes of both types are found to have comparable gaps of only a few meV, and lie completely inside the mean-field bandgap. The entire set of soft modes emerge as Goldstone modes of a different idealized model in which Berry flux is limited to a solenoid, which enjoys an enlarged U(8) symmetry. 
     In addition, it is shown that certain ground states admit nearly gapless modes despite the absence of a broken symmetry -- an unusual property arising from  the special form of the anisotropy terms in the Hamiltonian. Furthermore, we separately discuss the number of Goldstone modes for each symmetry-broken state, distinguishing the linearly vs quadratically dispersing modes at long wavelengths. Finally, we present a general analysis of the symmetry representations of the soft modes for all possible  insulating  Slater determinant states at integer fillings that preserve translation symmetry,  independent of the energetic details. The resulting soft mode degeneracies and symmetry quantum numbers provide a fingerprint of the different insulting states enabling their experimental identification from a measurement of their soft modes.
\end{abstract}

\maketitle

\section{Introduction}

When two graphene sheets are placed on top of each other with a small in-plane rotational mismatch close to the so-called magic angle $\theta_M \sim 1.1^\circ$, the dispersion of electrons moving through the resulting superlattice is strongly reduced and Coulomb interaction effects become important. In particular, when the Fermi level is inside the flat bands, a variety of interesting states such as correlated insulators \cite{PabloMott,Dean-Young,efetov,EfetovScreening,PabloNematic,LiVafekscreening}, superconductors \cite{PabloSC,Dean-Young,efetov,EfetovScreening,YoungScreening,NadjPergeSC,LiVafekscreening}, orbital ferromagnets and quantum anomalous Hall states \cite{sharpe2019emergent,YoungQAH,efetov,YazdaniChern,AndreiChern,EfetovChern,YoungImaging,YoungHofstadter} have been observed. Furthermore, tunneling spectroscopy and compressibility measurements have revealed that the density of states is significantly reconstructed by the Coulomb interaction not only at the integer fillings where correlated insulators are observed, but for almost all fillings inside the flat bands \cite{ColombiaSTM,CaltechSTM,RutgersSTM,PrincetonSTM,CascadeYazdani,Tomarken,CascadeShahal}. Interestingly, the physics of twisted bilayer graphene (TBG) is also found to be very sensitive to external perturbations such as substrate alignment or small out-of-plane magnetic fields, which can select different ground states and thus completely change the phase diagram \cite{sharpe2019emergent,YoungQAH,NadjPergeSC,YazdaniChern,AndreiChern,EfetovChern,YoungHofstadter}. This suggests that TBG is characterized by a small intrinsic energy scale by which the different interaction-driven orders at low temperatures are separated.

Previous studies of TBG \cite{MacdonaldHF, CaltechSTM, ShangHF,CeaGuinea} have indeed identified a large family of closely competing symmetry-broken states at the different integer fillings. Combining different methods including an exactly soluble parent Hamiltonian, effective sigma model field theory and Hartree Fock calculations, Ref.~\cite{KIVCpaper} systematically investigated these competing orders. For integer fillings, a large manifold of low-energy insulating states associated with an approximate $\U(4)\times\U(4)$ symmetry was identified (see also Ref. \cite{KangVafekPRL}). For realistic model parameters, the states in this manifold are split by only a few meV. This hints at the existence of many low-energy bosonic modes corresponding to rotations within the $\U(4)\times\U(4)$ manifold. An explicit non-linear sigma model describing these soft modes was derived by the authors in Ref.~\cite{SkPaper}. 

In addition to the insulators, numerical studies \cite{CaltechSTM, MacdonaldHF, ShangHF, KIVCpaper} have also identified a $C_3$-symmetry breaking nematic semimetal as one of the competitive states. This suggests that nematic fluctuations associated with this semimetal will also be important for understanding the physics of TBG. Further evidence for the importance of nematic fluctuations was provided by recent experiments \cite{PabloNematic} indicating strong nematicity in both the insulating and superconducting phases close to $\nu = -2$, as well as by the observed $C_3$ symmetry breaking in scanning tunneling spectroscopy measurements \cite{RutgersSTM,PrincetonSTM}. In addition, several theoretical works proposed a central role for nematic fluctuations in mediating superconducting pairing \cite{KoziiNematic, FernandesVanderbos, Chubukov2020}. However, the origin of these nematic modes remained enigmatic so far. In addition, it was unclear how these two fundamentally different types of collective modes -- those associated with the spin-valley flavor symmetry and those nematic modes associated with spatial rotation symmetry -- can be unified in a single framework.

In this work, we provide such a unifying theoretical framework to systematically investigate all possible soft modes in TBG (and broadly in other Moire systems). We show that the soft modes in TBG indeed fall under two categories: (i) approximate Goldstone modes associated with $\U(4) \times \U(4)$ symmetry breaking and (ii) nematic modes associated with rotational symmetry breaking (possibly combined with flavor symmetry breaking). We study two different aspects of these soft modes. First, we examine their energy spectrum and its dependence on the different parameters such as twist angle and interaction strength. Second, we study their transformation properties and quantum numbers under different symmetries. The quantum numbers of the soft modes are decoupled from the details of their energetics, and provide information about the symmetry-breaking orders which can be directly accessed experimentally. These quantum numbers also have the distinct advantage of being universal, in the sense that they do not depend on the details of the theoretical model used. Our results serve to unify and elucidate several findings from theory \cite{YouVishwanath, Kozii, KangFernandes, KoziiNematic, FernandesVanderbos, Chubukov2020, KangVafekPRL, KIVCpaper}, numerics \cite{ShangHF, MacdonaldHF, KIVCpaper} and experiment \cite{PabloNematic, CascadeShahal, CascadeYazdani, ShahalEntropy, YoungEntropy} which all point to the existence of a large number of soft modes that play an important role in the physics of TBG. 

The results of our study of the soft modes can be summarized as follows. In accordance with the above discussion, we find that there are two different types of soft modes. First, there are the approximate Goldstone modes which correspond to broken-symmetry generators of the approximate $\U(4) \times \U(4)$ symmetry identified in Ref. \cite{KIVCpaper}. These include true gapless Goldstone modes associated with broken exact symmetries as well as gapped pseudo-goldstone modes associated approximate symmetry generators. The mass of the latter increases as a function of $\kappa = w_0/w_1$, the ratio of the sublattice-diagonal inter-layer hopping over the sublattice off-diagonal inter-layer hopping, as this ratio governs the strength of the $\U(4)\times\U(4)$-breaking anisotropies in the Hamiltonian \cite{KIVCpaper}. The second type of soft modes are the nematic modes, which carry a non-trivial quantum number associated with three-fold rotations. As we will see below, the mass of the nematic modes decreases as a function of $\kappa$, which follows from the concentration of the Berry curvature of the flat bands near the Gamma point in the mini-Brillouin zone. Interestingly, the realistic value of $\kappa$ is such that both the pseudo-Goldstone and the nematic modes have a small mass on the order of $3 - 4$ meV.

\begin{figure*}
    \centering
    \includegraphics[width = 0.65 \textwidth]{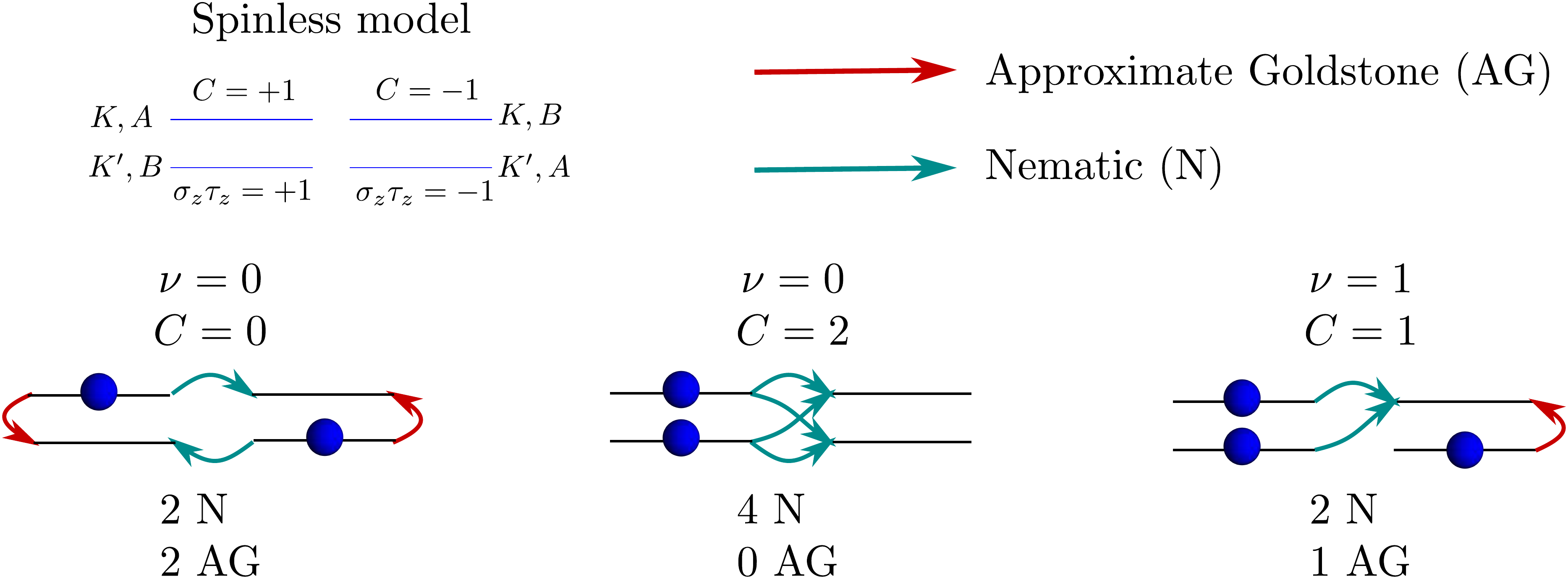}
    \caption{{\bf Schematic illustration of the soft modes in the spinless model}: The spinless model consists of four bands in all: $K, A$ and $K',B$ with Chern number $+1$ and $K, B$ and $K', A$ with Chern number $-1$. The approximate Goldstone (solid red) and nematic (dashed red) modes are shown for the $C = 2$ insulator (anomalous quantum Hall), $C = 0$ insulators (valley polarized, valley Hall or inter-valley coherent) at $\nu = 0$ as well as the $\nu = 1$ insulators which always have $|C| = 1$. The total number of modes is $4 - \nu^2$ with $\frac{4 - \nu^2 + C^2}{2}$ nematic modes and $\frac{4 - \nu^2 - C^2}{2}$ approximate Goldstone modes.}
    \label{fig:SpinlessSM}
\end{figure*}

\begin{figure*}
    \centering
    \includegraphics[width = 0.9 \textwidth]{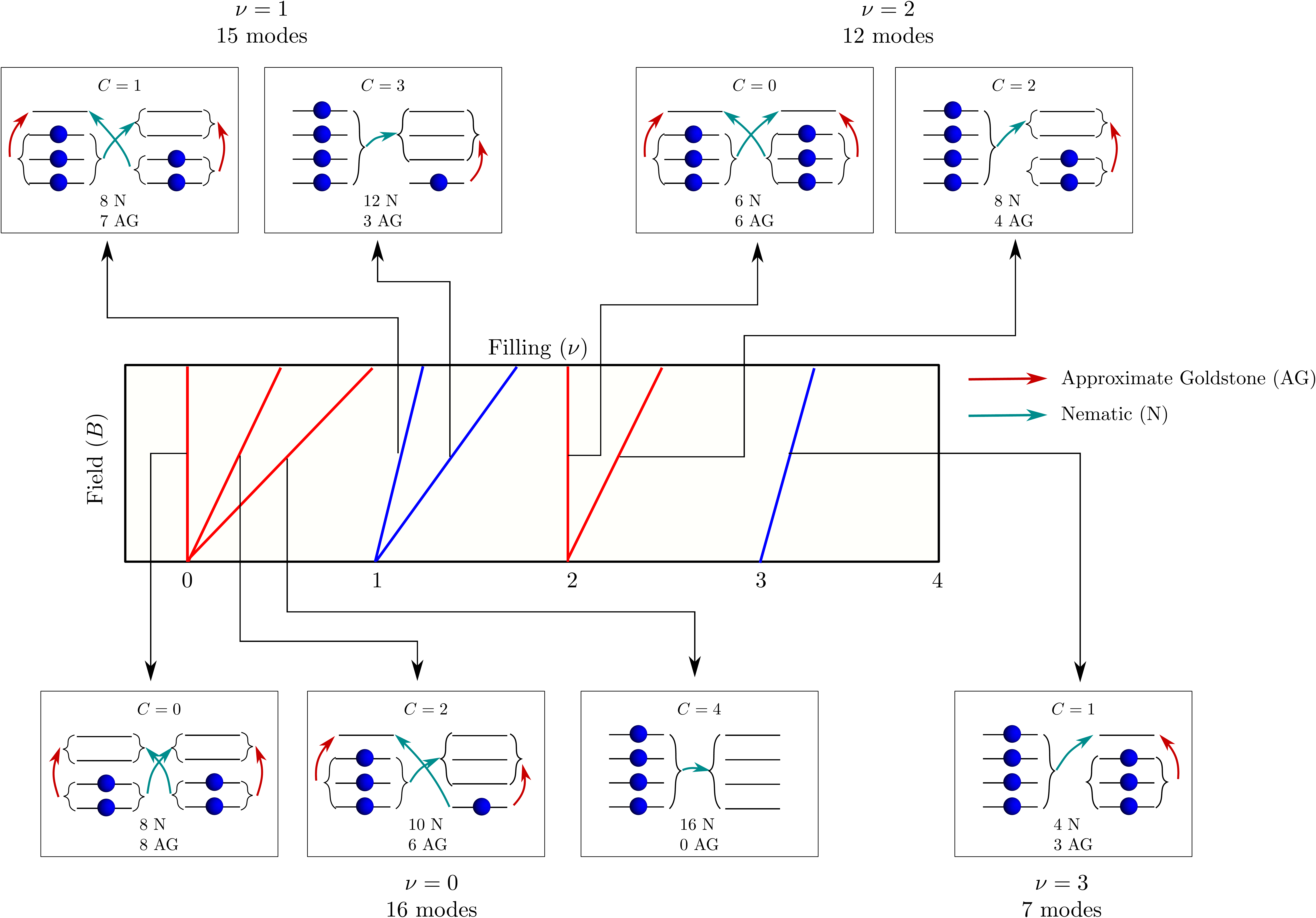}
    \caption{{\bf Soft mode count in MATBG (spinful model)}: A schematic illustration of the count of nematic and approximate Goldstone modes for the different insulating states (including Chern insulators potentially stabilized by an orbital magnetic field)  at different filling $\nu$. The insulating regime in the magnetic field versus filling phase diagram reflects the Chern number. The total soft mode count depends only on the total filling and is given by $16 - \nu^2$. These are split into $\frac{16 - \nu^2 - C^2}{2}$ approximate Goldstone modes (solid) and $\frac{16 - \nu^2 + C^2}{2}$ nematic modes (dashed) as shown in the figure for the different Chern states. }
    \label{fig:SoftModeSpinful}
\end{figure*}

\begin{figure}[h]
    \centering
    \includegraphics[scale=0.6]{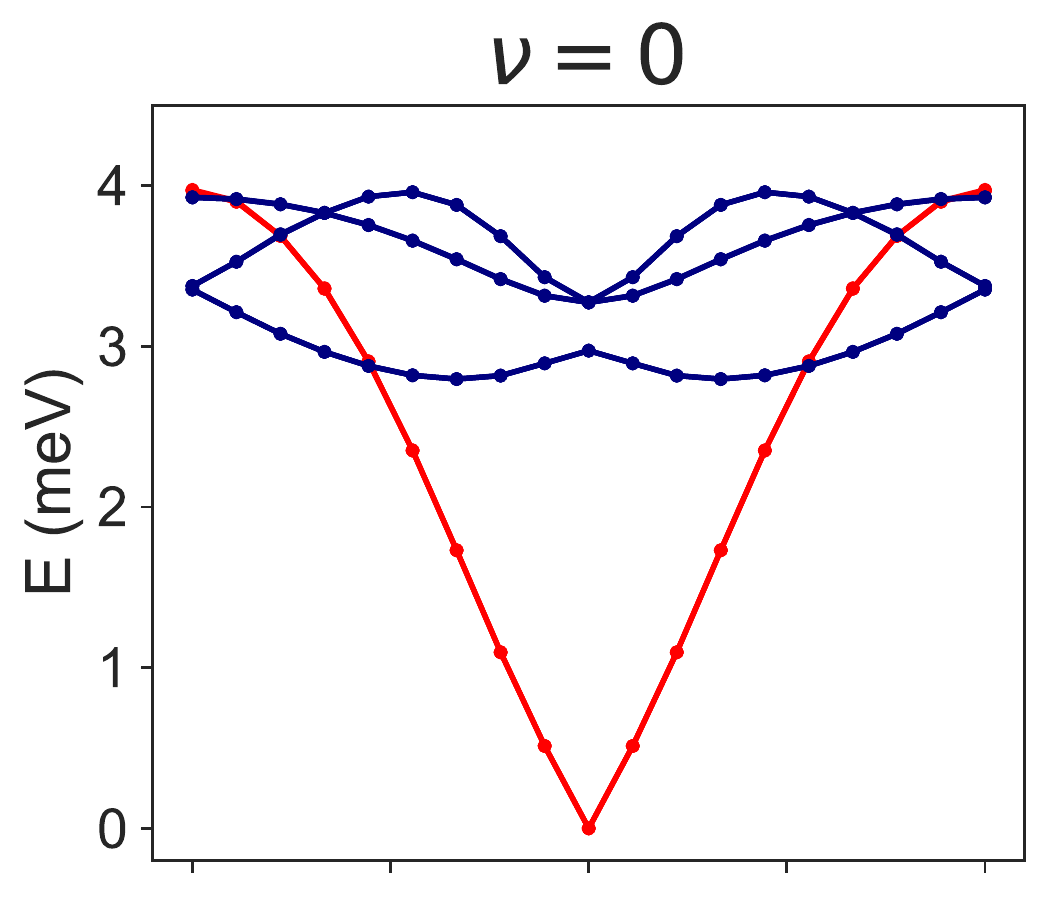}
    \includegraphics[scale=0.6]{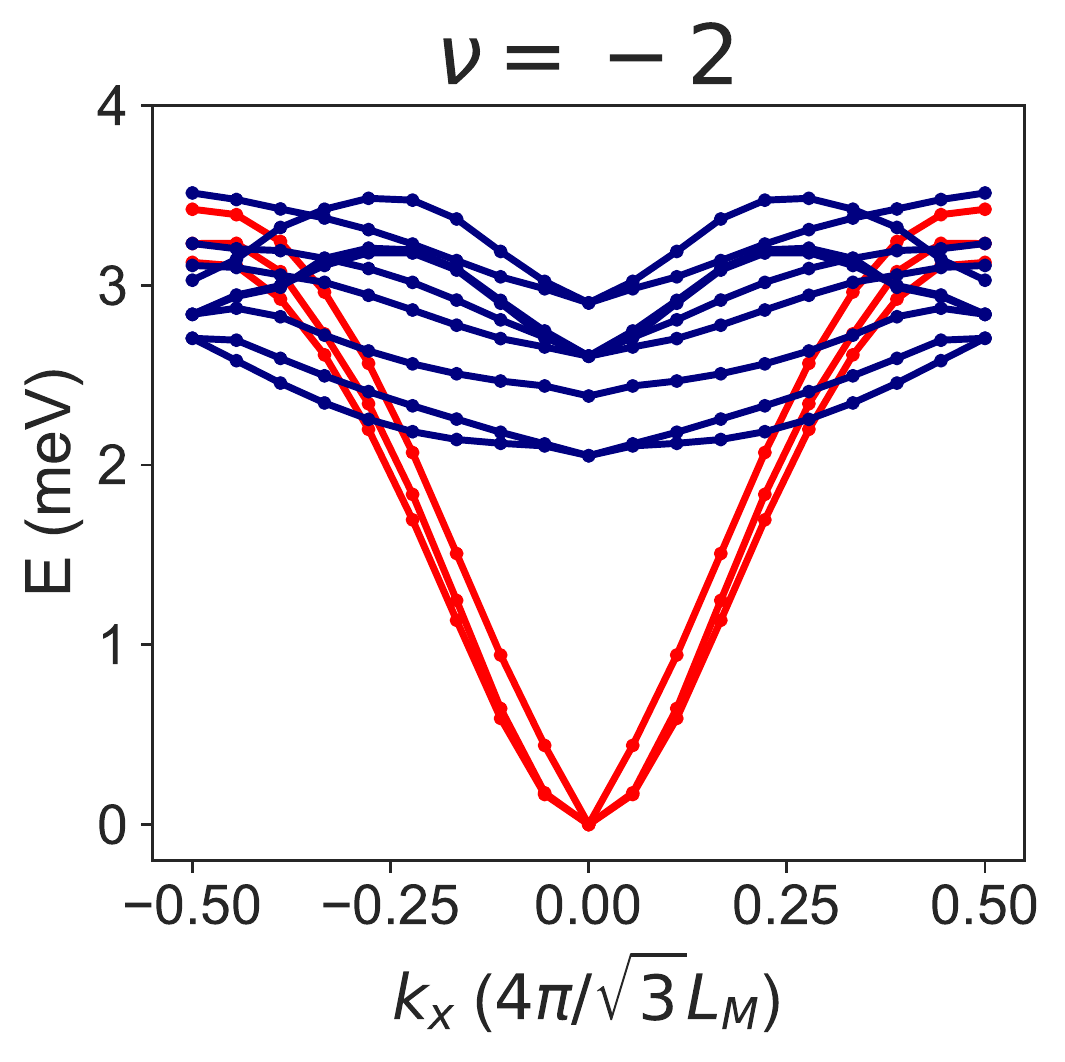}
    \caption{{\bf Time-dependent Hartree-Fock (TDHF) spectra}: TDHF spectra of the insulating K-IVC states at neutrality ($\nu = 0$), and half-filling ($\nu = -2$) along a BZ cut through $x$-direction, connecting $\Gamma$ and $M$ points, in units of $4\pi/\sqrt{3}L_M$, with $L_M$ the moir\'e lattice constant. Goldstone modes are shown in red, while gapped pseudo-Goldstone and nematic modes are shown in blue. For comparison, we note that the mean-field band gaps at $\nu = 0$ and $\nu = -2$ are respectively given by $25$ and $14$ meV, such that all soft modes are well inside these gaps. At $\nu = 0$, all modes are fourfold degenerate leading to 16 modes in total. At $\nu = -2$, there are 12 modes including 2 quadratically dispersing and one linearly dispersing Goldstone mode in addition to 9 gapped modes with the degeneracy pattern $2-1-4-2$. The spectra were obtained using an $(N_x,N_y) = (18, 18)$ grid for $\nu = 0$ and $(N_x,N_y) = (18, 12)$ at $\nu = -2$ using a dielectric constant $\epsilon_r = 12.5$, gate distance $d_s = 40$ nm and $\kappa = 0.75$ at $\theta = 1.08^o$.}
    \label{fig:TDHF}
\end{figure}

We also introduce a simple and intuitive counting rule for obtaining the number of soft modes on top of a particular symmetry-breaking state which is summarized in Figs.~\ref{fig:SpinlessSM} and \ref{fig:SoftModeSpinful}. Our counting rule shows that the total number of soft modes depends only on the total filling, whereas the relative number of approximate Goldstone and nematic modes depends in addition on the Chern number of the insulating ground state. Using the results of Refs. \cite{Nielsen,Watanabe,Watanabe2013,Watanabe2014,Watanabe2017}, we also discuss how to separately count the number of true Goldstone modes with linear and quadratic dispersion at long wavelengths. The counting rule and energetics of the soft modes are substantiated via time-dependent Hartree-Fock (TDHF), which we use to obtain the complete soft-mode spectrum for the inter-valley coherent insulating ground states at charge neutrality $\nu = 0$ and half-filling $\nu = -2$ (cf.~Fig.~\ref{fig:TDHF}). In both cases, we find that the TDHF spectrum produces the expected number of soft modes, and their energies are in good agreement with the anticipated mass scales.

Next, we derive an effective field theory containing only a few parameters which reproduces the soft mode spectrum at long wavelengths, and which allows us to consider the effects of different perturbations, such as for example the inter-valley Hund's coupling. The parameters in the field theory can be fixed by input from Hartree-Fock -- in particular, by calculating the energy splittings between the different symmetry-breaking orders.

In the last section, we switch to a more general symmetry analysis of the soft modes. We provide a general procedure to extract their symmetry transformation properties which is independent of the model details. The results for a selection of experimentally relevant states at different fillings are summarized in Tables \ref{tab:Spinless}, \ref{tab:Spinful}, and \ref{tab:Chern}.

\section{Review}
Let us first recall some basic facts about TBG. It consists of two graphene layers twisted relative to each other by a small angle $\theta$ which generates a long wavelength Moir\'e potential on the scale $a_M \sim a/\theta \sim$ 10 nm. Since the Moir\'e energy scale $\hbar v_F/a_M \sim$ 100 meV is much smaller than the graphene bandwidth, a continuum approximation for the graphene dispersion can be employed where valleys are taken to be decoupled. Combined with the weakness of spin-orbit coupling in graphene, this leads to an effective $\U(2) \times \U(2)$ corresponding to independent $\U(1)$ charge and $\SU(2)$ spin conservation in each valley. As a result, each electron can be labelled by a spin $s = \uparrow, \downarrow$ and a valley $\tau = K, K'$ indices. At angles $\theta \approx 1^o$, a pair of flat bands appears at neutrality for each of the four spin-valley flavor which are connected together by two Dirac points. This leads to 8 bands in total describing the low energy physics. 

\subsection{Band topology and sublattice basis}

The connection between the two flat bands within each flavor is protected by a combination of spinless time-reversal symemtry $\T$ and two-fold rotation symmetry $C_2$ which leave the valley index invariant and ensures the existence of two Dirac points. Three-fold rotation symmetry $C_3$ further pins these Dirac points to the Moire $K$ and $K'$ points. The nontrivial topology of the nearly flat bands which is protected by $C_2 \T$ symmetry can be inferred in different ways. For a  single spin and valley, the two Dirac points have the same chirality -  indicating an obstruction to constructing symmetric Wannier functions for the two bands \cite{Po2018,Zou2018} . The $C_2 \T$ protected band topology is also  captured by the so-called Stiefel-Whitney invariant \cite{StiefelWhitney,  MagicTopologicalBernevig,  Po2018faithful,BernevigTBGTopology}, which leads to a Wannier obstruction and fragile topology when realized in a two band model. This topology is made  explicit as follows \cite{Tarnopolsky,KIVCpaper,XiDai}. In Ref. \cite{Tarnopolsky} it was noticed that in the chiral limit, the flat bands separate into sublattice polarized bands with quantum Hall like wavefunctions. The flat bands can therefore be viewed as a pair of opposite  Chern bands with $C=\pm 1$ living on opposite sublattices which holds even away from the chiral limit. This Chern basis served as the starting point for investigating interaction effects in Ref.~\cite{KIVCpaper}, following which we will label the bands in this new basis by the sublattice $\sigma = A,B$ where they have the largest weight. This results in the picture of Fig.~\ref{fig:Spinful} \cite{KIVCpaper}  consisting of four $C = +1$ Chern bands and four $C=-1$ with approximate $\U(4)$ rotation symmetry in each sector. Note that, in this basis, band dispersion is not neglected. Instead, it appears as momentum dependent tunneling connecting $C_2 \T$ related pairs of opposite Chern bands which breaks the approximate $\U(4) \times \U(4)$ symmetry down to $\U(4)$. 

An important parameter in describing the low energy physics is the ratio between $w_0$, the interlayer tunneling amplitude within the same sublattice (AA/BB)  to $w_1$, the interlayer tunneling amplitude between opposite sublattices (AB/BA)  which we denote by $\kappa =\frac{w_0}{w_1}$. This parameter controls the amount of sublattice polarization ranging from perfect sublattice polarization in the limit $\kappa = 0$ \cite{Tarnopolsky} to a finite but small sublattice polarization for the realistic value $\kappa \approx 0.7-0.8$ \cite{Carr2019}. A non-zero value of $\kappa$ breaks the approximate $\U(4)$ symmetry discussed above down to the physical $\U(2) \times \U(2)$.

\subsection{Ground State at $\nu = 0,\,\pm 1,\,\pm 2,\,\pm 3$}

The hierarchy of approximate symmetries identified in Ref.~\cite{KIVCpaper} is captured by a simple energy expression provided in Ref.~\cite{SkPaper}:
\begin{gather}
    E[Q] = \frac{\rho}{8} \tr (\nabla Q)^2 + \frac{J}{4} \tr (Q \sigma_x)^2 - \frac{\lambda}{4} \tr (Q \sigma_x \tau_z)^2 \nonumber \\
    Q^2 = 1, \qquad [Q, \sigma_z \tau_z] = 0, \qquad \tr Q = 2\nu
    \label{EQz}
\end{gather}
where $Q$ is an $8 \times 8$ matrix describing the filling of the 8 Chern bands. The eigenvalues of $Q$ are $+1$ and $-1$ denoting full and empty bands, respectively. For an insulator at filling $\nu$, $Q$ satisfies $\tr Q = 2\nu$. In addition, $Q$ commutes with the Chern matrix $\sigma_z \tau_z$ such that the Chern number $C = \frac{1}{2} \tr Q \sigma_z \tau_z$ is well defined. It is easy to read off the ground state at different fillings from (\ref{EQz}). At $\nu = 0$, all three terms are minimized by the so-called Krammers intervalley coherent (K-IVC) order given by $Q = \sigma_y \tau_{x,y} s_0$ or any state related to it via the physical $\U(2) \times \U(2)$ symmetry \cite{KIVCpaper}. At $\nu = \pm 2$, all terms are minimized by a spin-polarized version of the K-IVC order $Q = \sigma_y \tau_{x,y} P_\uparrow + \sigma_0 \tau_0 P_\downarrow$ (here $P_{\uparrow/\downarrow}$ denotes the projector on the $\uparrow/\downarrow$ sector) or any state related to it via $\U(2) \times \U(2)$ as explained in Ref.~\cite{KIVCpaper}. At odd filling, all insulating ground states have a finite Chern number \cite{KIVCpaper}. At $\nu = \pm 1$, there are two degenerate manifolds of states with Chern numbers $\pm 1$ or $\pm 3$ including valley symmetry preserving (valley polarized) states or valley symmetry breaking states (IVC). All these states are degenerate on the level of (\ref{EQz}) and their energy competition is likely determined by smaller anisotropies. Finally, at $\nu = \pm 3$, the ground states have Chern number $\pm 1$ which can be either VP or IVC.
 
\section{Physical picture and soft mode count}
\label{sec:count}
We begin by providing a simple physical picture for the soft modes of MATBG yielding a simple rule for their number, symmetry properties and energetics. This will be substantiated by more rigorous analytical and numerical discussions later in the manuscript.

\begin{figure*}
\centering
    \includegraphics[width = 0.5 \textwidth]{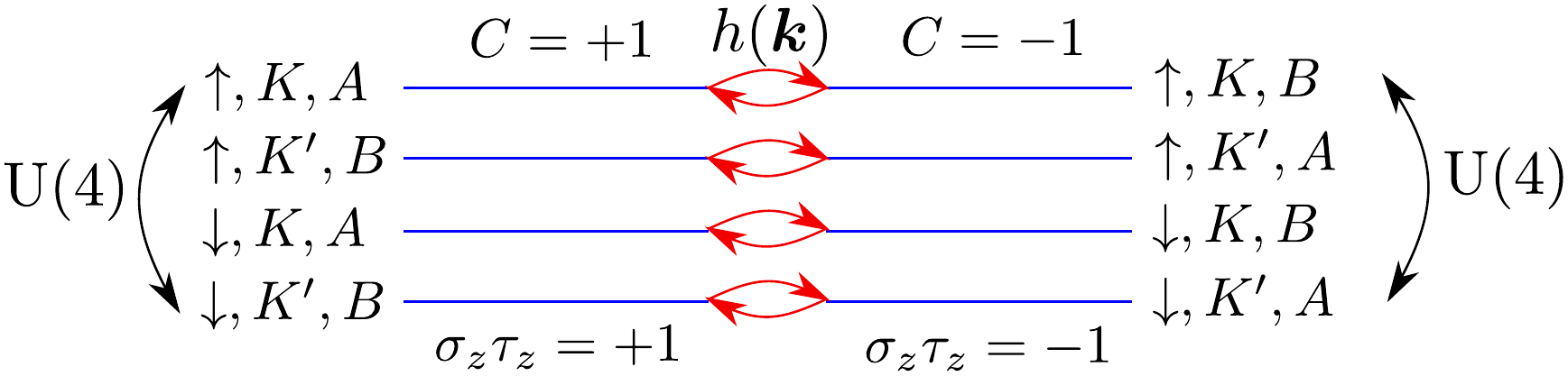}\\
    \vspace{1em}
    \bgroup
    \setlength{\tabcolsep}{0.5 em}
\setlength\extrarowheight{0.3em}
    \begin{tabular}{c|c|c|c|c}
         \hline \hline
        \multirow{4}{*}{Symmetry} & \multicolumn{3}{c|}{Approximation} & \multirow{4}{*}{\makecell{Number of \\ gapless modes}} \\
        \cline{2-4}
        & \multirow{3}{*}{\makecell{Solenoid flux \\ Berry phase \\ $\alpha \rightarrow 0$}}  & \multirow{3}{*}{\makecell{Flat band \\ $J \rightarrow 0$}} & \multirow{3}{*}{\makecell{Sublattice\\ polarized \\ $\lambda \rightarrow 0$}} \\
        & & & & \\
        & & & & \\
        \hline
        $\U(8)$ & \cmark & \cmark & \cmark & $16 - \nu^2$ \\
        $\U(4) \times \U(4)$ & \xmark & \cmark & \cmark & $\frac{16 - \nu^2 - C^2}{2}$ \\
        $\U(4)$ & \xmark & \xmark & \cmark & $\frac{16 - \nu^2 - C^2}{2}$ \\
        $\U(2) \times \U(2)$ & \xmark & \xmark & \xmark & Depends on the state \\
        \hline \hline
    \end{tabular}
    \egroup
    \caption{{\bf Spinful model and hierarchy of symmetries}: Schematic illustration of the spinful model in the sublattice basis with 4 $C = +1$ and 4 $C = -1$ bands (upper pabel) together with a table summarizing the different approximate symmetries, the corresponding approximation, and the number of 'gapless' modes arising from each approximation (lower panel). In the limit when the Berry curvature is concentrated at a single momentum point and can be implemented via a solenoid flux in momentum space, the distinction between the $\pm$ Chern sectors is lost and we have a full $\U(8)$ symmetry. Taking the finite Chern flux density into account but neglecting band dispersion and assuming perfect sublattice polarization leads to a $\U(4) \times \U(4)$ symmetry \cite{KIVCpaper} which is broken down to $\U(4)$ in the presence of dispersion and further down to the physical $\U(2) \times \U(2)$ by including sublattice off-diagonal interaction matrix elements. The parameter $\alpha$, $J$, and $\lambda$ parametrize the strength of explicit symmetry breaking and are defined in Eq.~\ref{EQ} and plotted in Fig.~\ref{fig:NLSMpars}. Each approximate symmetry is associated with a number of soft modes specified in the last column.}
    \label{fig:Spinful}
\end{figure*}

\subsection{Soft modes in the insulating phases}
The soft modes can be understood in a simple manner using the picture of Fig.~\ref{fig:Spinful}. In the limit of strong interactions, the ground state at any integer filling $\nu$ (measure relative to charge neutrality) is obtained by completely filling $4 + \nu$ of the 8 bands \footnote{Note that the approximate $\U(4)$ symmetry allows also for filling arbitrary linear combinations of bands within the same Chern sector.} describing a generalized Chern ferromagnet. The soft modes are simply given by the long wavelength limit of particle-hole excitations where the particle and the hole live within the same Chern sector:
\beq
\hat \phi_{\gamma,\gamma;\alpha\beta;\bq} = \sum_\bk \phi^{\alpha\beta}_{\gamma,\gamma,\bq}(\bk) c_{\gamma,\alpha,\bk}^\dagger c_{\gamma,\beta,\bk + \bq}
\label{aq}, \qquad \gamma = \pm
\eeq
where $c_{\pm,\alpha,\bk}$ denotes the annihilation operator for a flat band state in the $\pm$ Chern sector labelled by an index $\alpha$ going over the 4 bands in a given Chern sector and $\phi^{\alpha,\beta}_{\gamma,\gamma',\bq}(\bk)$ are some functions describing the momentum space profile of the soft modes. These modes correspond to generators of rotations in the spin-valley flavor space and thus carry non-trivial spin or valley quantum numbers. Furthermore, they are gapless goldstone modes in the limit of unbroken $\U(4)$ symmetry within each Chern sector. Realistic anisotropies which break the $\U(4) \times \U(4)$ symmetry down to the physical $\U(2) \times \U(2)$ induce a gap of 1-3 meV depending on the ground state. In the following, we will refer to these modes collectively as approximade goldstone (AG) modes. These includes both the perfectly gapless goldstone (G) modes as well as the gapped soft modes due to the anisotropies which we will refer to as pseudo-goldstone (PG) modes. 

To determine the number of these modes, we note that an insulating ground state is described by filling $n_\pm$ bands in the $\pm$ Chern sector such that the total filling $4 + \nu$ is $n_+ + n_-$ and the total Chern number $C$ is $n_+ - n_-$. Thus, for a given state, the indices $\alpha$ and $\beta$ in $\hat \phi_{\alpha \beta}$ go over the empty and filled states, respectively, yielding a total of $n_+(4 - n_+) + n_-(4 - n_-)$ soft modes. As an example, consider the possible insulating states at $\nu = 0$. If the total Chern number is zero, this corresponds to filling two bands within each sector leading to 4 soft modes per sector; 8 in total. Another example is the zero Chern insulating states at $\nu = 2$ with 3 filled bands in each sector leading to 3 soft modes per sector; 6 in total.

What about particle-hole excitations between different Chern sectors? These can be defined very similarly to Eq.~\ref{aq}:
\beq
\hat \phi_{\gamma, - \gamma;\alpha\beta;\bq} = \sum_\bk \phi_{\gamma,-\gamma,\bq}^{\alpha\beta}(\bk) c_{\gamma,\alpha,\bk}^\dagger c_{-\gamma,\beta,\bk + \bq}, \qquad \gamma = \pm
\label{bq}
\eeq
To understand their energetics, it is useful to perform a particle-hole transformation in one of the Chern sectors \cite{Bultinck19} by defining
\beq
f_{+,\bk} = c_{+,\bk}, \qquad f_{-,\bk} = c_{-,-\bk}^\dagger
\eeq
where we have omitted the $\U(4)$ index within each Chern sector. This transformation flips the Chern number in the $-$ sector so that $f_+$ and $f_-$ are both living in a $+1$ Chern band. In addition, it maps a inter-Chern particle-hole excitation $c_{+,\bk}^\dagger c_{-,\bk}$ into a Cooper pair $\Delta_\bk = f_{+,\bk}^\dagger f_{-,-\bk}^\dagger$. The energetics of inter-Chern particle-hole excitations can be understood in the transformed basis as the energetics of a Cooper pair in a Chern band which is characterized by two main features. First, due to the band topology, the phase of $\Delta_\bk$ winds by $4\pi$ around the Brillouin zone (BZ) which means that $\Delta_\bk$ has at least two $2\pi$ vortices \footnote{Here, we assume a smooth gauge choice}. Second, the Berry curvature of the bands acts like a magnetic field in momentum space. The energy of such Cooper pair can be estimated as \cite{Bultinck19, ShangHF}
\beq
E_\Delta \sim E_C \int d^2 \bk |(\nabla_\bk - 2 i A_\bk) \Delta_\bk|^2
\label{ECooper}
\eeq
where $E_C$ is of the order of the Coulomb energy scale $\sim 10$ meV and $A_\bk$ is the Berry connection. There are two different regimes for the behavior of this energy. In the limit of relatively uniform Berry curvature, the vortices of $\Delta_\bk$ are unscreened yielding a large energy contribution. On the other hand, if the Berry curvature is strongly concentrated at single point in the Brillouin zone, we can evade the large energy penalty by placing the two vortices at this point where they get  almost completely screened.

The question of energetics of the inter-Chern particle-hole excitations is then determined by the distribution of the Berry curvature in momentum space. As shown in Refs.~\cite{ShangHF, Ledwith2019}, the Berry curvature is relatively uniform in the chiral limit $\kappa = 0$ where the wavefunctions resemble the quantum Hall wavefunctions but become more concentrated at the $\Gamma$ point with increasing $\kappa$. At the physical value of $\kappa \approx 0.7-0.8$, the Berry curvature is very sharply peaked at $\Gamma$ leading to a relatively small energy for the inter-Chern modes of the order of a few meVs. The relatively low energy of these modes will be explicitly verified by the time-dependent Hartree-Fock study in Sec.~\ref{sec:TDHF}.

One important property of the inter-Chern modes is that they carry non-zero angular momentum under $C_3$. This can be seen by recalling that the action of $C_3$ in the valley-sublattice space as $C_3 = e^{\frac{2\pi i}{3} \sigma_z \tau_z}$ \cite{Po2018, BernevigTBGTopology, Hejazi}. Thus, electrons in the $\pm$ Chern sector carry a $C_3$ angular momentum of $e^{\pm \frac{2\pi i}{3}}$. As a result, the intra-Chern pseudogoldstone modes (\ref{aq}) always carry zero angular momentum under $C_3$ whereas the inter-Chern modes (\ref{bq}) always carry a non-zero angular momentum at $\bq = 0$ \footnote{Note that there is an ambiguity in defining the phase in intervalley excitations since the $\U_V(1)$ valley charge may in principle transform non-trivially under $C_3$. This can be resolved by taking $C_3$ and $\U_V(1)$ to commute which is assumed throughout this paper}. This can also be seen from the discussion of the energetics where that wavefunction $\Delta(\bk)$ has a winding of $\pm 4 \pi$ around the $\Gamma$ point and as a result transforms as $e^{\pm 2\pi i/3}$ under $C_3$. 
Thus, we will henceforth refer to the inter-Chern soft modes as the nematic (N) modes. We note that the condensation of these modes yields the nematic semimetal identified in previous numerical studies \cite{CaltechSTM, MacdonaldHF, ShangHF, Bultinck19} where the two gapless Dirac cones migrate to the vicinity of the $\Gamma$ point.

The total number of soft modes can then be understood as follows: denoting the number of soft modes between the Chern sector $\gamma = \pm$ and the Chern sector $\gamma' = \pm$ by $N_{\gamma, \gamma'}$, it is easy to see that
\beq
N_{\gamma,\gamma'} = n_\gamma (4 - n_{\gamma'})
\label{Ngg}
\eeq
Then, the number of approximate goldstone (intra-Chern) modes relating the different insulating states is $N_{++} + N_{--}$ whereas the number of nematic (inter-Chern) modes is $N_{+-} + N_{-+}$ leading to a total of $n (8 - n) = 16 - \nu^2$ soft modes. It is worth emphasizing that although the number of nematic and AG modes depends on the given state, the total number of soft modes depends only on the total filling.

As an example, consider the spinless limit where the problem is reduced to a pair of bands in each Chern sector. In this case, the $\U(4) \times \U(4)$ symmetry of the ideal limit reduces to $\U(2) \times \U(2)$ and the physical symmetry is $\U(1) \times \U(1)$ corresponding to total and valley charge conservation. At half-filling, there are several insulating low energy states obtained by filling two out of the four bands. First, we can fill two bands in the same Chern sector leading to a Quantum anomalous Hall state. In this state, there are no AG modes since the $\U(2) \times \U(2)$ symmetry is not spontaneously broken, but there are four nematic modes illustrated in Fig.~\ref{fig:SpinlessSM}. Second, we can fill one band in each sector. This breaks the $\U(2)$ symmetry in each sector to $\U(1) \times \U(1)$. As a result, there are two AG modes in addition to two nematic modes. In the presence of physical anisotropies which break the symmetries down to the valley $\U(1)$ (charge $\U(1)$ is always assumed), some of the AG modes acquire a gap. If the filled bands are valley eigenstates, leading to a valley polarized or valley Hall state, the two AG modes are are gapped PG modes since the physical valley $\U(1)$ symmetry is unbroken. On the other hand, if the filled band is not a valley eigenstate, e.g. a superposition of bands in $K$ and $K'$ valley, this results in an inter-valley coherent state which spontaneously breaks $\U(1)$ valley symmetry. As a result, one of the two AG modes will be a gapless goldstone mode whereas the other will be a gapped PG mode. 

The same discussion can be applied for the spinful case as shown in Fig.~\ref{fig:SoftModeSpinful}. The total number of soft modes depends only on the filling $\nu$ and is given by $16 - \nu^2$. The number of intra-Chern approximate goldstone (AG) and inter-Chern nematic modes  depends in addition on the Chern number $C$ but not on any other detail of the state and is given by
\beq
n_{\rm AG} = \frac{16 - \nu^2 - C^2}{2}, \quad n_{\rm N} = \frac{16 - \nu^2 + C^2}{2}
\label{PGNcount}
\eeq

In summary, this section presented a unified picture for understanding the number and energetics of the soft modes in any insulating symmetry breaking state which will be substantiated by more rigorous numerical and analytical arguments in the following sections. Broadly speaking, we can split the soft modes into intra-Chern approximate goldstone (AG) and inter-Chern nematic modes. The energy of the former is controlled by anisotropies in the manifold of insulating states which break the approximate $\U(4) \times \U(4)$ down to $\U(2) \times \U(2)$, while the energy of the latter is controlled by the distribution of the Berry curvature in momentum space. In the continuum model of TBG close to the magic angle, two types of modes happen to have comparable energies for the physically realistic value $\kappa \approx 0.7$.

\begin{table*}[t]
    \centering
    \small
    \bgroup
\setlength{\tabcolsep}{0.5 em}
\setlength\extrarowheight{0.5em}
    \begin{tabular}{c|c|c|c|c|c|c|c|c|c|c|c}
    \hline \hline
        \multirow{2}{*}{$\nu$} & \multirow{2}{*}{$C$} & \multirow{2}{*}{\makecell{State \\ $Q$}} & \multirow{2}{*}{Symmetry} & \multirow{2}{*}{Type} & \multicolumn{7}{c}{Irrep characters}  \\
        \cline{6-12}
        & & & & & $d$ & $\chi(C_3)$ & $\chi(e^{i \varphi \eta_z})$ & $\chi(C_2)$ & $\chi(M_y)$ & $\chi(M_x)$ & $\chi(\eta_z M_y)$ \\
        \hline
        \multirow{14}{*}{0} & \multirow{11}{*}{0} & \multirow{3}{*}{\makecell{VP \\ $\gamma_0 \eta_z$}} & \multirow{3}{*}{$\{C_3, \U_V(1), M_y, C_2 \T \}$} & \multirow{2}{*}{PG} & 1 & 1 & $e^{2 i \varphi}$ & - & 1 & - & $-1$  \\
        & & & & & 1 & 1 & $e^{2 i \varphi}$ & - & $-1$ & - & 1 \\
        \cline{5-12}
        & & & & N & 2 & $-1$ & $2e^{2 i \varphi}$ & - & 0 & - & 0 \\
        \cline{3-12}
        & & \multirow{2}{*}{\makecell{VH \\ $\gamma_z \eta_z$}} & \multirow{2}{*}{$\{C_3, \U_V(1), M_x, \T \}$} & PG & 2 & 2 & $2 \cos 2 \varphi$ & - & - & 0 & -  \\
        \cline{5-12}
        & & & & N & 2 & $-1$ & $2$ & - & - & 0 & - \\
        \cline{3-12}
        & & \multirow{3}{*}{\makecell{K-IVC \\ $\gamma_z \eta_{x,y}$}} & \multirow{3}{*}{$\{C_3, C_2, i \eta_z M_y, i \eta_z \T \}$} & G & 1 & 1 & - & $-1$ & - & - & 1 \\
        \cline{5-12}
        & & & & PG & 1 & 1 & - & $-1$ & - & - & $-1$  \\ 
        \cline{5-12}
        & & & & N & 2 & $-1$ & - & $-2$ & - & - & 0 \\
        \cline{3-12}
        & & \multirow{3}{*}{\makecell{$\T$-IVC \\ $\gamma_0 \eta_{x,y}$}} & \multirow{3}{*}{$\{C_3, C_2, M_y, \T \}$} & G & 1 & 1 & - & $-1$ & $-1$ & - & 1 \\
        \cline{5-12}
        & & & & PG & 1 & 1 & - & $-1$ & $1$ & - & $-1$ \\
        \cline{5-12}
        & & & & N & 2 & $-1$ & - & 2 & 0 & 0 & 0 \\
        \cline{2-12}
        & \multirow{3}{*}{2} & \multirow{3}{*}{\makecell{QAH \\ $\gamma_z \eta_0$}} & \multirow{3}{*}{$\{C_3, \U_V(1), C_2, M_y \T \}$} & \multirow{3}{*}{N} & 1 & $e^{-2\pi i/3}$ & 1 & $1$ & - & - & - \\
        & & & & & 1 & $e^{-2\pi i/3}$ & 1 & $-1$ & - & - & - \\
        & & & & & 2 & $2e^{-2\pi i/3}$ & $2 \cos 2 \varphi$ & $0$ & - & - & - \\ \hline
        \multirow{6}{*}{1} & \multirow{6}{*}{1} & \multirow{3}{*}{\makecell{VP-QAH \\ $P_+ \eta_0 + P_- \eta_z$}} & \multirow{3}{*}{$\{C_3, \U_V(1), M_y \T \}$} & PG* & 1 & 1 & $e^{2i \varphi}$ & - & - & - & - \\
        \cline{5-12}
         & & & & \multirow{2}{*}{N} & 1 & $e^{-2\pi i/3}$ & 1 & - & - & - & - \\
          & & & & & 1 & $e^{-2\pi i/3}$ & $e^{2i \varphi}$ & - & - & - & - \\
          \cline{3-12}
          & & \multirow{3}{*}{\makecell{IVC-QAH \\ $P_+ \eta_0 + P_- \eta_{x,y}$}} & \multirow{3}{*}{$\{C_3, C_2, M_y \T \}$} & G & 1 & 1 & - & $-1$ & - & - & - \\
        \cline{5-12}
         & & & & \multirow{2}{*}{N} & 1 & $e^{-2\pi i/3}$ & - & $-1$ & - & - & - \\
          & & & & & 1 & $e^{-2\pi i/3}$ & - & 1 & - & - & - \\
          \hline \hline
    \end{tabular}
    \egroup
    \caption{{\bf Symmetry representations for the spinless model}: Detailed symmetry properties for the bosonic soft modes in the spinless for all possible insulators at integer fillings. The labels 'VP', 'VH', 'IVC', and 'QAH' denote valley-polarized, valley Hall, intervalley coherent and quantum anomalous Hall states, respectively. The projector $P_\pm$ projects onto the $\pm$ Chern sector and is defined as $P_{\uparrow/\downarrow} = \frac{1 \pm \gamma_z}{2}$. The soft modes are divided into true goldstone modes (G) corresponding to continuous symmetry breaking, gapped pseudo-goldstone modes (PG) which only become gapless in the absence of anisotropies (i.e. in the $\U(4) \times \U(4)$ limit) and nematic modes (N) which are gapped and transform non-trivially under $C_3$. PG* denote pseudo-goldstone modes which do not correspond to breaking a continuous physical symmetry yet have a very small gap due to the absence of some symmetry allowed terms in the theory. The irrep character corresponding to a symmetry g is denoted by $\chi(g)$ with the representation dimension denoted by $d = \chi(\mathbbm{1})$.}
    \label{tab:Spinless}
\end{table*}

\begin{table*}[t]
    \centering
    \small
    \bgroup
\setlength{\tabcolsep}{0.3 em}
\setlength\extrarowheight{0.3 em}
    \begin{tabular}{c|c|c|c|c|c|c|c|c|c|c|c}
    \hline \hline
        \multirow{2}{*}{$\nu$} & \multirow{2}{*}{State} & \multirow{2}{*}{Symmetry} & \multirow{2}{*}{Type} & \multirow{2}{*}{$n$} & \multicolumn{6}{c}{Irrep characters}  \\
        \cline{6-12}
         & & & & & $d$ & $\chi(e^{i \alpha s_z})$ & $\chi(e^{i \beta P_K s_z})$ & $\chi(e^{i \vartheta P_\downarrow s_z})$ & $\chi(C_3)$ & $\chi(e^{i \varphi \eta_z})$ & $\chi(e^{i \zeta P_\downarrow \eta_z})$ \\
        \hline
         \multirow{22}{*}{0} & \multirow{6}{*}{\makecell{K-IVC\\ $\gamma_z \eta_x s_0$}} &  \multirow{6}{*}{ \makecell{$\{C_3, \SU(2)_S, \eta_z M_y,$\\ $ i \eta_z \T\}$}} &  \multirow{2}{*}{G-I} & 1 & 1 & 1 & - & - & 1 & - & - \\
        &&& & 1 & 3 & $1 + 2 \cos 2\alpha$ & - & - & 3 & - & - \\
        \cline{4-12}
        &&&  \multirow{2}{*}{PG} & 1 & 1 & 1 & - & - & 1 & - & - \\
        &&&  & 1 & 3 & $1 + 2 \cos 2\alpha$ & - & - & 3 & - & - \\
        \cline{4-12}
        &&&  \multirow{2}{*}{N} & 1 & 2 & 2 & - & - & $-1$ & - & - \\
        &&&  & 1 & 6 & $2 + 4 \cos 2\alpha$ & - & - & $-3$ & - & - \\
         \cline{2-12}
         & \multirow{2}{*}{\makecell{VP \\ $\gamma_0 \eta_z s_0$}} & \multirow{2}{*}{ \makecell{$\{C_3, \SU(2)_K, \SU(2)_{K'}, $\\ $U_V(1), M_y ,C_2 \T\}$}} &   PG & 2 & 4 & $2 + 2 \cos 2\alpha$ & $4 \cos \beta$ & $4 \cos \vartheta$ & 4 & $4 e^{2i \varphi}$ & $(1 + e^{i \zeta})^2$ \\
        \cline{4-12}
        &&&  N & 1 & 8 & $4 + 4 \cos 2\alpha$ & $8 \cos \beta$ & $8 \cos \vartheta$ & $-4$ & $4 e^{2i \varphi}$ & $2(1 + e^{i \zeta})^2$ \\
        \cline{2-12}
         & \multirow{3}{*}{\makecell{VH \\ $\gamma_z \eta_z s_0$}} & \multirow{3}{*}{ \makecell{$\{C_3, \SU(2)_K, \SU(2)_{K'}, $\\ $U_V(1), M_x ,\T\}$}} &  PG & 1 & 8 & $4 + 4 \cos 2\alpha$ & $4 \cos \beta$ & $4 \cos \vartheta$ & 8 & $8 \cos 2 \varphi$ & $8 \cos^2 \! \frac{\zeta}{2} \cos \zeta$ \\
        \cline{4-12}
        &&&  \multirow{2}{*}{N} & 1 & 2 & 2 & 2 & 2 & $-1$ & 2 & 2 \\
        &&& & 1 & 6 & $2 + 4 \cos 2\alpha$ & $4 + 2 \cos 2\beta$ & $4 + 2 \cos 2\vartheta$ & $-3$ & $6$ & $2 + 4 \cos \zeta$ \\
        \cline{2-12}
         & \multirow{5}{*}{\makecell{SP \\ $\gamma_0 \eta_0 s_z$}} & \multirow{5}{*}{ \makecell{$\{C_3, S^z_K, S^z_{K'}, $\\ $U_V(1), C_2, M_y ,\T\}$}} & G-II & 1 & 2 & $2 e^{2 i \alpha}$ & $1 + e^{2 i \beta}$ & $1 + e^{2 i \vartheta}$ & 2 & 2 & $2 \cos \zeta$ \\
        \cline{4-12}
        & & & \multirow{2}{*}{PG} & 1 & 2 & $2 e^{2 i \alpha}$ & $1 + e^{2 i \beta}$ & $1 + e^{2 i \vartheta}$ & 2 & 2 & $2 \cos \zeta$ \\
         & & & & 2 & 2 & $2 e^{2 i \alpha}$ & $2 e^{i \beta}$ & $2 e^{i \vartheta}$ & 2 & $2 \cos 2\varphi$ & $2 \cos \zeta$ \\
        \cline{4-12}
         & & & \multirow{2}{*}{N} & 1 & 4 & $4 e^{2 i \alpha}$ & $4 e^{i \beta}$ & $4 e^{i \vartheta}$ & $-2$ & $4 \cos 2\varphi$ & $4 \cos \zeta$ \\
         & & & & 1 & 4 & $4 e^{2 i \alpha}$ & $2 (1 + e^{2i \beta})$ & $2 (1 + e^{2i \vartheta})$ & $-2$ & $4$ & $4 \cos \zeta$ \\
         \cline{2-12}
         & \multirow{6}{*}{\makecell{$\T$-IVC \\ $\gamma_0 \eta_x s_0$}} &  \multirow{6}{*}{ \makecell{$\{C_3, \SU(2)_S, M_y,$\\ $\T\}$}} &  \multirow{2}{*}{G-I} & 1 & 1 & 1 & - & - & 1 & - & - \\
        &&& & 1 & 3 & $1 + 2 \cos 2\alpha$ & - & - & 3 & - & - \\
        \cline{4-12}
        &&&  \multirow{2}{*}{PG} & 1 & 1 & 1 & - & - & 1 & - & - \\
        &&&  & 1 & 3 & $1 + 2 \cos 2\alpha$ & - & - & 3 & - & - \\
        \cline{4-12}
        &&&  \multirow{2}{*}{N} & 1 & 2 & 2 & - & - & $-1$ & - & - \\
        &&&  & 1 & 6 & $2 + 4 \cos 2\alpha$ & - & - & $-3$ & - & - \\
         \hline
         \multirow{23}{*}{2} & \multirow{6}{*}{\makecell{SP K-IVC \\ $\gamma_0 \eta_0 P_\downarrow + \gamma_z \eta_x P_\uparrow$}} &  \multirow{6}{*}{ \makecell{$\{C_3, S^z, \eta_z M_y,$\\ $e^{i \zeta P_\downarrow \eta_z}, \eta_z \T\}$}} & G-I & 1 & 1 & 1 & - & - & 1 & - & 1 \\
         & & & G-II & 1 & 2 & $2e^{-2i \alpha}$ & - & - & 2 & - & $2 \cos \zeta$\\
        \cline{4-12}
        &&&  \multirow{2}{*}{PG} & 1 & 1 & 1 & - & - & 1 & - & 1 \\
        &&&  & 1 & 2 & $2e^{-2i \alpha}$ & - & - & 2 & - & $2 \cos \zeta$ \\
        \cline{4-12}
        &&&  \multirow{2}{*}{N} & 1 & 2 & 2 & - & - & $-1$ & - & 2 \\
        &&&  & 1 & 4 & $4 e^{-2 i \alpha}$ & - & - & $-2$ & - & $4 \cos \zeta$ \\
        \cline{2-12}
        & \multirow{5}{*}{\makecell{SP VP \\ $\gamma_0 \eta_0 P_\downarrow + \gamma_0 \eta_z P_\uparrow$}} &  \multirow{5}{*}{ \makecell{$\{C_3, \SU(2)_K, S^z_{K'},$\\ $U_V(1), M_y,\T\}$}} &  G-II & 1 & 1 & $e^{-2i \alpha}$ & 1 & $e^{-2i \vartheta}$ & 1 & 1 & $e^{-i \zeta}$\\
        \cline{4-12}
        &&&  \multirow{2}{*}{PG} & 1 & 1 & $e^{-2i \alpha}$ & 1 & $e^{-2i \vartheta}$ & 1 & 1 & $e^{-i \zeta}$ \\ &&&  & 2 & 2 & $1 + e^{-2i \alpha}$ & $2 \cos \beta$ & $2e^{-2i \vartheta}$ & 2 & $2 e^{2 i \varphi}$ & $1 + e^{i \zeta}$ \\
        \cline{4-12}
        &&&  \multirow{2}{*}{N} & 1 & 2 & $2 e^{-2i \alpha}$ & 2 & $1 + e^{-2i \vartheta}$ & $-1$ & 2 & $2 e^{-i \zeta}$\\
        &&&  & 1 & 4 & $2 + 2 e^{-2 i \alpha}$ & $4 \cos \beta$ & $4 e^{-2i \vartheta}$ & $-2$ & $4 e^{2i \varphi}$ & $2(1 + e^{i \zeta})$ \\
        \cline{2-12}
        & \multirow{6}{*}{\makecell{SP VH \\ $\gamma_0 \eta_0 P_\downarrow + \gamma_z \eta_z P_\uparrow$}} &  \multirow{6}{*}{ \makecell{$\{C_3, S^z_K, S^z_{K'},$\\ $U_V(1), M_x, \T\}$}} & G-II & 1 & 2 & $2 e^{-2i \alpha}$ & $1 + e^{-2i \beta}$ & $1 + e^{-2i \vartheta}$ & 2 & 2 & $2 \cos \zeta$ \\
        \cline{4-12}
        &&&  PG* & 1 & 2 & $2 e^{-2i \alpha}$ & $1 + e^{-2i \beta}$ & $1 + e^{-2i \vartheta}$ & 2 & $2 \cos 2 \varphi$ & $2 \cos \zeta$ \\ 
        &&& PG & 1 & 2 & $2$ & $2 \cos \beta$ & $2 \cos \vartheta$ & 2 & $2 \cos 2 \varphi$ & 2\\
        \cline{4-12}
        &&&  \multirow{3}{*}{N} & 1 & 2 & 2 & 2 & 2 & $-1$ & 2 & 2 \\
        &&&  & 1 & 2 & $2 e^{-2 i \alpha}$ & $2 e^{-i \beta}$ & $2 e^{-i \vartheta}$ & $-1$ & $2 \cos 2 \varphi$ & $2 \cos \zeta$ \\
        &&&  & 1 & 2 & $2 e^{-2 i \alpha}$ & $1 + e^{-2i \beta}$ & $1 + e^{-2i \vartheta}$ & $-1$ & $2$ & $2 \cos \zeta$ \\
        \cline{2-12}
        & \multirow{6}{*}{\makecell{SP $\T$-IVC \\ $\gamma_0 \eta_0 P_\downarrow + \gamma_0 \eta_x P_\uparrow$}} &  \multirow{6}{*}{ \makecell{$\{C_3, S^z, M_y,$\\ $e^{i \theta P_\downarrow \eta_z}, \T\}$}} &  G-I & 1 & 1 & 1 & - & - & 1 & - & 1 \\
        & & &  G-II & 1 & 2 & $2e^{-2i \alpha}$ & - & - & 2 & - & $2 \cos \zeta$\\
        \cline{4-12}
        &&&  \multirow{2}{*}{PG} & 1 & 1 & 1 & - & - & 1 & - & 1 \\
        &&&  & 1 & 2 & $2e^{-2i \alpha}$ & - & - & 2 & - & $2 \cos \zeta$\\
        \cline{4-12}
        &&&  \multirow{2}{*}{N} & 1 & 2 & 2 & - & - & $-1$ & - & 2\\
        &&&  & 1 & 4 & $4 e^{-2 i \alpha}$ & - & - & $-2$ & - & $4 \cos \zeta$ \\
        \hline \hline
\end{tabular}
    \egroup
    \caption{{\bf Symmetry representations for the soft modes for the $C = 0$ insulators at even integer fillings}: The labels 'VP', 'VH', and 'IVC' have the same meaning as in Table \ref{tab:Spinless} with the additional label 'SP' denoting spin-polarized states. The projector $P_{\uparrow/\downarrow}$ projects on the spin $\uparrow/\downarrow$ sector. As in Table \ref{tab:Spinless}, the soft modes are divided into goldstone (G) which can be of type I (linear dispersion) or type II (quadratic dispersion), pseudo-goldstone (PG), and nematic modes (N). PG* denotes pseudo-goldstone modes which do not correspond to breaking a continuous physical symmetry yet are gapless to a very good approximation. The irrep characters $\chi(g)$ corresponding to $S^z, S^z_{K,K'}$, $U_V(1)$ as well as $C_3$ are given, with $n$ denotes the count of irreps with these characters. The total number of modes is $16 - \nu^2$, half of which are nematic, in agreement with the expression derived in the main text.}
    \label{tab:Spinful}
\end{table*}

\begin{table*}[t]
\hspace{-0.3em}
    \centering
    \small
    \bgroup
\setlength{\tabcolsep}{0.25 em}
\setlength\extrarowheight{0.3em}
\resizebox{\textwidth}{!}{
    \begin{tabular}{c|c|c|c|c|c|c|c|c|c|c|c|c}
    \hline \hline
        \multirow{2}{*}{$\nu$} & \multirow{2}{*}{$C$} & \multirow{2}{*}{State} & \multirow{2}{*}{Symmetry} & \multirow{2}{*}{Type} & \multirow{2}{*}{$n$} & \multicolumn{7}{c}{Irrep characters}  \\
        \cline{7-13}
         & & & & & & $d$ & $\chi(e^{i \alpha s_z})$ & $\chi(e^{i \beta P_K s_z})$ & $\chi(e^{i \vartheta P_{K'} s_z})$ & $\chi(C_3)$ & $\chi(e^{i \varphi \eta_z})$ & $\chi(e^{i \zeta P_\downarrow \eta_z})$ \\
        \hline
        \multirow{3}{*}{0} &\multirow{3}{*}{4} & \multirow{3}{*}{$\gamma_z$} & \multirow{3}{*}{ \makecell{$\{C_3, \SU(2)_K, \SU(2)_{K'}, $\\ $U_V(1), C_2 ,M_y \T\}$}} & \multirow{3}{*}{N} & 2 & 1 & 1 & 1 & 1 & $e^{-\frac{2\pi i}{3}}$ & 1 & 1 \\
         &&&&  & 1 & 6 & $2 + 4 \cos 2\alpha$ & $4 + 2 \cos 2\beta$ & $4 + 2 \cos 2\vartheta$ & $6 e^{-\frac{2\pi i}{3}}$ & $6$ & $2 + 4 \cos \zeta$\\
         &&&& & 1 & 8 & $4 + 4 \cos 2\alpha$ & $8 \cos \beta$ & $8 \cos \vartheta$ & $8 e^{-\frac{2\pi i}{3}}$ & $8 \cos 2 \varphi$ & $8 \cos^2 \! \frac{\zeta}{2} \cos \zeta$ \\
         \hline
        \multirow{13}{*}{1} & \multirow{13}{*}{3} & \multirow{8}{*}{$P_+ + P_- (P_\uparrow \eta_z - P_\downarrow)$} & \multirow{8}{*}{ \makecell{$\{C_3, S^z_{K}, \SU(2)_{K'},$\\ $\U_V(1),M_x \T\}$}} & G-II & 1 & 1 & $e^{2 i \alpha}$ & $e^{2 i \beta}$ & 1 & 1 & 1 & $e^{-i \zeta}$\\
        \cline{5-13}
        & & & & PG* & 1 & 2 & $1 + e^{2 i \alpha}$ & $2 e^{i \beta}$ & $2 \cos \vartheta$ & 2 & $2 e^{2 i \varphi}$ & $1 + e^{i \zeta}$ \\
        \cline{5-13}
        & & & & \multirow{6}{*}{N} & 2 & 1 & 1 & 1 & 1 & $e^{-\frac{2\pi i}{3}}$ & 1 & 1 \\
        & & & & & 1 & 1 & $e^{2 i \alpha}$ & $e^{2 i \beta}$ & 1 & $e^{-\frac{2\pi i}{3}}$ & 1 & $e^{-i \zeta}$ \\
        & & & & & 1 & 2 & $1 + e^{2 i \alpha}$ & $2e^{ i \beta}$ & $2 \cos \vartheta$ & $2e^{-\frac{2\pi i}{3}}$ & $2 e^{-2 i \varphi}$ & $e^{-i \zeta}(1 + e^{-i \zeta})$ \\
        & & & & & 1 & 2 & $1 + e^{2 i \alpha}$ & $2e^{ i \beta}$ & $2 \cos \vartheta$ & $2e^{-\frac{2\pi i}{3}}$ & $2 e^{2 i \varphi}$ & $1 + e^{-i \zeta}$ \\
        & & & & & 1 & 2 & $1 + e^{-2 i \alpha}$ & $2e^{- i \beta}$ & $2 \cos \vartheta$ & $2e^{-\frac{2\pi i}{3}}$ & $2 e^{2 i \varphi}$ & $e^{-i \zeta}(1 + e^{-i \zeta})$ \\
        & & & & & 1 & 3 & $1 + 2 \cos 2 \alpha$ & $3$ & $1 + 2 \cos 2 \vartheta$ & $3 e^{-\frac{2\pi i}{3}}$ & $3$ & $1 + 2 \cos \zeta$ \\
        \cline{3-13}
        & & \multirow{6}{*}{$P_+ + P_- (P_\uparrow \eta_x - P_\downarrow)$} & \multirow{6}{*}{$\{C_3, S_z, e^{i \theta P_\downarrow \eta_z}, M_y \T\}$} & G-I & 1 & 1 & 1 & - & - & 1 & - & 1\\
        & & & & G-II & 1 & 2 & $2e^{2 i \alpha}$ & - & - & 2 & - & $2 \cos \zeta$ \\
         \cline{5-13}
          & & & & \multirow{4}{*}{N} & 4 & 1 & 1 & - & - & $e^{-\frac{2\pi i}{3}}$ & - & 1 \\
          & & & & & 1 & 2 & 2 & - & - & $2 e^{-\frac{2\pi i}{3}}$ & - & $2 \cos 2 \zeta$ \\
         & & & & & 2 & 2 & $2e^{2 i \alpha}$ & - & - & $2e^{-\frac{2\pi i}{3}}$ & - & $2 \cos \zeta$\\
          & & & & & 1 & 2 & $2e^{-2 i \alpha}$ & - & - & $2e^{-\frac{2\pi i}{3}}$ & - & $2 \cos \zeta$ \\
          \hline 
          \multirow{14}{*}{2} & \multirow{14}{*}{2} & \multirow{4}{*}{$P_+ + P_- \eta_z$} & \multirow{4}{*}{ \makecell{$\{C_3, \SU(2)_{K}, \SU(2)_{K'},$\\ $\U_V(1),M_x \T\}$}} & PG* & 1 & 4 & $2 + 2 \cos 2\alpha$ & $4 \cos \beta$ & $4 \cos \vartheta$ & 4 & $4 e^{2 i \varphi}$ & $(1 + e^{i \zeta})^2$\\
          \cline{5-13}
        & & & & \multirow{3}{*}{N} & 1 & 1 & 1 & 1 & 1 & $e^{-\frac{2\pi i}{3}}$ & 1 & 1\\
        & & & & & 1 & 3 & $1 + 2 \cos 2 \alpha$ & 3 & $1 + 2 \cos 2 \vartheta$ & $e^{-\frac{2\pi i}{3}}$ & 3 & $1 + 2 \cos \zeta$ \\
        & & & & & 1 & 4 & $2 + 2 \cos 2\alpha$ & $4 \cos \beta$ & $4 \cos \vartheta$ & $4 e^{-\frac{2\pi i}{3}}$ & $4 e^{2 i \varphi}$ & $(1 + e^{i \zeta})^2$ \\
        \cline{3-13}
        & & \multirow{6}{*}{$P_+ + P_- s_z$} & \multirow{6}{*}{ \makecell{$\{C_3, S^z_{K}, S^z_{K'},$\\ $\U_V(1), C_2, M_x \T\}$}} & G-II & 1 & 2 & $2 e^{2 i \alpha}$ & $1 + e^{2i \beta}$ & $1 + e^{2i \vartheta}$ & 2 & 2 & $2 \cos \zeta$ \\ \cline{5-13}
        & & & & PG* & 1 & 2 & $2 e^{2 i \alpha}$ & $2e^{i \beta}$ & $2 e^{i \vartheta}$ & 2 & $2 \cos 2\varphi$ & $2 \cos \zeta$ \\ \cline{5-13}
        & & & & \multirow{4}{*}{N} & 2 & 1 & 1 & 1 & 1 & $e^{-\frac{2\pi i}{3}}$ & 1 & 1\\
        & & & & & 1 & 2 & 2 & $2 \cos \beta$ & $2 \cos \vartheta$ & $2 e^{-\frac{2\pi i}{3}}$ & $2 \cos 2 \varphi$ & $2 \cos 2 \zeta$\\
        & & & & & 1 & 2 & $2 e^{2 i \alpha}$ & $1 + e^{2 i \beta}$ & $1 + e^{2 i \vartheta}$ & $2 e^{-\frac{2\pi i}{3}}$ & $2$ & $2 \cos \zeta$ \\
        & & & & & 1 & 2 & $2 e^{2 i \alpha}$ & $2e^{ i \beta}$ & $2e^{i \vartheta}$ & $2 e^{-\frac{2\pi i}{3}}$ & $2 \cos 2 \varphi$ & $2 \cos \zeta$ \\
        \cline{3-13}
        & & \multirow{4}{*}{$P_+ + P_- \eta_x$} & \multirow{4}{*}{ $\{C_3, \SU(2),M_y \T\}$} & G-I & 1 & 1 & 1 & - & - & 1 & - & - \\ \cline{5-13}
        & & & & PG & 1 & 3 & $1 + 2 \cos 2 \alpha$ & - & - & 3 & - & - \\ \cline{5-13}
       &  & & & \multirow{2}{*}{N} & 2 & 1 & 1 & - & - & $e^{-\frac{2\pi i}{3}}$ & - & - \\
        & & & & & 2 & 3 & $1 + 2 \cos 2 \alpha$ & - & - & $e^{-\frac{2\pi i}{3}}$ & - & - \\
        \hline
        \multirow{9}{*}{3}  & \multirow{9}{*}{1} & \multirow{5}{*}{$P_+ + P_- (P_\uparrow + P_\downarrow \eta_z)$} & \multirow{5}{*}{ \makecell{$\{C_3, S^z_{K'}, \SU(2)_{K},$\\ $\U_V(1),M_x \T\}$}} & G-II & 1 & 1 & $e^{-2 i \alpha}$ & 1 & $e^{-2 i \vartheta}$ & 1 & 1 & $e^{-i \zeta}$\\ \cline{5-13}
       & & & & PG* & 1 & 2 & $1 + e^{-2i \alpha}$ & $2 \cos \beta$ & $2 e^{-2 i \vartheta}$ & 2 & $2 e^{2 i \varphi}$ & $1 + e^{i \zeta}$ \\ \cline{5-13}
        & & & & \multirow{3}{*}{N} & 1 & 1 & 1 & 1 & 1 & $e^{-\frac{2\pi i}{3}}$ & 1 & 1 \\
        & & & & & 1 & 1 & $e^{-2i \alpha}$ & 1 & $e^{-2 i \vartheta}$ & $e^{-\frac{2\pi i}{3}}$ & 1 & $e^{-i \zeta}$ \\
        & & & & & 1 & 2 &  $1 + e^{-2i \alpha}$ & $2 \cos \beta$ & $2e^{-2 i \vartheta}$ & $2e^{-\frac{2\pi i}{3}}$ & $2 e^{2 i \varphi}$ & $1 + e^{i \zeta}$ \\
         \cline{3-13}
        & & \multirow{4}{*}{$P_+ + P_- (P_\uparrow \eta_x + P_\downarrow)$} & \multirow{4}{*}{\makecell{$\{C_3, S^z, e^{i \theta P_\downarrow \eta_z},$ \\ $M_y \T \}$}} & G-I & 1 & 1 & 1 & - & - & 1 & - & 1\\
       & & & & G-II & 1 & 2 & $2e^{-2 i \alpha}$ & - & - & 2 & - & $2 \cos \zeta$ \\
        \cline{5-13}
        & & & & \multirow{2}{*}{N} & 1 & 2 & $2e^{-2 i \alpha}$ & - & - & $2e^{-\frac{2\pi i}{3}}$ & - & $2\cos \zeta$\\
        & & & & & 2 & 1 & 1 & - & - & $e^{-\frac{2\pi i}{3}}$ & - & 1\\
         \hline \hline
        \end{tabular}
        }
    \egroup
    \caption{{\bf Symmetry representation of the soft modes for the Chern insulators}: Insulators with maximal Chern number $C = |4 - \nu|$ at integer filling $\nu$, which are expected to be stabilized at finite out-of-plane field, are considered. The state $Q$ is described in terms of the projectors $P_\pm = \frac{1 \pm \gamma_z}{2}$ and $P_{\uparrow/\downarrow} = \frac{1 \pm s_z}{2}$. As in Table \ref{tab:Spinless}, the soft modes are divided into goldstone (G), pseudo-goldstone (PG) which can be of type I (linear dispersion) or type II (quadratic dispersion), and nematic modes (N). PG* denotes pseudo-Goldstone modes which do not correspond to breaking a continuous physical symmetry yet are gapless to a very good approximation. The irrep characters $\chi(g)$ corresponding to $S^z, S^z_{K,K'}$, $U_V(1)$ as well as $C_3$ are given, with $n$ denotes the count of irreps with these characters. The total number of modes is $16 - \nu^2$ with $4 (4 - \nu)$ nematic modes in agreement with the expression derived in the main text.}
    \label{tab:Chern}
\end{table*}

\section{Soft mode spectrum}

\subsection{Time-dependent Hartree-Fock}
\label{sec:TDHF}

In this section we discuss how the soft mode spectrum for magic angle graphene can be obtained at the mean-field level via the time-dependent Hartree-Fock (TDHF) formalism. Here, we keep the discussion general and we will not rely on any exact or approximate symmetries of the Hamiltonian. We simply outline how to obtain the TDHF equation, and solve it numerically. The interpretation of the soft mode spectrum in terms of the approximate symmetries and the connection to the non-linear sigma model will be discussed in the next section. We also note that TDHF has previously been used to study collective excitations of quantum anomalous Hall states in moir\'e systems in Refs. \cite{KwanParameswaran,WuDasSarmaTBG,WuDasSarmaTDBG}.

\subsubsection{Formalism}
We start from the following interacting continuum Hamiltonian for magic angle graphene in the BM band basis:

\beq
\hat{H} = \sum_\bk c_\bk^\dagger h(\bk) c_\bk + \frac{1}{2A} \sum_\bq V_\bq \delta \rho_\bq \delta \rho_{-\bq}
\label{Hproj}
\eeq
Here and throughout, we employ a matrix notation where $c_\bk$ denotes a vector of annihilation operators whose components are labelled by the flavor and the band index. In writing this Hamiltonian, we implicitly assume that we have projected the full Hamiltonian into the subspace where most or all of the remote BM valence bands are completely filled, and most or all of the remote BM conduction bands are completely empty. For our numerics, we project out all but two remote valence and conduction bands per spin and valley. For the analytical discussion later on, we project out all remote valence and conduction bands.

The single-particle term $h(\bk)$ in Eq. \eqref{Hproj} contains not only the BM band energies, but also contributions from both the remote valence bands which have been projected out, and from a subtraction term to avoid double counting of certain interaction effects \cite{MacdonaldHF, KIVCpaper}. The second term in Eq. \eqref{Hproj} corresponds to the Coulomb interaction, for which we use a dual-gate screened potential $V_{\bq} = \tanh(d_sq)/2\epsilon_0\epsilon_rq$ with dielectric constant $\epsilon_r$ and gate distance $d_s$. The Fourier components of the projected charge density operator are given by 

\beq
\delta \rho_\bq = \sum_\bk c_\bk^\dagger \Lambda_\bq(\bk) c_{\bk + \bq}\,,
\eeq
where $\Lambda_{\bq}(\bk)$ is the matrix of form factors in the band and flavor index defined in terms of overlaps between the cell-periodic parts of the BM Bloch states:

\begin{equation}
    [\Lambda_\bq(\bk)]_{\alpha,\beta} = \langle u_{\alpha,\bk} | u_{\beta,\bk + \bq} \rangle
\end{equation}
with $\alpha$, $\beta$ ranging over flavor and band indices. 
The starting point of TDHF is a solution of the Hartree-Fock self-consistency equation, described by the following correlation matrix

\begin{equation}
    P_{\alpha\beta}(\bk) = \langle c^\dagger_{\beta,\bk}c_{\alpha,\bk}\rangle\, ,
    \label{Pk}
\end{equation}
which projects onto the occupied states in the mean-field band spectrum. If we write the Hartree-Fock Hamiltonian constructed from $P(\bk)$ as $H_{\rm SC}\{P\}(\bk)$ (see App. \ref{app:TDHF} for a definition of this Hamiltonian), then self-consistency means that the following equation should be true:

\begin{equation}\label{selfconst}
    [P(\bk),H_{\rm SC}\{P\}(\bk)] = 0
\end{equation}
The self-consistency condition has an intuitive physical interpretation, which becomes clear after using Eq. \eqref{selfconst} to show that the following equality holds:

\begin{equation}\label{twopart}
   \sum_{\lambda\gamma} P_{\lambda\alpha}(\bk)P^\perp_{\beta\gamma}(\bk)\langle c^\dagger_{\lambda,\bk}c_{\gamma,\bk}\hat{H}\rangle_{HF} = 0\, ,
\end{equation}
where $P^\perp(\bk) = \mathds{1}-P(\bk)$, and $\langle \cdot \rangle_{HF}$ means that we take the expectation value with respect to the Hartree-Fock ground state Slater determinant with correlation matrix $P(\bk)$. Eq. \eqref{twopart} implies that self-consistency is equivalent to the condition that $\hat{H}$ should create at least two particle-hole excitations when acting on the ground state Slater determinant.

Let us now define the following bosonic operators:
\begin{equation}
    \hat{\phi}_\bq = \sum_\bk c^\dagger_\bk \phi_{\bq}(\bk)c_{\bk+\bq}
    \label{HatPhi}
\end{equation}
where $\phi_\bq(\bk)$ is a matrix in the flavor and band indices. The goal of TDHF is to find those operators $\hat{\phi}_\bq$ which (1) are a superposition of creation and annihilation operators of particle-hole excitations of the mean-field band spectrum, and (2) which correspond to eigenmodes satisfying $i \partial_t \hat{\phi}_\bq = [\hat{H},\hat{\phi}_\bq] = \omega_\bq \hat{\phi}_\bq$ at the mean-field level. In general, the commutator $[\hat{H},\hat{\phi}_\bq]$ will contain both terms with two and four fermion operators. So working at the mean-field level means in practice that we map the four-fermion terms in the commutator to two-fermion terms by performing all partial Wick contractions with $P(\bk)$ which leave precisely two fermion operators uncontracted. As discussed in detail in App. \ref{app:TDHF}, the resulting eigenvalue problem for obtaining the soft mode spectrum $\omega_\bq$ is equivalent to diagonalizing a quadratic boson Hamiltonian. The fact that this mean-field boson Hamiltonian is quadratic is consistent with the fact that $\hat{H}$ creates at least two particle-hole excitations.

\subsubsection{Results}

In Fig. \ref{fig:TDHF} we show the TDHF spectra for the Kramers inter-valley coherent (K-IVC) insulators at charge neutrality, and at a flat band filling of two electrons per moir\'e unit cell. The K-IVC state was introduced in Ref. \cite{KIVCpaper}, and we will discuss its properties in more detail below. For now, it suffices to mention that both at $\nu = 0$ and $\nu = -2$ the K-IVC state breaks the $\U_V(1)$ symmetry, and that the K-IVC state at $\nu = -2$ is a spin polarized version of the one at neutrality. The mean-field band gaps for the K-IVC states at $\nu = 0$ and $\nu = -2$ are respectively given by $25$ and $14$ meV. From Fig. \ref{fig:TDHF} it is clear that all soft modes lie well below these band gaps.   

The TDHF spectrum at $\nu = 0$ is shown in the top panel of Fig. \ref{fig:TDHF}. Note that the modes are exactly four-fold degenerate at every momentum point, such that there are 16 soft modes in total, as predicted by the counting rule discussed in Sec. \ref{sec:count}. To understand the four-fold degeneracy, we assume without loss of generality that the K-IVC state at neutrality does not break the global spin rotation symmetry (we will come back to this point in more detail in Sec. \ref{sec:softmodefieldth}). In this case, the soft modes are labeled by their spin quantum number, and every mode appears both as a spin singlet and a spin triplet, because microscopically it consists of two spin-$1/2$ fermion operators. For the most general interaction compatible with the symmetries, this implies that the soft mode spectrum will consist of 4 singlet modes and 4 three-fold generate triplet modes (see Sec.~\ref{sec:symmreps}). However, for a density-density interaction we find that the singlet and triplet modes are not split, resulting in a four-fold degeneracy. In particular, this implies that there are four degenerate Goldstone modes, which are shown in red in the top panel of Fig. \ref{fig:TDHF}. For the spin-singlet K-IVC state, these Goldstone modes are associated with the broken $\U_V(1)$ symmetry (spin-singlet mode corresponding to generator $\tau_z$), and with the broken symmetry of opposite spin rotations in the different valleys (spin-triplet mode corresponding to generators $\tau_zs_x, \tau_zs_y$ and $\tau_zs_z$).  

Fig. \ref{fig:TDHF} also shows an additional degeneracy at the $\Gamma$ point between the two upper branches of the $\nu = 0$ soft mode spectrum. These modes correspond to the nematic modes, and as we explain in Sec. \ref{sec:symmreps}, the additional degeneracy at $\Gamma$ is caused by the valley-diagonal mirror symmetry which interchanges the modes with opposite $C_3$ angular momenta. We also want to point out that beyond the four-fold degeneracies associated with the global spin rotation symmetry there are no additional degeneracies at the $M$ points, even though the two lower and upper soft mode branches are very close in energy there.

For the K-IVC state at $\nu = -2$, we find 12 different soft modes, which again agrees with the counting rule discussed in Sec. \ref{sec:count}. Compared to the spectrum at neutrality, which has four linearly dispersing Goldstone modes, one of the main differences is that at $\nu = -2$ there are three Goldstone modes, two of which are quadratically dispersing at small $\bq$, and one which has a linear dispersion. The linearly dispersing mode is again associated with the broken $\U_V(1)$ symmetry (generator $\tau_z$), while the two quadratically dispersing modes are the result of broken spin rotation symmetry (generators $s_x$ and $s_y$). At the $\Gamma$-point, the degeneracies of the gapped soft modes are $2 - 1 - 4 -2$. We will explain the origin of these degeneracies in Sec. \ref{sec:symmreps}.

\subsection{Flat band projection and soft mode energetics}
\label{sec:Energetics}

In this section, we provide an analytical understanding for the soft mode energetics by projecting the Hamiltonian (\ref{Hproj}) onto the flat bands and employing the approximate $\U(4) \times \U(4)$ symmetry to understand its energetics. For the analytic treatment, we will find it more convenient to switch from the Hamiltonian approach of the previous section to a Lagrangian approach as explained below.

 We start by restricting ourselves to the 8 flat bands labelled by a sublattice $\sigma = \pm = A/B$,  valley $\tau = \pm = K/K'$ and spin $s = \uparrow, \downarrow$ indices, with each band having Chern number $\sigma \tau$. It is more convenient to define an alternative basis where the bands are labelled by a Chern index $\gamma = \pm$ and a pseudospin index $\eta$ within each Chern sector \cite{SkPaper}:
\beq
\gamma_{x,y,z} = (\sigma_x, \sigma_y \tau_z, \sigma_z \tau_z), \qquad \eta_{x,y,z} = (\sigma_x \tau_x, \sigma_x \tau_y, \tau_z)
\eeq
The projector $P(\bk)$ defined in (\ref{Pk}) is now an $8 \times 8$ matrix. We will find it convenient to define the matrix $Q(\bk)$ as  $Q(\bk) = 2P(\bk) - 1$, which can be written directly in terms of the band-projected creation/annihilation operators as
\beq
Q_{\alpha \beta}(\bk) = \langle [c_{\alpha,\bk}^\dagger, c_{\beta,\bk}] \rangle
\label{Qk}
\eeq
Any Slater determinant state is completely characterized by the $Q$ matrix which satisfies $Q(\bk)^2 = 1$. This condition means that the eigenvalues of $Q(\bk)$ are $\pm 1$ at each $\bk$, such that there exists a basis where each state is either full $(+1)$ or empty $(-1)$. If we further restrict to insulating or semimetallic states at an integer filling $\nu$, then the number of $+1$ and $-1$ eigenvalues is independent of $\bk$ leading to the additional condition $\tr Q = 2\nu$. In the following, we will additionally assume that $Q$ is $\bk$-independent which is true in the $\U(4) \times \U(4)$ limit and holds to a good approximation in the realistic limit \cite{KIVCpaper}. We will discuss later how this assumption can be lifted when considering the effective field theory.

We can parametrize the fluctuations around a given Slater determinant state $|\psi_Q \rangle$ described by the matrix $Q$ by writing
\beq
|\psi_Q(\phi) \rangle = e^{i \sum_\bq \hat \phi_\bq} |\psi_Q \rangle, 
\eeq
where $\hat \phi_\bq$ is defined in (\ref{HatPhi}) in terms of a matrix-valued function $\phi_\bq(\bk)$ which anticommutes with $Q$ and acts in the space of flat bands. $\phi_\bq(\bk)$ satisfies $\phi^\dagger_{\bq}(\bk) = \phi_{-\bq}(\bk + \bq)$ and can be expanded in terms of the $2 (16 - \nu^2)$ generators of $\U(8)$ which anticommute with $Q$, which we denote by $t_\mu$, as
\beq
\phi_\bq(\bk) = \sum_{\mu=1}^{2(16 - \nu^2)} \phi^\mu_\bq(\bk) t_\mu, \qquad \{t_\mu, Q\} = 0
\eeq
where $t_\mu^\dagger = t_\mu$.
The energy of the state $|\psi_Q(\phi) \rangle$ is
\beq
E(Q, \phi) = \langle \psi_Q(\phi)| \hat H | \psi_Q(\phi) \rangle
\eeq
where $\hat H$ is the flat-band projected Hamiltonian. Expanding $E(Q, \phi)$ in powers of $\phi$, we find that the linear term vanishes if and only if $Q$ is a solution to the Hartree-Fock self-consistency equation, which we will assume here. The quadratic term gives the leading contribution in $\phi$
\beq
E(Q,\phi) = \sum_{\bq,\bk,\bk',\mu,\nu} \phi^\mu_{\bq}(\bk)^* \H^{\mu,\nu}_\bq(\bk,\bk') \phi^\nu_{\bq}(\bk')
\label{EQphi}
\eeq
We can expand $\phi^\mu_\bq(\bk)$ in terms of eigenfunctions of $\H_\bq$ as
\begin{gather}
    \phi_\bq^\mu(\bk) = \sum_n a_{n, \bq} \phi_{n,\bq}^\mu(\bk), \label{PhiExp} \\ \sum_{\bk', \nu} H_\bq^{\mu, \nu}(\bk, \bk') \phi_{n,\bq}^\nu(\bk') = \varepsilon_{n, \bq} \phi_{n,\bq}^\mu(\bk)
    \label{Hphi}
\end{gather}
Here, $n$ and $\bq$ are labels for the wavefunctions whereas $\mu$ and $\bk$ are internal indices such that the wavefunction can be understood as a vector in $\mu$ and $\bk$. Substituting (\ref{PhiExp}) in (\ref{EQphi}) leads to
\beq
E(Q,\phi) = \sum_{\bq,n} \varepsilon_{n,\bq} a_{n, \bq}^* a_{n, \bq}, 
\label{EQW}
\eeq
The quantum theory is obtained by promoting the fluctuations $a_{n,\bq}$ to be dynamical and defining the Lagrangian
\beq
\L = \langle \psi(\phi)| i\frac{d}{dt} - \hat H | \psi(\phi) \rangle
\label{Lphi}
\eeq
The second term yields the energy $E(Q,\phi)$ (\ref{EQW}), whereas the first yields
\begin{gather}
     \langle \psi(\phi)| i\frac{d}{dt} | \psi(\phi) \rangle = i\sum_{\bq,n, m} \rho_{nm,\bq} a^*_{n,\bq}  \partial_\tau a_{m, \bq}, \label{rho} \\ \rho_{n m,\bq} = \frac{1}{2} \sum_{\mu,\nu} \tr Q t_\mu t_\nu \sum_{\bk}  [\phi_{n,\bq}^\mu(\bk)]^* \phi_{m,\bq}^\nu(\bk)
\label{Eqn:rho}
\end{gather}
The matrix $\rho$ satisfies $\rho^T_{-\bq} = -\rho_\bq$ and thus defines a symplectic structure that pairs up different modes as canonically conjugate variables. By going to Fourier space and writing the action as
\begin{gather}
    S = \int d\tau \L = \sum_{n,m,\bq,\omega} a_{n, \bq, \omega}^* M_{nm, \bq, \omega} a_{m,\bq,\omega}, \label{SM} \\ M_{nm,\bq, \omega} =  \rho_{nm,\bq} \omega - \delta_{n,m} \varepsilon_{n,\bq}
\end{gather}
The soft mode spectrum is obtained by taking the positive solutions $\omega = \omega_\bq$ of the equation $\det M_{\bq,\omega} = 0$ \cite{Auerbach}. It is worth noting that up to this point, we have not made any assumption about exact or approximate symmetries. 

Let us now consider the projection of the Hamiltonian \eqref{Hproj} onto the flat bands. In this limit, the effect of the remote bands is included only through the renormalization of the band dispersion $h(\bk)$ \cite{MacdonaldHF, Repellin19, ShangHF, Bultinck19}. 
Following Ref.~\cite{KIVCpaper}, we can start by considering the $\U(4) \times \U(4)$ symmetric limit where the form factor has the simple form
\beq
\Lambda_\bq(\bk) = F_\bq(\bk) e^{i \Phi_\bq(\bk) \gamma_z}
\label{LambdaSymm}
\eeq
As a result, the ground states of the Hamiltonian $\H$ are Slater determinant states characterized by $\bk$-independent $Q$ satisfying $[Q, \gamma_z] = 0$ \cite{KIVCpaper}. 

To make further progress, we switch to a different basis for the generators of $\U(8)$ that makes the form of the soft mode Hamiltonian as simple as possible. This basis, which we denote by $\{r^{\gamma \gamma'}_\alpha\}$, is labelled by $\gamma, \gamma' = \pm$ and $\alpha = 1,\dots, 2 N_{\gamma,\gamma'}$. The generators $\{r^{\gamma, \gamma'}_\alpha \}$ correspond to particle-hole excitations from Chern sector $\gamma$ to Chern sector $\gamma'$. We note that the new basis of generators is not hermitian since the hermitian conjugate of $r^{\gamma ,\gamma'}_\alpha$ belongs to the set $\{r^{\gamma',\gamma}_\alpha\}$. At the end of the calculation, we can transform back to the Hermitian basis $\{t_\mu\}$. 

As shown in Appendix \ref{app:DerivationHq}, the condition $[Q, \gamma_z] = 0$ implies that $\H_\bq$ is block diagonal in the $(++,+-,-+,--)$ space with blocks denoted by $\H_{\gamma,\gamma';\bq}$. 
Furthermore, the generators $r^{\gamma,\gamma'}_\alpha$ can be chosen such that $\H_{\gamma,\gamma'}^{\alpha,\beta}$  is proportional to $\delta_{\alpha,\beta}$ for all $\gamma$ and $\gamma'$ (see Appendix \ref{app:DerivationHq}). This means that we can relabel the soft mode wavefunctions by splitting the index $n$ into $(\gamma_n, \gamma'_n, \alpha_n, l_n)$ where $\gamma_n, \gamma'_n$, and $\alpha_n$ label the generator corresponding to the eignfunction $\phi_n$ and $l_n$ is an integer $l_n \geq 0$ labelling the set of eigenfunctions corresponding to the same generators such that the energy $\epsilon_{\gamma, \gamma', \alpha, l_n}(\bq)$ is an increasing function of $l_n$. Due to the simple form of the Hamiltonian (appendix \ref{app:DerivationHq}), these wavefunctions have the form:
\beq
\phi_{\gamma_1,\gamma_2,\alpha,l,\bq}^{\gamma_3, \gamma_4, \beta}(\bk) = \delta_{\gamma_1, \gamma_2} \delta_{\gamma_3, \gamma_4} \delta_{\alpha,\beta} \psi_{\gamma_1,\gamma_2;l,\bq}(\bk)
\label{phinq}
\eeq
where $\psi_{\gamma,\gamma';l,\bq}(\bk)$ is a scalar wavefunction (with no vector indices) labelled only by the generator Chern sector $\gamma, \gamma'$ and an integer $l$ for a given $\bq$. Eq.~\ref{phinq} means that $\phi_{\gamma_1,\gamma_2,\alpha,l,\bq}$ which is a vector in the index $\mu = (\gamma_3, \gamma_4, \beta)$ has only one non-vanishing component. As we will show later, the low energy soft modes are obtained by restricting to the lowest lying eigenfunctions $l = 0$.

\begin{figure}
    \centering
    \includegraphics[width = 0.4 \textwidth]{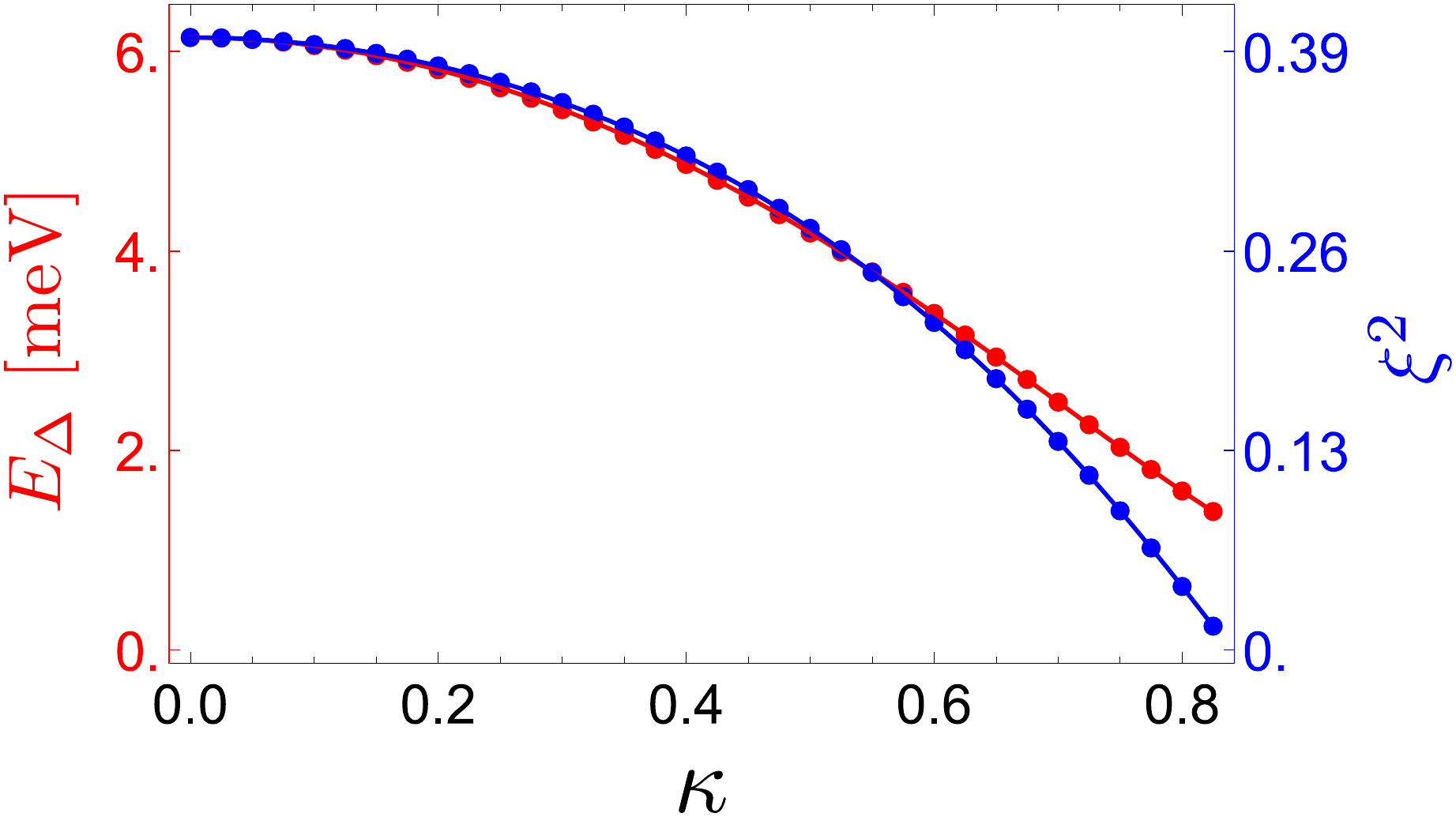}
    \caption{{\bf Gap of the inter-Chern nematic soft mode}: Plot of $E_\Delta$, the smallest eigenvalue of $\H^{+-}$ as defined in Eq.~\ref{Em0} (red) together with the quantity $\xi^2 = \frac{1}{2\pi} \langle \Omega(\bk) \bk^2 \rangle$ (blue) which measures the spread of Berry curvature, as a function of $\kappa = w_0/w_1$. Here, $\Omega(\bk)$ denotes the Berry curvature and $\langle \cdot \rangle$ denote the BZ average with $\frac{1}{2\pi} \langle \Omega(\bk) \rangle = 1$. $E_\Delta$ decreases with increasing $\kappa$ as the Berry curvature becomes more concentrated \cite{ShangHF, Ledwith2019}. Here, we used the dielectric constant $\epsilon = 12.5$ and gate distance $d = 20$ nm. }
    \label{fig:EDelta}
\end{figure}

The expressions for $\H_{\gamma,\gamma'}$ can be further simplified in the limit $\bq = 0$. To simplify the notation, we will drop the $\bq$ dependence of the Hamiltonian $\H_{\gamma,\gamma';\bq}$ and the wavefunctions $\psi_{\gamma,\gamma';l,\bq}$ whenever $\bq = 0$. The Hamiltonian $\H_{\gamma,\gamma'}$ is given by:
\begin{multline}
\H_{\gamma,\gamma'}(\bk,\bk') =  \frac{1}{A} \sum_{\bq} V_{\bq} F_{\bq}(\bk)^2 [\delta_{\bk,\bk'} \\ - \delta_{\bk',[\bk + \bq]} e^{i(\gamma-\gamma') \Phi_{\bq}(\bk)}]
\label{Hgamma}
\end{multline}
where the sum over $\bq$ extends over all momenta, and,  $[\bk]$ equals $\bk$ modulo a reciprocal lattice vector and lies in the first Brillouin zone. For the intra-Chern fluctuations, $\gamma = \gamma'$, we can see immediately that a constant function is a zero eigenfunction of $\H_{\gamma, \gamma}$ for $\gamma = \pm$. This corresponds to the Goldstone modes of the $\U(4) \times \U(4)$ symmetry breaking. Furthermore, we can verify by an explicit calculation that all other eigenvalues of $\H_{\gamma, \gamma}$ have a relatively large mass $\sim$ 10-20 meV and can thus be neglected at low energies. For inter-Chern fluctuations, $\gamma = -\gamma'$, the situation is different. Due to the phase factor in the second term, the eigenstates of $\H_{\gamma,-\gamma}$ always have a finite gap. This was shown in the supplemental material of Ref.~\cite{KIVCpaper} and we will reproduce this argument below. For definiteness, let us focus on the $+-$ sector. We begin by noting that the lowest energy state $\Delta(\bk) = \psi_{+-,l_n = 0}(\bk)$ is obtained by minimizing the expectation value:
\begin{multline}
    E_\Delta = \langle \Delta| \H_{+-} | \Delta \rangle = \frac{1}{AN} \sum_{\bq,\bk} V_{\bq} F_{\bq}(\bk)^2 \\ \times [\Delta(\bk)^2 - \Delta(\bk) \Delta(\bk + \bq) e^{2i \Phi_{\bq}(\bk)}]
    \label{Em0}
\end{multline}
This can be further simplified by assuming the magnitude of the form factor decays relatively quickly with the relative momentum $\bq$ which enables us to expand the expression inside the sum in $\bq$ leading to
\begin{gather}
   E_\Delta = \frac{1}{N} \sum_\bk E_C(\bk) |(\nabla_\bk - 2 i A_\bk) \Delta(\bk)|^2 \\
    E_C(\bk) = \frac{1}{A} \sum_\bq \bq^2 V_{\bq} F_{\bq}(\bk)^2
\end{gather}
Here, we used $\Phi_\bq \approx \bq \cdot A_\bk + O(\bq^2)$ with $A_\bk$ denoting the Berry connection $A_\bk = -i \langle u_{+,\bk}|\nabla_\bk|u_{+,\bk} \rangle$ where $u_{+,\bk}$ denotes the wave-function for any of the bands within the $+$ Chern sector (which are all equal due to $\U(4) \times \U(4)$ symmetry). If we further assume $E_C(\bk)$ depends weakly on $\bk$, we can pull it out of the $\bk$ sum and get an expression identical to the energy of a Cooper pair in magnetic field as in Eqn. (\ref{ECooper}). This energy expression can be understood as the energy of a superconducting vortex if we identify the momentum $\bk$ with the real coordinate, $E_C$ with the superfluid stiffness, and identify $\xi^2 = \frac{1}{2} \langle \Omega(\bk) \bk^2 \rangle$ with the area of the vortex core in momentum space (here $\Omega(\bk)$ is the Berry curvature and $\langle \cdot \rangle$ denote BZ average). As a result, we expect the smallest eigenvalues of $\H_{+-}$ to decrease with decreasing the vortex area $\xi^2$ as the Berry curvature becomes more concentrated. This is verified by a numerical calculation of the gap of $\H_{+-}$ as a function of $\kappa$, which controls the Berry curvature distribution, shown in Fig.~\ref{fig:EDelta}. We can also verify that all other eigenstates of $\H^{+-}$ have a relatively large gap and can be integrated out.

Thus, for every generator $r_\alpha^{\gamma, \gamma'}$, we can restrict ourselves only to the lowest lying state $\phi_{\gamma, \gamma';\alpha,l = 0}$ by integrating out the massive modes with $l>0$. The resulting low energy bosonic modes can be divided into two categories as anticipated in Sec.~\ref{sec:count}: (i) approximate goldstone modes acting within the same Chern sector which correspond to generators $r^{\gamma, \gamma}_\alpha$ commuting with $\gamma_z$ and (ii) nematic modes acting between Chern sectors which correspond to the generators $r^{\gamma,-\gamma}_\alpha$ anticommuting with $\gamma_z$.  

So far the analysis has been restricted to the $\U(4) \times \U(4)$ limit. To see what happens in the realistic limit, we need to include the effect of the sublattice off-diagonal part of the form factor and the dispersion which were discussed in detail in Ref.~\cite{KIVCpaper}. The former induces an extra energy cost $\lambda \simeq $ 0.6-0.8 meV \cite{SkPaper} to out-of-plane pseudospin fluctuations leading to a gap of a few meV for some of the approximate goldstone modes. The latter appears as a linear coupling to the fluctuations $\sim \tr Q^T h(\bk) \phi_{\bq = 0}(\bk)$ which is only non-vanishing for the inter-Chern component of $\phi$ since $h(\bk)$ acts predominantly between Chern sectors, $h(\bk) \approx h_x(\bk) \gamma_x + h_y(\bk) \gamma_y$. Expanding $\phi_{\bq = 0}(\bk)$ into eigenmodes (Eq.~\ref{PhiExp}) and integrating out the massive ones $l>0$ leads to an energy contribution $J \simeq h_+ H_{+-}^{-1} h_- \simeq$ 0.5 - 1 meV, where $h_\pm(\bk) = h_x(\bk) \pm i h_y(\bk)$ \footnote{Strictly speaking, we should remove the lowest mode $\psi_{+-,l=0} = \Delta$ when computing $h_+ H_{+-}^{-1} h_-$. In practice, it makes little difference since the overlap of $h_\pm(\bk)$ and $\Delta(\bk)$ is relatively small}. This contribution favors spin and pseudospin antiferromagnetic coupling between the Chern sector and plays an important role in the skyrmion pairing mechanism proposed in Ref.~\cite{SkPaper}. Both corrections are much smaller than the gap to higher energy modes $l > 0$ and only induces a gap of the order of a few meV, meaning that our restriction to the $l = 0$ modes remains valid in the realistic limit.

Finally, we can express the matrix $\rho$ defined in (\ref{rho}) in the $r$-basis where it has a simple $\bq$-independent and block-diagonal form with the blocks given by
\beq 
\rho^{\gamma,\gamma'}_{\alpha \beta} = -\frac{1}{2} \tr Q [r^{\gamma,\gamma'}_\alpha]^\dagger r^{\gamma,\gamma'}_\beta
\eeq
Note that $\rho$ is not antisymmetric in the $r$-basis since $r^{\gamma,\gamma'}$ are not hermitian and are thus related by a non-orthogonal transformation to the hermitian generators $t_\mu$. To highlight the symplectic structure of the theory, we can now go back to the Hermitian basis $t_\mu$ (where $\H$ is not diagonal) in which $\rho$ is antisymmetric and has the simple form given by
\beq
\rho_{\mu \nu} = -\frac{1}{2} \tr Q t_\mu t_\nu
\label{rhot}
\eeq
Thus, $\rho$ defines a symplectic structure of the theory by pairing up different generators as canonically conjugate variables. As shown in Appendix \ref{app:RankRho}, $\rho$ is a full rank matrix, which means that all generators are paired in canonically conjugate pairs. Thus, the number of soft modes is equal to half the number of generators yielding $16 - \nu^2$. Furthermore, since $Q$ commutes with $\gamma_z$, the matrix $\rho$ does not mix the intra-Chern ($[t_\mu, \gamma_z] = 0)$ and inter-Chern ($\{t_\mu, \gamma_z \} = 0$) generators. As a result, the count of approximate goldstone and the nematic modes can be identified with half the number of generators $t_\mu$ which commute or anticommute with $\gamma_z$, respectively, leading to the expression in Eq.~\eqref{PGNcount}.

\subsection{Goldstone mode count}
\label{sec:GoldstoneCount}
The analysis of the soft modes above does not distinguish the true gapless Goldstone modes which correspond to breaking the continuous physical symmetry, given to an excellent approximation by independent charge and spin rotations in the two valleys hence: $\U(2) \times \U(2)$ symmetry,  from the pseudo-Goldstone modes which only break the more approximate $\U(4) \times \U(4)$ symmetry and have a gap of a few meV. 

To make this distinction, let us review some recent results related to counting Goldstone modes in systems without Lorentz invariance \cite{Watanabe, Watanabe2013, Watanabe2014}. These works derived a general expression for the count of the Goldstone modes in terms of the number of broken symmetry generators $n_{\rm BG}$ and the rank of the matrix $\rho$ defined in (\ref{rhot}) given by
\beq
n_{G} = n_{\rm BG} - \frac{1}{2} {\rm Rank} \rho
\label{nG}
\eeq
This expression also enables us to extract the Goldstone mode dispersion by noting that modes corresponding to canonically conjugate variable have a linear time derivative term in the effective action leading to a quadratic dispersion, $\omega_\bq \sim \bq^2$ whereas the remaining modes have a quadratic time derivative time leading to a linear dispersion i.e. $\omega_\bq \sim |\bq|$. Thus, we can identify the number of linearly dispersing and quadratically dispersing Goldstone modes as:
\beq
n_{\rm G\text{-}I} = n_{\rm BG} - {\rm Rank} \rho, \qquad n_{\rm G\text{-}II} = \frac{1}{2} {\rm Rank} \rho
\label{nGIII}
\eeq
 where $n_{G,\rm I}$ and $n_{G, \rm II}$ denote the count of so-called type I and type II Goldstone modes introduced in Ref.~\cite{Nielsen} according to whether the leading power in the soft mode dispersion at small $\bq$ is odd or even, respectively. In our theory, type I and II correspond to linearly and quadratically dispersing soft modes, respectively, since higher order dispersions are not possible.

Let us first apply these results to the approximate $\U(4) \times \U(4)$ symmetry. In this limit, we can identify the broken symmetry generators by the generators $t_\mu$ of $\U(4) \times \U(4)$ which anticommute with $Q$. The number of such generators is $16 - \nu^2 - C^2$, where $\nu$ is the filling as measured from charge neutrality and $C$ is the Chern number of the ground state. Next, we note that the matrix $\rho$ in this case is always a full rank matrix (see Appendix \ref{app:RankRho}) so its rank is equal to its dimension which is precisely the number of broken generators $n_{\rm BG, app}$. Thus, using Eq.~\ref{nG}, the number of approximate Goldstone modes is equal to half the number of broken symmetry generators leading to Eq.~\ref{PGNcount}, i.e. $n_{\rm AG} = \frac{16-\nu^2-C^2}{2}$. In the $\U(4) \times \U(4)$ limit, all such modes will be type II (quadratically dispersing). Note, for the remaining $n_{\rm N} = \frac{16-\nu^2+C^2}{2}$ nematic modes,  we do not need to discuss the form of their low energy dispersion since they are generically gapped. 

The Goldstone mode count corresponding to the physical $\U(2) \times \U(2)$ symmetry is similar. The main difference is that we need to restrict ourselves to the broken symmetry generators corresponding to the physical $\U(2) \times \U(2)$ symmetry generated by $s_{0,x,y,z}$ and $\eta_z s_{0,x,y,z}$. Denoting these generators by $t_\mu^{\rm phys}$, we can define $\rho^{\rm phys}$ as in (\ref{rhot}) but using only $t_\mu^{\rm phys}$ leading to
\beq
n_G = n_{\rm BG, phys} - \frac{1}{2} {\rm Rank} \rho^{\rm phys}
\eeq
The rank of $\rho^{\rm phys}$ counts the number of physical symmetry generators which are canonically conjugate variables, and which thus give rise to a single Goldstone mode. 

The count of Goldstone modes for different possible states is provided in Tables \ref{tab:Spinless}, \ref{tab:Spinful}, and \ref{tab:Chern}.  To understand the results of these tables, let us start with the spinless limit which serves to illustrate the idea before considering more complicated scenarios. At neutrality, the K-IVC state spintaneously breaks $\U(1)$ valley symmetry generated by $\eta_z$. In addition, it is easy to verify that $\rho^{\rm phys} = 0$ leading to a single linearly dispersing (type I) mode as seen in Table \ref{tab:Spinless}. As a more complicated example, let us now consider the spinful K-IVC state at neutrality. In addition to breaking the $\U(1)$ valley symmetry, this state also breaks independent spin rotations within each valley, $\SU(2)_K \times \SU(2)_{K'}$ down to a single $\SU(2)$. Thus, there are 4 broken symmetry generators $\eta_z s_{0,x,y,z}$. The rank of $\rho^{\rm phys}$ vanishes leading to 4 linearly dispersing (type I) Goldstone modes as seen in Tables \ref{tab:Spinful} and \ref{tab:NLSM}. Finally, let us consider the spin-polarized K-IVC state at $\nu = 2$. This state breaks $\U_V(1) \times \SU(2)_K \times \SU(2)_{K'}$ down to the group generated by $s_z$ and $P_\downarrow \eta_z$ (assuming the filled spin flavor is up). Thus, there are 5 broken symmetry generators. The rank of $\rho^{\rm phys}$ in this case is equal to 4 so that the number of Goldstone modes is $5-2 = 3$ with 2 quadratically dispersing (type II) modes and one linearly dispersing (type I) mode. In Table \ref{tab:NLSM}, we also include the count in the presence of intervalley Hund's coupling which breaks the $\U(2) \times \U(2)$ explicitly down to $\U(1) \times \U(1) \times \SU(2)$ as explained in the next section.

\section{Effective field theory of soft modes}
The energetics of the soft modes can be conveniently captured by deriving an effective field theory in the form of a non-linear sigma model. Such field theory reproduces the soft mode Hamiltonian derived in the previous section but it can also allow us to go beyond the quadratic approximation and include interactions between the soft modes. We note that a non-linear sigma model describing the intra-Chern soft modes was already derived in Ref.~\cite{SkPaper}. The main difference here is the additional inclusion of the inter-Chern nematic modes.

\subsection{Non-linear sigma model}
\label{sec:NLSM}
To derive a non-linear sigma model, we need to identify a large manifold of soft modes related by symmetries which are weakly broken. For the intra-Chern soft modes, this is the $\U(4) \times \U(4)$ symmetry identified in Ref.~\cite{SkPaper} which is broken by the dispersion and the sublattice off-diagonal form factor. The approximate symmetry related to the nematic inter-Chern modes is more subtle. To understand this symmetry, we notice that these modes are only low in energy when the Berry curvature is strongly concentrated at a point. In this case, we can approximate the Berry curvature by a delta function representing the flux of a solenoid at the $\Gamma$ point such that the approximate energy expression in (\ref{Em0}) vanishes for an appropriate choice of $\Delta(\bk)$. One such choice is
\beq
A_\bk = \frac{(-k_y, k_x)}{|\bk|^2}, \quad \Delta(\bk) = e^{2i \varphi_\bk}, \quad \varphi_\bk = \arg (k_x + i k_y)
\label{AGauge}
\eeq
In this limit, the inter-Chern nematic modes become true Goldstone modes. To see the corresponding symmetry explicitly, we note that under the gauge transformation $c_\bk \mapsto \tilde c_\bk = e^{i \varphi(\bk) \gamma_z} c_\bk$, the sublattice-diagonal form factor (\ref{LambdaSymm}) changes as
\begin{align}
   \Lambda_\bq(\bk) &\mapsto \tilde \Lambda_\bq(\bk) = e^{i \varphi(\bk) \gamma_z} \Lambda_\bq(\bk) e^{-i \varphi(\bk + \bq) \gamma_z} \nonumber \\
   &\approx F_\bq(\bk) e^{i \bq \cdot (A_\bk - \nabla_\bk \varphi_\bk)} = F_\bq(\bk)
   \label{LambdaApp}
\end{align}
Thus, under such an approximation, the symmetry of the sublattice-diagonal part of the interaction is enhanced to $\U(8)$. Physically, this can be understood as follows. Recall, the initial $\U(4)\times \U(4)$ symmetry does not permit rotation between opposite Chern  sectors since the single particle wavefunctions of the opposite Chern bands are rather different. However, if all the Chern flux is concentrated into a solenoid, it can be eliminated by a singular gauge transformation, and this obstacle is circumvented. The extra symmetry generators correspond to the nematic modes and relate the insulating quantum Hall ferromagnets to the nematic semimetals. To see this, we note that \emph{any} $\bk$-independent $\tilde Q$ matrix (in terms of the transformed variable $\tilde c$) is a ground state for the $\U(8)$ symmetric interaction. The corresponding states in the original basis are given by the $\bk$-dependent $Q$ matrix:
\beq
Q_\bk = e^{i \varphi(\bk) \gamma_z} \tilde Q e^{-i \varphi(\bk) \gamma_z}
\eeq
For $\tilde Q$ commuting with $\gamma_z$, this transformation does nothing and we get the same insulating states as before $Q = \tilde Q$. On the other hand, for $\tilde Q$ anticommuting with $\gamma_z$, we get
\beq
Q_\bk = \tilde Q \cos 2 \varphi(\bk) - i \tilde Q \gamma_z \cos 2 \varphi(\bk)
\eeq
which describes a semimetal where $Q_\bk$ winds around the zeros of $\varphi(\bk)$. For example, $\tilde Q = \gamma_x$ yields the order parameter $Q_\bk = \gamma_x \cos 2 \varphi(\bk) + \gamma_y \sin 2 \varphi(\bk)$ which describes the nematic semimetal identified in Ref.~\cite{ShangHF}.

\begin{figure}
     \centering
     \includegraphics[width = 0.45 \textwidth]{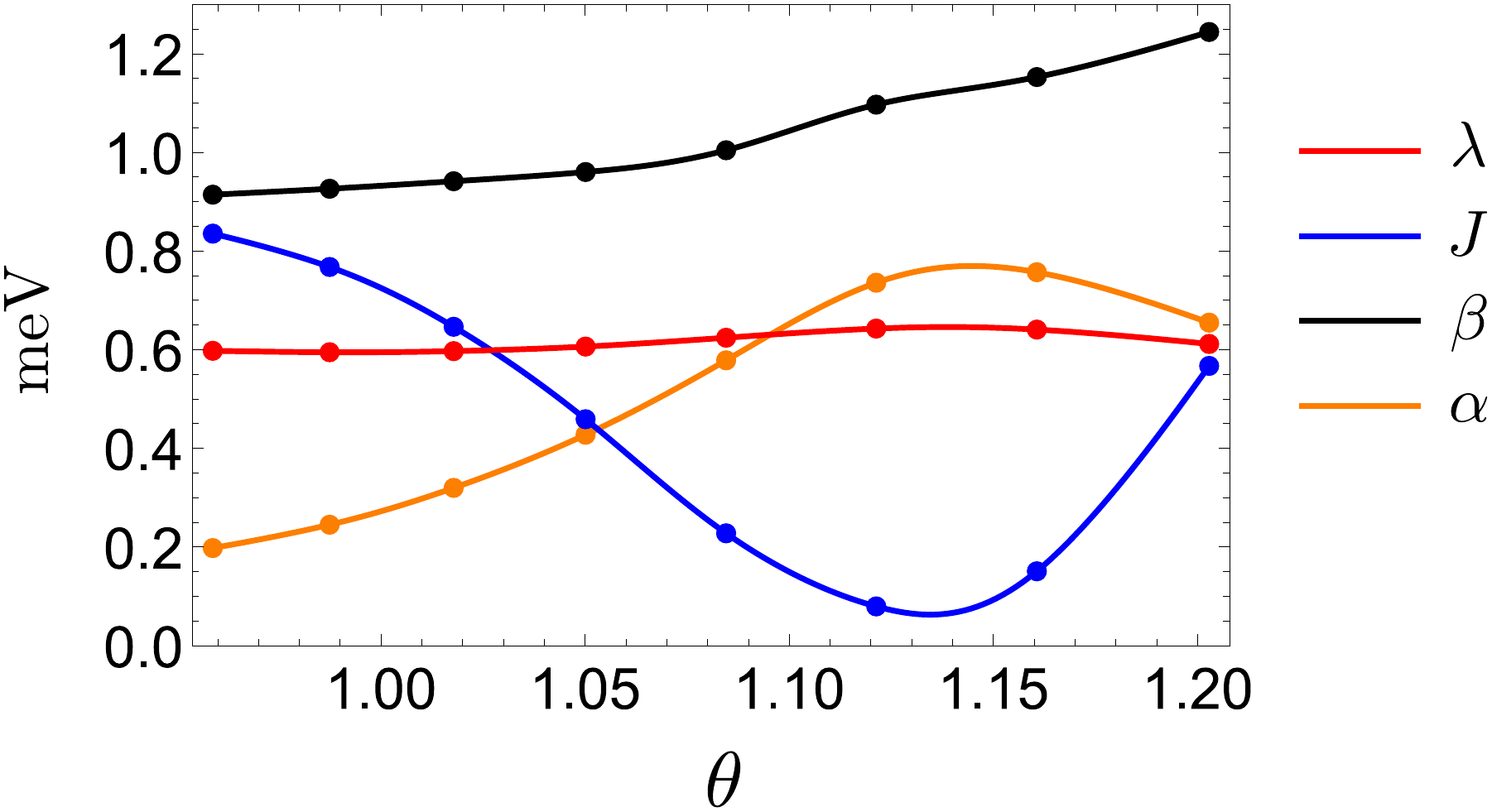}
     \caption{{\bf Field theory parameters}: Parameters of the non-linear sigma model defined by Eqs. (\ref{LQ}) and (\ref{EQ}) as a function of the twist angle $\theta$. The parameters are extracted using the energy splitting between the different self-consistent Hartree-Fock solutions in accordance with Eq.~\ref{NLSMpars} at $\kappa = 0.7$ with a grid size of $16 \times 16$, 6 bands included per spin and valley and with interaction parameters $\epsilon = 12.5$ and screening length $d = 20$ nm.}
     \label{fig:NLSMpars}
 \end{figure}

This naturally leads to a $\U(8)$ sigma model description which unifies the insulating and semimetallic order parameters incorporating both nematic and approximate Goldstone modes. The sigma model Lagrangian can be written as
\beq
    \L[\tilde Q] = \frac{1}{2} \tr T(\br)^\dagger \tilde Q \partial_t T(\br) - \frac{\rho}{8} \tr [\nabla \tilde Q(\br)]^2 - E[\tilde Q(\br)]
    \label{LQ}
\eeq
where the spatially dependent $\tilde Q(\br)$ is obtained from the spatially constant ground state by applying the unitary rotation $T(\br) = e^{i \phi(\br)}$ through
\begin{gather}
\tilde Q(\br) = T^\dagger(\br) \tilde Q T(\br), \quad T(\br) = e^{i \phi(\br)}, \nonumber \\
\phi(\br) = \sum_{\bq,\bk} e^{i \bq \cdot \br} \phi_\bq(\bk)
 \label{Qr}
\end{gather}
 $E[\tilde Q]$ contains the anisotropy terms which break the $\U(8)$ symmetry down to the physical $\U(2) \times \U(2)$ which can be summarized as follows. First, there is the energy penalty associated with deviations from the approximations leading to (\ref{LambdaApp}) which disfavors the inter-Chern states (semimetals which anticommute with $\gamma_z$) relative to intra-Chern states (insulators which commute with $\gamma_z$). This can be captured by a term of the form $-\tr (Q \gamma_z)^2$. In addition, there is the anisotropy term due to the sublattice off-diagonal form factor $-\tr (\gamma_{x,y} \eta_z Q)^2$ and the antiferromagnetic coupling obtained from the dispersion $+ \tr (\gamma_{x,y} Q)^2$ already derived in Refs.~\cite{KIVCpaper, SkPaper}. This leads to the energy expression
\begin{multline}
    E[\tilde Q] = -\frac{\alpha}{4} \tr (\tilde Q \gamma_z)^2 + \frac{J}{8} \tr [(\tilde Q \gamma_x)^2 + (\tilde Q \gamma_y)^2] \\ - \frac{\lambda}{8} \tr [(\tilde Q \gamma_x \eta_z)^2 + (\tilde Q \gamma_x \eta_z)^2]
    \label{EQ}
\end{multline}
The parameter $J$ and $\lambda$ were derived from the microscopic theory in Ref.~\cite{SkPaper}. The new parameter $\alpha$ can be identified with $\frac{E_\Delta}{8}$ which can be seen by substituting (\ref{Qr}) in (\ref{EQ}) and expanding to quadratic order in $\phi$ ignoring the $J$ and $\lambda$ terms. An alternative approach is to consider these as phenomenological parameters that can be fit to numerics. This is done by noting that the energy per particle obtained from $E[\tilde Q]$ for the different states relative to the minimum K-IVC state at $\nu = 0$ is
\beq
\Delta E_{\rm VP} = J, \quad \Delta E_{\rm VH} = \lambda, \quad \Delta E_{\rm SM} = \alpha + \frac{J + \lambda}{2} = \beta
\label{NLSMpars}
\eeq
where 'VP', 'VH', and 'SM' denote the valley polarized $\tilde Q = \eta_z$, valley Hall $\tilde Q = \gamma_z \eta_z$ and semimetal $\tilde Q = \gamma_z$, respectively. Comparing to the Hartree-Fock numerics these parameters can be extracted as shown in Fig.~\ref{fig:NLSMpars}. It is worth noting that the field theory derived in Ref.~\cite{SkPaper} can be obtained from (\ref{LQ}) and (\ref{EQ}) in the limit of large $\alpha$. 
 
A notable aspect about the field theory (\ref{LQ}) is that it does not include all possible symmetry allowed terms. In particular, notice the absence of terms of the form $\tr (Q \eta_z)^2$ and $\tr (Q \gamma_z \eta_z)^2$ which, despite being symmetry allowed, turn out to be significantly smaller than the other terms and can be neglected. To understand what these terms represent, it is instructive to focus on the insulating states $[Q, \gamma_z] = 0$ for which $Q$ can be split into $Q_\pm$ corresponding to the $\pm$ Chern sectors. In this limit, the $\lambda$ term above takes the form $\tr Q_\pm \eta_z Q_\mp \eta_z$ whereas the neglected symmetry allowed terms take the form $\tr Q_\pm \eta_z Q_\pm \eta_z$. Clearly both types of terms represent anisotropy terms of the pseudospin ($\eta$) favoring easy axis or easy plane orientation. However, the $\lambda$ term is an inter-Chern term which only yields a non-vanishing contribution when both $Q_+$ and $Q_-$ are non-trivial (not equal to a multiple of the identity) whereas the other terms are intra-Chern terms. This can be seen more clearly in spinless limit where $Q_\pm = \bn_\pm \cdot {\boldsymbol \eta}$ such that the $\lambda$ term has the form $n_{+,z} n_{-,z}$ whereas the other terms have the form $n_{\pm,z}^2$. This has important implications for the soft mode dispersion and occasionally leads to cases where the sigma model  (\ref{LQ}) yields a gapless mode which is {\it not} associated with any continuous symmetry breaking (and is thus not a true Goldstone mode). As a simple example consider the spinless model at $\nu=1$, and $C=1$. In this case, we fill both bands in $C=1$ and a single band in the opposite Chern sector, which we will take to be valley polarized (VP). Now, the physical valley symmetry would not predict any gapless modes, however the mechanism described above implies the anisotropy $\lambda$ that usually lifts the SU(2) valley rotation symmetry is absent in this particular case leading to a nearly gapless mode which we denote by PG*. In the TDHF numerics, these correspond to modes with a very small gap $\sim 0.02-0.1$ meV. In tables \ref{tab:Spinless}, \ref{tab:Spinful}, and \ref{tab:Chern}, we see several instances of PG* which are pseudo-Goldstone modes which are approximately gapless. 

Another type of symmetry allowed term not included in the field theory (\ref{LQ}) are possible anisotropies in the stiffness $\rho$. Such terms will assign slightly different energy cost to spatial deformations of $\tilde Q$ in different directions. These likely affect the overall soft mode dispersion but will be unimportant if we are only interested in leading contribution to the soft mode dispersion at small $\bq$.

\subsection{Soft mode spectrum from the field theory}\label{sec:softmodefieldth}
In this section, we explain how the soft mode spectrum at small $\bq$ can be obtained from the sigma model given by Eqs.~\ref{LQ} and \ref{EQ}.

\subsubsection{Spinless model}
\begin{figure}
     \centering
     \includegraphics[width = 0.4 \textwidth]{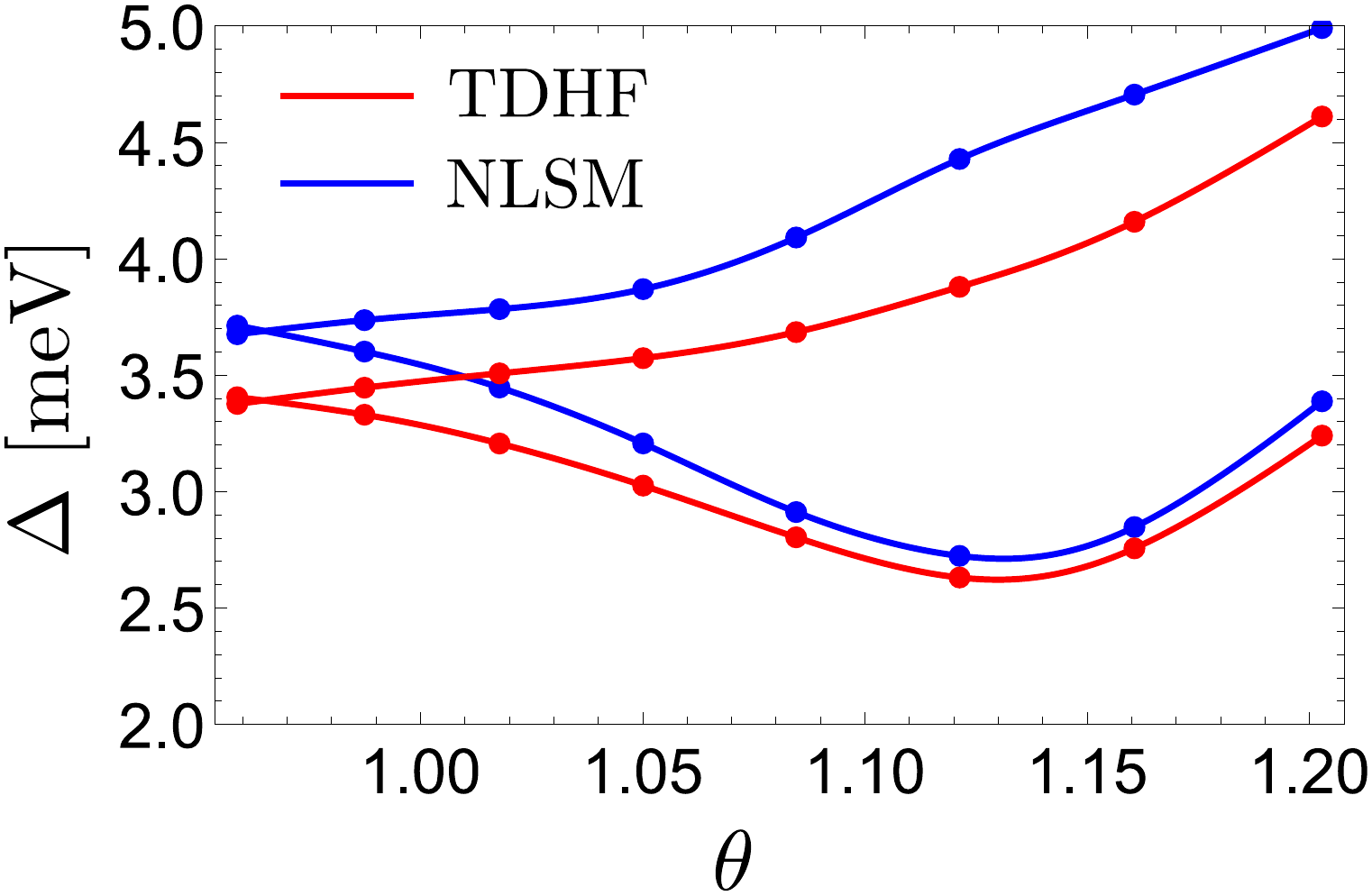}
     \caption{{\bf Comparison of TDHF and NLSM results}: The gap for the pseudo-Goldstone mode(s) (lower branch) and the nematic modes (upper branch) for the K-IVC state at neutrality as a function of the twist angle $\theta$ computed from the time-dependent Hartree-Fock TDHF (red) and the non-linear sigma model NLSM (blue). The NLSM gaps are given by $4 \sqrt{\lambda (J + \lambda)}$ and $4\beta$ (cf.~Eq.~\ref{omegaq}) with $J$, $\lambda$, and $\beta$ extracted from the energy spliting between different self-consistent Hartree-Fock states (cf.~Eq.~\ref{NLSMpars}). To calculate both the field theory parameters and the TDHG gaps, we used a $12 \times 12$ momentum space grid with 6 bands per spin and valley with at $\kappa = 0.7$, $\epsilon = 12.5$ and gate distance $d = 20$ nm.}
     \label{fig:TDHFvsNLSM}
 \end{figure}

We start by considering the spinless limit which serves as an illustration for the computation. At $\nu = 0$, the ground state minimizing all three terms in (\ref{EQ}) can be chosen without loss of generality to be $Q = \gamma_z \eta_x$ Which is nothing but the K-IVC of reference \cite{KIVCpaper}. A basis of hermitian generators anticommuting with $Q$ can be chosen as $t_\mu = {\gamma_{x,y} \eta_{0,x}, \gamma_{0,z} \eta_{y,z}}$. Substituting $Q = e^{-i \sum_\mu \phi_\mu t_\mu} Q_0 e^{i \sum_\mu \phi_\mu t_\mu}$ and expanding to quadratic order in $\phi$ yields an action of the form (\ref{SM}) from which we can extract the matrix $M$. The dispersion of the Goldstone modes is obtained by solving the equation $\det M(\bq,\omega) = 0$ which has four positive solutions:
\begin{align}
    \omega_{\rm G,I}(\bq) &= 2 \sqrt{2 J \rho} |\bq| + O(\bq^3) \nonumber \\
    \omega_{\rm PG}(\bq) &= 4 \sqrt{\lambda (J + \lambda)} + O(\bq^2) \nonumber \\
    \omega_{\rm N;1,2}(\bq) &= 4 \beta + O(\bq^2)
    \label{omegaq}
\end{align}
 We see that $\omega_{\rm G,I}(\bq)$ represents the linearly dispersing Goldstone mode, $\omega_{\rm PG}(\bq)$ represents the gapped pseudo-Goldstone mode which becomes gapless in the limit of perfect sublattice polarization $\lambda = 0$, and $\omega_{\rm N}(\bq)$ represent the two gapped nematic modes which are exactly degenerate at $\bq = 0$. The existence of a single Goldstone mode is consistent with Eq.~\eqref{nG} with a single broken symmetry generator $\eta_z$ and with the rank of $\rho^{\rm phys}$ equal to 0.  In Fig.~\ref{fig:TDHFvsNLSM}, we compare the gaps of the pseudo-Goldstone mode $\Delta_{\rm PG} = \omega_{\rm PG}(0)$ and the nematic modes $\Delta_N = \omega_{\rm N}(0)$ obtained from the TDHF with the field theory result (\ref{omegaq}) and we can see that the two agree reasonably well. This serves as a justification for the field theory description in the limit of small $\bq$. 
 
\subsubsection{Spinful model at $\nu = 0$}
The spectrum of the spinful model at charge neutrality is the same as the for the spinless model with four copies of each mode which correspond to a spin-singlet and three spin-triplet modes for each mode of the spinless model. To see this, we note that the spin unpolarized K-IVC state of the spinful model, $Q = \gamma_z \eta_x s_0$ breaks the $\U(2) \times \U(2)$ symmetry down to $\U(2)$ which corresponds to overall spin and charge conservation. This means that there is a manifold of degenerate K-IVC ground states given by $\frac{\U(2) \times \U(2)}{\U(2)} \simeq \U(2)$. This manifold includes both spin singlet and spin triplet K-IVC states and can be generated by acting on the singlet K-IVC state with different spin rotations in the two valleys. Since all K-IVC states are symmetry related, their spectra are identical and we can focus on a single representative which we take to be the spin-singlet K-IVC state. Since this state preserves the overall $\SU(2)$ symmetry, we can label the soft modes by a spin quantum number such that each mode in the spinless model corresponds to a singlet and three triplet modes which are degenerate. The existence of 4 linearly dispersing Goldstone modes arises due to breaking the continuous symmetries generated by the 4 generators $\eta_z$ and $s_{x,y,z} \eta_z$, which result in a matrix $\rho^{\rm phys}$ that vanishes. The detailed symmetry properties of the different soft modes will be discussed in detail in the next section.

We can distinguish the spin-singlet and triplet K-IVC state by including an intervalley Hund's coupling term \cite{Zhang2018} which has the form:
\begin{multline}
    \L_H = \frac{J_H}{4} \sum_{i=x,y,z} \left\{\tr \frac{1 + \eta_z}{2} s_i Q \tr \frac{1 - \eta_z}{2} s_i Q \right. \\ \left. - \tr \frac{1 + \eta_z}{2} s_i Q \frac{1 - \eta_z}{2} s_i Q \right\}
\end{multline}
where $|J_H|$ is given roughly by the Coulomb scale divided by the ratio of the Moir\'e to lattice length scale yielding a value of 0.1-0.2 meV. This term breaks the separate $\SU(2)$ spin symmetry in each valley down to a single overall $\SU(2)$. 

The ground state for finite $J_H$ depends on the sign of $J_H$ with $J_H > 0$ favoring the spin-singlet state $Q = \gamma_z \eta_x s_0$ and $J_H < 0$ favoring the spin-triplet state $Q = \gamma_z \eta_x s_z$ \cite{KIVCpaper}. The soft mode spectrum at $\nu = 0$ for finite $J_H$ is summarized in Table \ref{tab:NLSM}. For $J_H > 0$, the three triplet Goldstone modes are gapped leaving the singlet mode gapless. This arises since the symmetries corresponding to the generators $s_{x,y,z} \eta_z$ are explicitly broken, leaving a single broken symmetry generator corresponding to $\U_v(1)$ which gives rise to a single linearly dispersing Goldstone mode. In contrast, for $J_H < 0$, the overall $\SU(2)$ spin symmetry is broken (down to $\U(1)$), in addition to the broken $\U_v(1)$ symmetry. As a result, three of the four Goldstone modes remain gapless, corresponding to the broken symmetry generators $\eta_z$ and $s_{x,y}$ (since $\rho^{\rm phys} = 0$ also for the spin-triplet K-IVC state).

 \begin{table*}[t]
     \centering
     \small
     \bgroup
\setlength{\tabcolsep}{0.5 em}
\setlength\extrarowheight{0.7em}
     \begin{tabular}{c|c|c|c|c|c|c|c}
     \hline \hline
       \multirow{2}{*}{$\nu$} & \multirow{2}{*}{Type} & \multicolumn{2}{c|}{$J_H = 0$} & \multicolumn{2}{c|}{$J_H > 0$} & \multicolumn{2}{c}{$J_H < 0$} \\
        \cline{3-8}
        & & $n$ & $\omega_\bq$ & $n$ & $\omega_\bq$ & $n$ & $\omega_\bq$\\
        \hline
        \multirow{7}{*}{0} & \multirow{3}{*}{G} & \multirow{3}{*}{4} & \multirow{3}{*}{$2 \sqrt{\rho J} |\bq|$} & 1 &  $\sqrt{\rho (4J + 3 J_H)} |\bq|$ & 1 & $\sqrt{\rho (4J - J_H)}|\bq|$ \\
        \cline{5-8}
        & & & & \multirow{2}{*}{3} & \multirow{2}{*}{$2\sqrt{J_H (4J - J_H)}$} & 2 & $2\sqrt{\rho (J + 3 J_H)}|\bq|$ \\
        \cline{7-8}
        & & & & & & 1 & $2\sqrt{-J_H (4J - 5 J_H)}$ \\
        \cline{2-8}
         & \multirow{2}{*}{PG} & \multirow{2}{*}{4} & \multirow{2}{*}{$4 \sqrt{\lambda (J + \lambda)}$} & 1 & $2 \sqrt{(4 \lambda + 3 J_H)(J + \lambda)}$ & 1 & $2 \sqrt{(4\lambda - J_H)(J + \lambda - J_H)}$ \\
        \cline{5-8}
        & & & & 3 & $2 \sqrt{(4\lambda + 3 J_H)(J + \lambda + J_H)}$ & 3 & $2 \sqrt{(4\lambda - J_H)(J + \lambda)}$ \\
        \cline{2-8}
        & \multirow{2}{*}{N} & \multirow{2}{*}{8} & \multirow{2}{*}{$4 \beta$} & 2 & $2\sqrt{ \beta (4 \beta + 3 J_H)}$ & 2 & $\sqrt{2(2\beta - 2 J_H)(4 \beta - J_H)}$ \\
        \cline{5-8}
        & & & & 6 & $4\sqrt{(\beta + J_H)(4 \beta + 3 J_H)}$ & 6 & $2\sqrt{\beta(4 \beta - J_H)}$ \\
        \hline 
        \multirow{9}{*}{2} & \multirow{3}{*}{G} & 1 & $2 \sqrt{\rho J} |\bq|$ & 1 & $\sqrt{\rho (4J + J_H)}|\bq|$ & 1 & $ \sqrt{\rho (4J - J_H)}|\bq|$ \\
        \cline{3-8}
        & & \multirow{2}{*}{2} & \multirow{2}{*}{$\rho \bq^2$} & \multirow{2}{*}{2} & \multirow{2}{*}{$\sqrt{2 \rho J_H}|\bq|$} & 1 & $\rho \bq^2$\\
        \cline{7-8}
        & & & & & & 1 & $-2 J_H + \rho \bq^2 $\\
        \cline{2-8}
        & \multirow{3}{*}{PG} & 1 & $4 \sqrt{\lambda (\lambda + J)}$ & 1 & $ 2\sqrt{(4\lambda - J_H)(\lambda + J)}$ & 1 & $2 \sqrt{(4\lambda + J_H)(\lambda + J)}$ \\
        \cline{3-8}
        & & \multirow{2}{*}{2} & \multirow{2}{*}{$4 \lambda$} & \multirow{2}{*}{2} & \multirow{2}{*}{$2\sqrt{2 \lambda (2\lambda + J_H)}$} & 1 & $4\lambda$\\
        \cline{7-8}
        & & & & & & 1 & $4\lambda - 2 J_H$\\
        \cline{2-8}
        & \multirow{3}{*}{N} & 2 & $4 \beta$ & 2 & $2\sqrt{\beta (4\beta - J_H)}$ & 2 & $2\sqrt{\beta (4 \beta + J_H)}$\\
        \cline{3-8}
        & & \multirow{2}{*}{4} & \multirow{2}{*}{$4 \beta - 2J$} & \multirow{2}{*}{4}  & \multirow{2}{*}{$2 \sqrt{(2 \beta - J) (2 \beta - J + J_H)}$} & 2 & $2 (2\beta - J)$ \\
        \cline{7-8}
        & & & & & & 2 & $2 (2 \beta - J - J_H)$ \\
        \hline \hline
     \end{tabular}
     \egroup
     \caption{{\bf Soft mode dispersion}: Dispersion of the bosonic soft modes for the spinful model at charge neutrality $\nu = 0$ and half-filling $\nu = 2$ obtained from the sigma model defined in Eqs.~\ref{LQ} and \ref{EQ} to leading order in the momentum $\bq$ in terms of the sigma model parameter $J$, $\lambda$, and $\beta = \alpha - \frac{J + \lambda}{2}$. At $\nu = 0$ and in the absence of intervalley Hund's coupling $J_H$, there are 4 degenerate linearly dispersing gapless goldstone (G) modes, four degenerate gapped psuedo-goldstone (PG) modes, and 8 degenerate nematic (N) modes. For $J_H > 0$, three of gapless modes acquire a gap $\sim \sqrt{J_H J}$ whereas for $J_H < 0$, three modes remain gapless and one acquires a gap $\sim \sqrt{|J_H| J}$. At $\nu = 2$ and in the absence of intervalley Hund's coupling $J_H$, there are 3 gapless goldstone (G) modes (one with linear and two with quadratic dispersion), 3 gapped psuedo-goldstone (PG) modes, and 6 nematic (N) modes. For $J_H > 0$, the three goldstone modes remain gapped but the dispersion of two of them become linear rather than quadratic whereas for $J_H < 0$m one of the two quadratic modes acquires a gap $\sim |J_H|$.}
     \label{tab:NLSM}
 \end{table*}

\subsubsection{Spinful model at $\nu = 2$}
At $\nu = 2$, the ground state is a spin-polarized K-IVC state. Similar to the case of $\nu = 0$, there is a manifold of states related by the action of $\U(2) \times \U(2)$ symmetry on the simple K-IVC ferromagnet $Q = P_\uparrow + P_\downarrow \gamma_z \eta_x$. This manifold consists of states where an arbitrary spin orientation is chosen independently for each valley. Since all these states are symmetry related, we can restrict ourselves to the K-IVC ferromagnet where the same spin direction is chosen in both valleys. Following the same procedure as in the spinless case, we can extract the soft mode spectrum which is given by
\begin{gather}
    \omega_{\rm G,I}(\bq) = 2 \sqrt{\rho J}|\bq|, \qquad \omega_{\rm G,II; 1,2}(\bq) = \rho \bq^2 \nonumber \\
    \omega_{\rm PG;1}(\bq) = 4 \sqrt{\lambda (\lambda + J)}, \qquad  \omega_{\rm PG;2,3}(\bq) = 4 \lambda, \nonumber \\
    \omega_{\rm N;1,2}(\bq) = 4 \beta, \qquad \omega_{\rm N;3,4,5,6}(\bq) = 4 \beta - 2 J
\end{gather}
The spectrum consists of 3 Goldstone, 3 pseudo-Goldstone and 6 nematic modes whose dispersion. Of the three Goldstone modes, only one is linearly dispersing and can be associated with the broken $\U(1)$ valley symmetry while the other two are quadratically dispersing and can be associated with spin symmetry breaking. This is compatible with Eqs.~\eqref{nG} and \eqref{nGIII} since there are 5 broken symmetry generators corresponding to $s_{x,y}$ and $\eta_z s_{0,x,y}$ with ${\rm Rank}\rho^{\rm phys} = 4$, leading to 2 type-II Goldstone modes and one type-I mode. The pseudo-Goldstone modes are split into 1 + 2 whereas the nematic modes are split into 2 + 4. These degeneracies match the ones seen in the numerical TDHF spectrum in Fig.~\ref{fig:TDHF} and will be explained using the symmetry analysis of Sec.~\ref{sec:symmreps}.

In the presence of intervalley Hund's coupling, the independent $\SU(2)$ spin rotation symmetry in each valley is explicitly broken down to a single overall $\SU(2)$ spin rotation symmetry. As a result, there are only two spontaneously broken spin symmetry generators $s_{x,y}$, in addition to the spontaneously broken $\U(1)$ valley symmetry. For a ferromagnetic Hund's coupling $J_H < 0$, the ground state is the ferromagnetic K-IVC state with ${\rm Rank} \rho^{\rm phys} = 2$ leading to one type-I mode and one type-II mode. In contrast, the antiferromagnetic Hund's coupling $J_H > 0$ selects a K-IVC state with opposite spin orientation in the two opposite valleys. This state has ${\rm Rank} \rho^{\rm phys} = 0$, leading to three type-I Goldstone modes. The modified dispersion for the soft modes in the presence of intervalley Hund's coupling is given in Table \ref{tab:NLSM}.

\subsubsection{Spinful model at $\nu = 1$ and $C=3$: Correlated Chern insulator}
\label{sec:CorrelatedChern}
Let us now consider the odd filling $\nu = 1$. In this case, there are several possible states which minimize the energy functional (\ref{EQ}). These are divided into two categories: (i) we can fill 3 bands in one Chern sector and 2 in the other Chern sector leading to a state with $|C| = 1$, or (ii) we can fill 4 bands in one Chern sector and 1 in the other leading to a state with $|C| = 3$. These states are degenerate on the level of the energy functional (\ref{EQ}) but we expect a small orbital magnetic field to favor the $|C| = 3$ state since it lowers the energy of one Chern sector relative to the other. Indeed, such states have recently been observed in experiments \cite{AndreiChern, YazdaniChern, YoungHofstadter}. For the $|C| = 3$, one Chern sector is fully filled and the other is quarter filled where we have to pick one spin-valley flavor out of the four flavors within this sector to fill. This necessarily breaks $\SU(2)$ spin rotation symmetry but it may or may not break $\U(1)$ valley rotation. The latter depends on whether we choose the filled state to be a valley eigenstate (VP) with $Q = P_+ + P_- (P_\uparrow \eta_z - P_\downarrow$) or to be an equal superposition of the two valleys (IVC) with $Q = P_+ + P_- (P_\uparrow \eta_{x,y} - P_\downarrow)$, where $P_\pm$ and $P_{\uparrow/\downarrow}$ denote the projectors unto the different Chern and spin sectors, respectively. The two possibilities are considered below.

First, we can consider the VP case where the spectrum can be computed leading to
\begin{gather}
    \omega_{\rm G;1,2,3}(\bq) = \rho \bq^2, \qquad 
     \omega_{\rm N;1,2,3}(\bq) = 2 (2\beta - J - \lambda), , \nonumber \\ \omega_{N;3,\dots,12}(\bq) = 4 (\beta - \lambda)
\end{gather}
Note that the expression of the Goldstone mode count (\ref{nGIII}) yields a single type II goldstone mode corresponding to the breaking of $\SU(2)$ spin symmetry in the $K$ valley down to $\U(1)$. On the other hand, the spectrum obtained above seems to have three quadratically dispersing gapless modes. To resolve this issue, recall the discussion of Sec.~\ref{sec:NLSM} that the sigma model does not include all possible symmetry allowed terms and thus can have 'accidential' gapless modes not associated with any continuous symmetry breaking. In the case considered here, the issue arises because the pseudospin 'easy plane anisotropy' term $\lambda$ which prevents us from performing an arbitrary  $\SU(2)$ valley rotation vanishes if one Chern sector is completely filled or completely empty. This can be verified by adding the symmetry allowed intra-Chern easy plane anisotropy term $\frac{\kappa}{4} \tr (Q \eta_z)^2$ to (\ref{EQ}). This perturbation splits the degeneracy between the VP and the IVC state by favoring the VP state for $\kappa > 0$ and the IVC state for $\kappa < 0$. If we now recompute the soft mode dispersion for the VP state with $\kappa > 0$, we find that the two of the three quadratically dispersing modes acquire a gap of $4 \kappa$ verifying that these are not true gapless goldstone modes. We will denote these modes by PG* which means they are pseudo-goldstone modes that are not associated with a broken physical symmetry yet is almost gapless due to the form of the Hamiltonian. 

Next, let us consider the IVC state with $Q = P_+ + P_- (P_\uparrow \eta_{x,y} - P_\downarrow)$. The soft mode spectrum for this state is given by
\begin{gather}
    \omega_{\rm G;1,2,3}(\bq) = \rho \bq^2, \qquad 
     \omega_{\rm N;1,\dots,6}(\bq) = 4 (\beta - \lambda) , \nonumber \\ \omega_{N;7,8,9}(\bq) = 2 (2\beta - \lambda), \qquad \omega_{N;7,8,9}(\bq) = 2 (2\beta - 2\lambda - J)
\end{gather}
Here, we also see a discrepancy with the Goldstone mode count (\ref{nGIII}) which yields two quadratically dispersing modes and one linearly dispersing mode rather than three quadratically dispersing modes as seen above. This again can be resolved by including the extra anisotropy term $\frac{\kappa}{4} \tr (Q \eta_z)^2$ (with $\kappa < 0$ in this case). Redoing the calculation, we find that one of the three gapless quadratic modes acquire linear dispersion given by $2 \sqrt{-\kappa \rho}|\bq|$ as expected.

\section{Symmetry representations of the soft modes}\label{sec:symmreps}

In this section, we present an analysis of the symmetry representations of the soft modes. The analysis will be done in full generality without assuming a particular ground state. Each symmetry breaking state, regardless of the detailed anisotropies or perturbations which selects for it, yields a set of soft modes which transform as a representation under the group of symmetries which leave this state invariant. The decomposition of this representation into irreducibles (irreps) only depends on the state in question and encodes the degeneracies and symmetry quantum numbers of the modes, representing a fingerprint of insulating state. Such fingerprint can be very useful in identifying the broken symmetry state in experiments by measuring the soft mode spectrum and response to different perturbations. 

\subsection{Formalism}

The symmetry group of MATBG, which we denote by $\G$, is generated by the following symmetries: $\U_V(1)$ valley charge conservation, $\SU(2)_K \times \SU(2)_{K'}$ corresponding to independent $\SU(2)$ spin conservation in each valley, time-reversal symmetry $\T$ and two-fold rotation symmetry $C_{2z}$ which exchanges valleys, in addition three-fold rotation $C_3$ and mirror symmetry $M_y$ which act within a given valley (the latter exchanges layers and is sometimes denoted by $C_{2x}$). A given state described by the matrix $Q$ breaks $\G$ down to a subgroup $\G_Q \subset \G$. The generators of $\G_Q$ for a selection of states at different integer fillings are given in Table \ref{tab:Spinless} for the spinless model and Tables \ref{tab:Spinful} and \ref{tab:Chern} for the spinful model.

To obtain the symmetry representations of the soft modes, we start by considering a general symmetry $\S$ whose action on the band-projected operator $c_\bk$ is given by
\beq
\S c_\bk \S^\dagger \equiv U_\bk c_{O \bk}, \qquad U_\bk U^\dagger_{O \bk} = 1 
\label{Sck}
\eeq
where $O$ is an element of the $O(2)$ group corresponding to the spatial action of the symmetry $\S$. The soft modes are defined through the operator $\hat \phi_{n,\bq}$ which creates a soft mode characterized by the eigenfunction $\phi_{n,\bq}$ at energy $\varepsilon_{n,\bq}$ (cf.~Eq.~\ref{Hphi}) given explicitly by
\beq
\hat \phi_{n,\bq} = \sum_\bk c_\bk^\dagger \phi^\mu_{n,\bq}(\bk) t_\mu c_{\bk + \bq}
\label{Wck}
\eeq
The soft modes $\hat \phi_{n,\bq}$ transform as a representation of $\G_Q$ given explicitly by
\beq
\S \hat \phi_{n,\bq} \S^{-1} = \sum_m S_{nm}(\bq) \hat \phi_{m,O \bq}
\label{SPhiHat}
\eeq
The matrix representation $S(\bq)$ can be obtained from symmetry action on the operators $c_\bk$ in Eq.~\ref{Sck} in addition to the knowledge of the soft mode wavefunctions $\phi_{n,\bq}$ as explained in detail in appendix (\ref{app:SymRep}).

So far the discussion has been completely general. To make further progress, we restrict ourselves to the flat bands where the action of the different symmetries in $\G$ in the Chern-pseudospin-spin basis was derived in Refs.~\cite{KIVCpaper, SkPaper} (see appendix \ref{app:SymRep} for details). Furthermore, we restrict our attention to the $16 - \nu^2$ lowest energy bosonic modes which are separated by a large gap ($\sim 10-15$ meV) from the remaining high energy modes (See discussion of Sec.~\ref{sec:Energetics}). One simplification which will enable us to obtain the symmetry representations explicitly is to consider $\U(4) \times \U(4)$ limit discussed in Sec.~\ref{sec:Energetics} where the eigenfunctions $\phi_{n,\bq}$ for the soft modes have a particularly simple form. Since the symmetry representations cannot change continuously, these results should equally apply for the realistic limit with all anisotropies included as long as there is no mixing with the higher energy modes.

Thus, given a state $Q$ which is invariant under a symmetry group $\G_Q \subset \G$, the full symmetry action of $\G_Q$ on the soft modes is specified by knowing the symmetry representations $S(\bq)$ and the symplectic matrix $\rho$ in a certain basis which enables us to construct them in any other basis. In a hermitian basis of generators, $\rho$ is antisymmetric and it pairs different generators as canonically conjugate position-momentum variables. However, it is more convenient to choose a non-hermitian basis which diagonalizes $\rho$. The corresponding basis transformation is non-orthogonal and corresponds to going from a pair of position-momentum canonical variables to creation-annihilation operators. Since $\rho$ in invariant under any symmetry $\S \in \G_G$, we can restrict the symmetry action to the $+1$ eigenvalue sector of $\rho$ (corresponding to, let's say, annihilation operators) which we will denote by $S_+$. Furthermore, at a given momentum $\bq$, we should only consider the 'little group' $\G_{Q,\bq} \subseteq \G_Q$ which leaves the point $\bq$ invariant, leading to a $16 - \nu^2$ dimensional matrix representation for $\G_{Q,\bq}$ which can be decomposed into irreps. 

\subsection{Results}

In the following, we will restrict our attention to the $\Gamma$ point, $\bq = 0$, since many experimental probes only couple to the small momentum modes but our analysis can be extended to any momentum. At the $\Gamma$ point $\G_{Q,\bq}$ is equal to the full group $\G_Q$ and the irreps are obtained by simultaneously block-diagonalizing the matrices $S_+$ for all symmetries in $\G_Q$. Such process can be automated yielding the results of tables \ref{tab:Spinless}, \ref{tab:Spinful}, and \ref{tab:Chern}. Table \ref{tab:Spinless} includes the results for all possible insulating state for the spinless model which serves to illustrate the concept. In Table \ref{tab:Spinful}, we include a selection of physically relevant states for the spinful model at even integer fillings $\nu = 0$ and $2$ where $C = 0$ insulators have been experimentally observed \cite{PabloMott, Dean-Young, efetov}. In addition, Table \ref{tab:Chern} includes the Chern insulators with maximal Chern number $|C| = 4 - |\nu|$ which have been recently observed to be stabilized for small out-of-plane \cite{AndreiChern, EfetovChern, YazdaniChern, YoungHofstadter}.

The tables specify the symmetry content of the soft modes as follows. The first columns specify the state $Q$ and the generators of the symmetry group $\G_Q$ at a given filling $\nu$ and Chern number $C$. The soft modes are divided into Goldstone modes (G) which correspond to breaking an exact continuous symmetry, pseudo-Goldstone modes (PG) which break the approximate $\U(4) \times \U(4)$ symmetry but none of the continuous physical symmetries. Both G and PG modes correspond to excitations within the same Chern sectors. Then there are nematic modes (N) which correspond to inter-Chern modes. The goldstone modes can be further divided into type I and type II depending on whether they have linear or quadratic dispersion at small momenta (see Sec.~\ref{sec:GoldstoneCount}). The last few columns specify the decomposition of the soft modes into irreducible representations of the symmetry group $\G_Q$ which are specified by their characters. Recall that a representation character denotes the traces of the symmetry elements in the given irrep providing a basis-independent characterization of the irrep. 

Given the characters, it is straightforward to deduce the symmetry quantum numbers of the soft modes. Let us begin with the $\U_V(1)$ symmetry generated by $e^{i \varphi \eta_z}$. A $d$-dimensional representation with character $d e^{\pm i m \varphi}$ means that all the $d$ modes in the representation carry the same valley charge of $m$. A character of $d \cos m\varphi$ describes a representation with $d/2$ modes of valley charge $+m$ and $d/2$ with valley charge $-m$. In general, we can extract the number of modes $n_m$ with valley charge $m$ within a given representation by projecting onto this charge sector using
\beq
n_m = \int_0^{2\pi} \frac{d\varphi}{2\pi} e^{-i m \varphi} \chi(e^{i \phi \eta_z})
\label{chiValley}
\eeq
Similarly for $C_3$ symmetry, if all the soft modes carry the same "angular momentum" $l$ under $C_3$, we get a character of $d e^{\frac{2\pi i}{3} l}$. On the other hand, a character of $-d/2 = d (e^{\frac{2\pi i}{3}} + e^{-\frac{2\pi i}{3}}
)$ corresponds to an irrep with an equal number of $l = +1$ and $l = -1$ modes. Generally, the number of modes with angular momentum $l = 0,\pm 1 \mod 3$ is
\beq
n_0 = \frac{d + 2 {\rm Re} \chi(C_3)}{3} , \quad
    n_{\pm 1} = \frac{d - {\rm Re} \chi(C_3)}{3} \pm \frac{{\rm Im} \chi(C_3)}{\sqrt{3}}
\eeq

This enables us to read of the symmetry properties of the soft modes in the spinless limit from Table \ref{tab:Spinless}. For example, let us start with the valley-polarized (VP) state where one band is filled in each Chern sector corresponding to the same valley. There are two intra-Chern PG and two inter-Chern N modes. Since all modes create electron-hole excitations between opposite valleys, they all carry the same valley quantum number $+2$. The two PG modes transform trivially under $C_3$ and each form a 1D irrep while the two N modes transform with opposite angular momenta $l = \pm 1$ which map to each other under $M_y$ forming a 2D irrep. Another example is the valley-polarized quantum anomalous Hall state (VP-QAH) at $\nu = 1$ where one Chern sector is completely filled whereas the other has one valley completely filled. There is one PG mode which carries a valley charge of $-2$ within the half-filled Chern sector. This mode is not associated with any continuous symmetry breaking but it will be gapless according to the field theory descibed by Eqs.~(\ref{LQ}) and (\ref{EQ}) due to the absence of symmetry allowed anisotropy terms within each Chern sector separately (see Sec.~\ref{sec:NLSM}). In numerics, such mode will have very small gaps and we denote them by PG*. In addition, there is two nematic modes which carry the same angular momentum $l = -1$ and transform as a 1D irrep each.

The representation content for the spinful case is more complicated due to the existence of the $\SU(2)$ spin symmetries in each valley. As usual, the irreps of $\SU(2)$ are labelled by a half-integer $S$ with the corresponding character given by
\beq
\chi(e^{i \alpha s_z}) = \sum_{l=-S/2}^{S/2} e^{2i \alpha l}
\eeq
which yields $1$, $2 \cos \alpha$, and $1 + 2 \cos 2 \alpha$ for $S = 0, 1/2, 1$ respectively. The $S_z$ charge can be extracted from $\chi(e^{i \alpha s_z})$ as in Eq.~\ref{chiValley}. Using this information, we can read the symmetry properties of the soft modes from Tables \ref{tab:Spinful} and \ref{tab:Chern}. 

\subsection{Examples}
Let us now consider a few examples:

\emph{Spin-singlet K-IVC at $\nu = 0$}: As an example, let us consider the spin-singlet K-IVC state. Here, individual spin rotation in each valley is broken but the overall spin symmetry remains unbroken. There are four Goldstone and four pseudo-Goldstone modes which split each into a singlet $S = 0$ and a triplet $S = 1$. In addition, there are 8 nematic modes which split into a 2D irrep and a 6D irrep. The former (latter) consists of two singlets (triplets) with opposite $C_3$ angular momenta tied together with $\eta_z M_y$. Note that from the point of view of the unbroken symmetry group $\G_Q$, there is no reason to expect the spin-singlet and spin-triplet modes to be degenerate as seen in the TDHF spectrum (Fig.~\ref{fig:TDHF}). This can be seen from the addition of inter-valley Hund's coupling which does not break any symmetry in $\G_Q$ but lifts the degeneracy between the singlet and triplet modes (Table \ref{tab:NLSM}). Thus, it is possible sometimes for the soft mode spectrum to have more symmetry than that of the state $Q$. Since in this section we are taking the most general viewpoint allowing for perturbations to the Hamiltonian which may break some of its symmetries to select a specific ground state, we are not going to consider such symmetries, i.e. we are going to assume the presence of the most general perturbation compatible with the symmetry group $\G_Q$ of the state. One should keep in mind though that there may be some extra unaccounted for degeneracies in the spectrum in the absence of these perturbations.

\emph{Spin-singlet VP at $\nu = 0$}: Another example at $\nu = 0$ is the valley-polarized state which is invariant under the full $\SU(2)_K \times \SU(2)_{K'} \simeq \mathrm{SO}(4)$. Here, there are 8 pseudo-Goldstone modes which split into two 4D irreps corresponding to $(S_K, S_{K'}) = (1/2, 1/2)$ (this is equivalent to the fundamental of $\mathrm{SO}(4)$) carrying the same valley charge $+2$ and angular momentum $l = 0$. In addition, there are 8 nematic modes transforming as an 8D irrep consisting of two 4D $\SU(2)_K \times \SU(2)_{K'}$ irreps with opposite $C_3$ angular momenta $l = \pm 1$ tied together by $M_y$. 

\emph{Spin polarized K-IVC at $\nu = 2$}: At Half-filling $\nu = 2$, we can consider the spin-polarized (SP) K-IVC which is found to be the minimum energy state within Hartree-Fock. In addition to the linearly dispersing and the pair of quadratically dispersing Goldstone modes, the dispersion for the gapped, shown in Fig.~\ref{fig:TDHF}, exhibits the pattern of degeneracies $2-1-4-2$ at $\Gamma$. This is in agreement of the symmetry irreps in Table \ref{tab:Spinful} with three PG modes split into a 1D and 2D irrep and 6 nematic modes split into a 2D and a 4D irrep. To understand the origin of these, we note that for the SP K-IVC, $\U_V(1)$ is broken only in the filled spin species, which we take to be $\uparrow$. Thus, the state is invariant under the $\U(1)$ symmetry generated by valley rotation acting only on the down spin, $P_\downarrow \eta_z$. The 2D irrep for the PG mode corresponds to a pair of modes transforming with opposite charge under this $\U(1)$ symmetry tied together with the combination $\eta_z \T$. For the nematic modes, there is an extra degeneracy due to the opposite $C_3$ angular momenta being tied together via $M_y$ which leads to 2D and 4D irreps.

\emph{Correlated $C = 3$ insulators at $\nu = 1$}: As a final example, let us consider the $C = 3$ insulating states at $\nu = 1$ discussed in Sec.~\ref{sec:CorrelatedChern}. For the VP state $Q = P_+ + P_- (P_\uparrow \eta_z - P_\downarrow)$, the $\SU(2)$ spin rotation symmetry in the $K$ valley is broken down to $\U(1)$ leading to a single quadratically dispersing goldstone mode. In addition, there is a pair of almost gapless pseudo-goldstone modes (PG*) which transform as a 2D irrep transforming as $S = 1/2$ under $\SU(2)_{K'}$ with both modes carrying the same valley and $S^z_K$ charges. In addition, there are 12 nematic modes which all carry the same angular momentum under $C_3$ and are split into 3 1D irreps, 2 2D irreps, and 1 3D irrep corresponding to $S = 0$, $1/2$, and $1$ representations under $\SU(2)_{K'}$. For the IVC state $Q = P_+ + P_- (P_\uparrow \eta_x - P_\downarrow)$, $\U_V(1) \times \SU(2)_K \times \SU(2)_{K'}$ is broken down to $\U(1) \times \U(1)$ generated by $s_z$ and $P_\downarrow \eta_z$. This results in one linearly dispersing and two quadratically dispersing goldstone modes which form a 2D irrep of opposite $e^{i \theta P_\downarrow \eta_z}$ eigenvalues paired via $M_y \T$. In addition, there are 12 nematic modes with the same $C_3$ angular momentum which are split into 4 1D irreps and 4 2D irreps which all correspond to the doublet of the group generated by $\{e^{i \theta P_\downarrow \eta_z}, M_y \T\}$ similar to the type-II Goldstone mode above. It is worth noting that symmetry irreps for the spin-valley polarized state at $\nu = 3$ (Table \ref{tab:Chern}) are compatible with earlier time-dependent Hartree-Fock studies at $\nu = 3$ \cite{MacdonaldTDHFTalk, MacdonaldTDHF}. 

\section{Discussion}

Let us now summarize our main findings. We began by providing a simple criterion to determine the count of soft modes based on the picture of Fig.~\ref{fig:Spinful} leading to $\frac{16 - \nu^2 - C^2}{2}$ approximate Goldstone modes associated with the enlarged $\U(4) \times \U(4)$ symmetry and $\frac{16 - \nu^2 + C^2}{2}$ nematic modes which correspond to excitations between Chern sectors. We then obtained the soft mode spectrum numerically in Sec.~\ref{sec:TDHF} for the candidate ground states at $\nu = 0$ and $\nu = 2$. In Sec.~\ref{sec:Energetics}, we provided a detailed analysis of the energetics based on approximate symmetries which identified the ratio $\kappa = w_0/w_1$ as the main parameter in determining the soft mode gaps. In particular, we find that increasing $\kappa$ has an opposite effect on the gaps of the approximate Goldstone and the nematic modes; Whereas the gap of the former increases with $\kappa$ which acts as an anisotropy term breaking the approximate $\U(4) \times \U(4)$ symmetry, the gap to the latter decreases with $\kappa$ as a result of the increased concentration of the Berry curvature in momentum space close to the $\Gamma$ point. This motivated the discussion of a simplified solenoid flux model for the Berry curvature where the symmetry is enhanced to $\U(8)$ and where the effects of different symmetry breaking anisotropies can be systematically included. The resulting $\U(8)$ non-linear sigma model is used to compute the soft mode gaps and shown to agree very well with the numerics (cf.~Fig.~\ref{fig:TDHFvsNLSM}).

In Sec.~\ref{sec:symmreps}, we switch to a more general discussion of the soft modes where we assumed by assuming the most general Slater determinant translationally symmetric insulating state and deriving the degeneracies and the symmetry quantum numbers for the soft modes. This, combined with the general discussion of the properties of the Goldstone modes in Sec.~\ref{sec:GoldstoneCount}, provides a complete picture of the soft modes in any given state which is independent on details of the energetics and model parameters.

We now discuss a few implications of our results in light of some recent works. Two very recent experimental works \cite{YoungEntropy, ShahalEntropy} have provided evidence for a Pomeranchuk Ising-type transition at finite temperature in twisted bilayer graphene close to $\nu = \pm 1$. In both cases, it was proposed that the transition is driven by a large entropic contribution arising from the existence of a large number of soft bosonic modes. In particular, Ref.~\cite{YoungEntropy} argued that such finite temperature state is the same state stabilized by a finite in-plane magnetic field pointing to ferromagnetic order. Similarly, Ref.~\cite{ShahalEntropy} also suggested a magnetic origin for the soft modes by showing that the large entropy gain at the 'cascade transition' at $\nu = 1$ is suppressed via an in-plane magnetic field. In principle, such scenarios are compatible with our finding of a large number of soft bosonic modes for the symmetry breaking insulators which become activated at a temperature of about 10-30 K. However, to make a more quantitative statement, we need a detailed quantitative comparison to the entropic contribution of the metallic state arising from filling the charge neutrality band structure at $\nu = 1$. This computation is more complicated than the corresponding one for the gapped insulating phases since we need to carefully distinguish between the contributions to the entropy coming from the particle-hole continuum and the soft modes. Thus, we leave such an analysis to future work.

The results presented here suggests an experimentally feasible way to investigate the properties of the correlated Chern insulators seen in recent experiment in the presence of small out-of-plane magnetic field \cite{AndreiChern, YazdaniChern, YoungHofstadter}. This can be achieved by employing the setting of Ref.~\cite{Yacouby} which used the edge modes to measure the magnon gap in graphene based quantum Hall ferromagnets.
The results of such experiment can be directly compared to the predictions of Table \ref{tab:Chern}. For example, at $\nu = 1$, there are 3 gapless (or almost gapless modes), two of which are magnons, i.e. carry $S_z$ quantum numbers. At higher energies of a few meV, there are 12 more nematic soft modes, half of which carry non-vanishing $S_z$. It is worth noting that the soft modes can also be directly probed via optical excitations \cite{Wang2018}.

On the theory side, coupling between electrons and soft modes have been proposed as a pairing mechanism for superconductivity in several works \cite{Po2018, YouVishwanath, Kozii, KangFernandes}. Refs.~\cite{YouVishwanath} and \cite{Kozii} focused on soft modes associated with broken valley symmetry for intervalley coherent orders whereas Ref.~\cite{KangFernandes} considered the soft modes associated with $\U(4)$ spin-valley flavor rotation symmetry. Our current work suggests an even larger set of soft modes which includes pseudo-Goldstone modes associated with an enlarged $\U(4) \times \U(4)$ symmetry in addition to a large number of nematic modes which transform non-trivially under three-fold rotation. The importance of the latter was pointed out in a recent experiment which observed pronounced nematicity in the insulator and superconducting states \cite{PabloNematic}. Our symmetry analysis summarized in Tables \ref{tab:Spinful} and \ref{tab:Chern} imposes important restrictions on the coupling of the soft modes to the electrons at small momenta which ultimately determines the favored pairing channels. This can be used as the basis of systematic study of superconductivity induced by coupling to the soft modes which will be the subject of future study. 

Finally, the field theory calculation whose results are summarized in Table \ref{tab:NLSM} enables us to bridge theory and experiment by providing a direct experimental probe to measure the field theory parameters. The field theory given by Eqs.~\eqref{LQ} and \eqref{EQ} can be seen as the simplest theory which captures the essentials of the symmetries and energetics of the different competing phases in terms of a few parameters. Although these parameters can be computed microscopically or numerically, they are likely to be sensitive to microscopic details such as strain and substrate alignment. Thus, a better approach is to view them as phenomenological fitting parameters to be directly compared to experiment. Thus, by using the soft mode gaps and Goldstone modes computed in Sec.~\ref{sec:NLSM} and summarized in Table \ref{tab:NLSM}, we can use experimental data about the soft modes to directly access the field theory parameters. This has important implications for the behavior of this field theory, in particular, in relation to the proposed skyrmion mechanism of superconductivity \cite{SkPaper, SkDMRG} which depends sensitively on the ratio between the parameters $J$ and $\lambda$.

\emph{Note added --} During the final stages of this work, Ref.~\cite{VafekKangRg}
appeared, which employed a promising RG approach and computed the soft mode spectrum in the strong coupling limit at charge neutrality. In addition, we note an independent related work by Kumar, Xie, and Macdonald \cite{MacdonaldTDHF} which also discusses collective modes in TBG. The results of both works are consistent with ours.

\section{Acknowledgments} 
We acknowledge stimulating discussions with Shahal Ilani,  Pablo Jarillo-Herrero and Andrea Young regarding their experimental data. NB would also like to thank Sid Parameswaran for helpful discussions. AV was supported by a Simons Investigator award and by the Simons Collaboration on Ultra-Quantum Matter, which is a grant from the Simons Foundation (651440, AV). EK was supported by a Simons Investigator Fellowship, and by the German National Academy of Sciences Leopoldina through grant LPDS 2018-02 Leopoldina fellowship.  MZ was supported by the Director, Office of Science, Office of Basic Energy Sciences, Materials Sciences and Engineering Division of the U.S. Department of Energy under contract no. DE-AC02-05-CH11231 (van der Waals heterostructures program, KCWF16)

%

\clearpage

\begin{widetext}

\appendix

\section{Time-dependent Hartree-Fock}\label{app:TDHF}

In this appendix we provide more details about the derivation of the TDHF equation. For convenience, we repeat here the interacting continuum model for MATBG:
\beq
\hat{H} = \sum_\bk c_\bk^\dagger h(\bk) c_\bk + \frac{1}{2A} \sum_\bq V_\bq \delta \rho_\bq \delta \rho_{-\bq}
\eeq
Before discussing the TDHF equation, we first define for future use the following generalized Hartree Hamiltonian functional constructed from $\hat{H}$:

\begin{equation}
    H_H\{\phi_{\bq}\}(\bk) =  \frac{1}{A}\sum_{\bG}V_{\bq+\bG} 
     \left[\sum_{\bk'} \text{tr} \left(\Lambda_{-\bq-\bG}(\bk') \phi_\bq(\bk')\right) \right]\Lambda_{\bq+\bG}(\bk)\nonumber\, .
\end{equation}
We also similarly define a corresponding generalized Fock Hamiltonian functional:

\begin{equation}
    H_F\{\phi_\bq\}(\bk) = -\frac{1}{A} \sum_{\bq'}V_{\bq'}\Lambda_{\bq'}(\bk) \phi_\bq(\bk+\bq') \Lambda_{-\bq'}(\bk+\bq+\bq')
\end{equation}
Note that $H_H\{\phi_\bq\}$ and $H_F\{\phi_\bq\}$ respectively become the conventional Hartree and Fock Hamiltonians if we take $\bq = 0$. The sum of the generalized Hartree and Fock Hamiltonians we write as

\begin{equation}
\label{genhfham}
    H_{\rm HF}\{\phi_\bq\}(\bk) = H_H\{\phi_\bq\}(\bk) + H_F\{\phi_\bq\}(\bk)
\end{equation}
Let us now consider a solution of the self-consistent Hartree-Fock equation described by the correlation matrix $[P(\bk)]_{\alpha\beta} = \langle c^\dagger_{\beta,\bk} c_{\alpha,\bk}\rangle$. If we define the corresponding Hartree-Fock Hamiltonian as

\begin{equation}\label{hfham}
    H_{\rm SC}\{P\}(\bk) = h(\bk) + H_{\rm HF}\{P\}(\bk) \, ,
\end{equation}
then self-consistency implies that $[P(\bk),H_{\rm SC}\{P\}(\bk)]=0$.

To derive the TDHF equation, we define the following bosonic operator:
\begin{equation}
    \hat{\phi}_\bq = \sum_\bk c^\dagger_\bk \phi_{\bq}(\bk)c_{\bk+\bq}\, ,
\end{equation}
and evaluate its commutator with $\hat{H}$. As discussed in the main text, this commutator has to be evaluated at the mean-field level by reducing the four-fermion terms to two-fermion terms by partial Wick contractions with $P(\bk)$. If we write the resulting partially contracted commutator as $\langle [\hat H,\hat{\phi}_\bq]\rangle_{\rm HF}$, then we find

\begin{eqnarray}\label{superL}
\langle [\hat H,\hat{\phi}_\bq] \rangle_{HF} & = & \sum_{\bk\bk'}\sum_{\alpha\beta\lambda\gamma} c^\dagger_{\alpha,\bk} c_{\beta,\bk+\bq} \tilde{\mathcal{L}}^{\alpha\beta,\lambda\gamma}_\bq(\bk,\bk') \phi_{\bq}^{\lambda\gamma}(\bk') \nonumber \\ 
& \equiv & \sum_{\bk}\sum_{\lambda\gamma} c^\dagger_{\alpha,\bk} c_{\beta,\bk+\bq} \tilde{\mathcal{L}}^{\alpha\beta}[\phi_\bq](\bk)
\end{eqnarray}
In the first line of Eq. \eqref{superL} we have written $\tilde{\mathcal{L}}_\bq$ as a matrix acting on the vector $\phi_\bq(\bk)$, while in the second line we have written $\tilde{\mathcal{L}}$ as an operator acting on matrices. Both notations will be used below. The explicit expression for $\tilde{\mathcal{L}}$ is most conveniently written via its action on matrices, and it takes the following form:

\begin{equation}
    \tilde{\mathcal{L}}[\phi_\bq](\bk) = [H_{\rm SC}, \phi_\bq](\bk) + H_{\rm HF}\{ [P, \phi_\bq] \}(\bk)\, ,
\end{equation}
where we have used the Hamilonians $H_{\rm SC}$ and $H_{\rm HF}$ respectively defined in Eqs. \eqref{hfham} and \eqref{genhfham}.

At this point, it is important to note that we have defined $\hat{\phi}_\bq$ using a general matrix $\phi^{\alpha\beta}_{\bq}(\bk)$ with both $\alpha$ and $\beta$ running over all band and flavor indices. However, we want to restrict ourselves to only those $\hat{\phi}_\bq$ which create or annihilate particle-hole excitations of the mean-field band spectrum. To this end, we define the projector $P_{\rm PH}$

\begin{equation}
    P_{\rm PH}^{\alpha\beta,\lambda\gamma}(\bk,\bk') = \delta_{\bk,\bk'}\left[P^\perp_{\alpha\lambda}(\bk)P_{\beta\gamma}(\bk) + P_{\alpha\lambda}(\bk)P^\perp_{\beta\gamma}(\bk) \right]
\end{equation}
Acting with $P_{\rm PH}$ on $\phi_\bq$ projects out all contributions in $\hat{\phi}_\bq$ which do not create or annihilate a particle-hole excitation. Correspondingly, we are also only interested in the part of $\tilde{\mathcal{L}}_\bq$ which acts within the particle-hole subspace, so we define

\begin{equation}
    \mathcal{L}_{\bq} = P_{PH}\tilde{\mathcal{L}}_{\bq}P_{PH}
\end{equation}
To find the soft mode spectrum $\omega_\bq$, we numerically solve for the smallest eigenvalues of $\mathcal{L}_\bq$ to obtain the bosonic operators which satisfy $i\partial_t \hat{\phi}_{\bq} = [\hat{H},\hat{\phi}_\bq] = \omega_\bq \hat{\phi}_\bq$ at the mean-field level. However, note that $\mathcal{L}_{\bq}$ is not hermitian. To investigate the properties of the eigenvalue problem at hand, we first define the following matrix

\begin{equation}
    \mathcal{Z}^{\alpha\beta,\lambda\gamma}(\bk,\bk') = \delta_{\bk,\bk'}\left[P_{\alpha\lambda}(\bk)\delta_{\gamma\beta} - \delta_{\alpha\lambda}P_{\gamma\beta}(\bk) \right]
\end{equation}
As an operator acting on matrices, the action of $\mathcal{Z}$ can be written as a simple commutator: $\mathcal{Z} \phi_\bq = [P, \phi_\bq]$. It is readily verified that this matrix satisfies $\mathcal{Z}^2 = P_{PH}$ and has eigenvalues $+1$ ($-1$) if $\alpha,\lambda$ lie in the subspace of occupied (unoccupied) mean-field states and $\beta,\gamma$ lie in the unoccupied (occupied) subspace. Let us now define the matrix $\mathcal{H}_\bq$ via the relation

\begin{equation}
    \mathcal{L}_{\bq} = \mathcal{Z}\mathcal{H}_{\bq}
\end{equation}
In contrast to $\mathcal{L}_\bq$, the matrix $\mathcal{H}_\bq$ is hermitian. In fact, $\mathcal{H}_\bq$ has one more important property -- namely, it has a particle-hole symmetry:

\begin{equation}
    X_\bq^\dagger \mathcal{H}_{\bq}^*X_\bq = \mathcal{H}_{-\bq} \;\;\text{ with }\;\; X^{\alpha\beta,\lambda\gamma}_\bq(\bk,\bk') = \delta_{[\bk+\bq],\bk'} \delta_{\alpha\gamma} \delta_{\beta\lambda}\, ,
\end{equation}
where $[\bk+\bq]$ lies in the first mini-BZ, and is equal to $\bk+\bq$ modulo a moire reciprocal lattice vector. From $X_\bq\mathcal{Z} = -\mathcal{Z}X_\bq$, we also immediately see that $\mathcal{L}_\bq$ satisfies

\begin{equation}
    X_\bq^\dagger \mathcal{L}_{\bq}^*X_\bq = - \mathcal{L}_{-\bq}
\end{equation}
From this we conclude that solving the TDHF equation is equivalent to solving the equation of motion of an effective quadratic boson Hamiltonian $\mathcal{H}_\bq$ obtained from $\hat{H}$ at the mean-field level. As discussed in the main text, this interpretation is consistent with the fact that the self-consistency condition implies that $\hat{H}$ creates at least two particle-hole excitations.

\section{Derivation of the soft mode Hamiltonian $\H_\bq$}
\label{app:DerivationHq}

In this appendix, we will provide details for the derivation of the soft mode Hamiltonian $\H_\bq$ in Sec.~\ref{sec:Energetics}. The derivation follows closely the related derivation in Ref.~\cite{SkPaper} and employ the same notation and conventions. $\H_\bq$ is obtained by expanding the energy defined in Eq.~\ref{EQW} to second order in $\phi$. Written more explicitly
\beq
E(Q, \phi) = \langle \psi(\phi)| \hat H| \psi(\phi) \rangle = \langle e^{-i \sum_\bq \hat \phi_\bq} \hat H e^{i \sum_\bq \hat \phi_\bq}  \rangle 
\eeq
The second order term in $\phi$ is given by
\beq
\label{E2}
E_2 = -\frac{1}{2A} \sum_{\bq,\bq'} V_{\bq'} \langle [\hat \phi_\bq, \rho_{\bq'}] [\hat \phi_{-\bq}, \rho_{-\bq'}] \rangle = -\frac{1}{2A} \sum_{\bq,\bq'} V_{\bq'} \Tr \tilde P^T [\tilde \phi_\bq, \tilde \Lambda_{\bq'}] [\tilde  \phi_{-\bq}, \tilde \Lambda_{-\bq'}]
\eeq
The second equality is obtained by evaluating the expectation value of the product of commutators of the two bilinear operators as explained in Ref.~\cite{SkPaper}. Here, the trace with capital T includes momentum summation in addition to tracing over the matrix index and the symbols with a tilde are matrices in momentum as well as internal indices defined as
\beq
[\tilde \phi_\bq]_{\bk,\bk'} = \phi_\bq(\bk) \delta_{\bk', \bk + \bq}, \qquad [\tilde \Lambda_\bq]_{\bk,\bk'} = \Lambda_\bq(\bk) \delta_{\bk', \bk + \bq}, \qquad [\tilde P]_{\bk,\bk'} = P \delta_{\bk',\bk}
\eeq
The projector $P$ is related to $Q$ via $P = \frac{1}{2}(1 + Q)$. We notice that since $\phi_\bq$ anticommutes with $Q$ while $\Lambda_\bq$ commutes with $Q$, the term containing $Q$ in the energy vanishes since
\beq
\sum_{\bq,\bq'} V_{\bq'} \Tr \tilde Q^T [\tilde \phi_\bq, \tilde \Lambda_{\bq'}] [\tilde  \phi_{-\bq}, \tilde \Lambda_{-\bq'}] = - \sum_{\bq,\bq'} V_{\bq'} \Tr \tilde Q^T [\tilde  \phi_{-\bq}, \tilde \Lambda_{-\bq'}] [\tilde \phi_\bq, \tilde \Lambda_{\bq'}] =  - \sum_{\bq,\bq'} V_{\bq'} \Tr \tilde Q^T [\tilde  \phi_{\bq}, \tilde \Lambda_{\bq'}] [\tilde \phi_{-\bq}, \tilde \Lambda_{-\bq'}] = 0
\eeq
Here, we have used the cyclic  property of the trace in the first line and made the replacements $\bq \mapsto -\bq$, $\bq' \mapsto -\bq'$ in the second line while also using $V_{-\bq} = V_\bq$. Thus, Eq.~\ref{E2} simplifies to
\begin{align}
    E_2 &= -\frac{1}{4A} \sum_{\bq,\bq'} V_{\bq'} \Tr  [\tilde \phi_\bq, \tilde \Lambda_{\bq'}] [\tilde  \phi_{-\bq}, \tilde \Lambda_{-\bq'}] \nonumber \\ 
    &= -\frac{1}{4A} \sum_{\bq,\bq', \bk} V_{\bq'} [\phi_\bq(\bk) \Lambda_{\bq'}(\bk + \bq) - \Lambda_{\bq'}(\bk) \phi_\bq(\bk + \bq')] [  \phi_{-\bq}(\bk + \bq + \bq') \Lambda_{-\bq'}(\bk + \bq') - \Lambda_{-\bq'}(\bk + \bq' + \bq) \phi_{-\bq}(\bk + \bq) ]
    \nonumber \\ 
    &= -\frac{1}{4A} \sum_{\bq,\bq', \bk} V_{\bq'} [\phi_\bq(\bk) \Lambda_{\bq'}(\bk + \bq) - \Lambda_{\bq'}(\bk) \phi_\bq(\bk + \bq')] [  \phi^\dagger_{\bq}(\bk + \bq') \Lambda^\dagger_{\bq'}(\bk) - \Lambda^\dagger_{\bq'}(\bk + \bq) \phi^\dagger_{\bq}(\bk) ]
    \label{E2k}
\end{align}
where we used the relations $\phi_\bq(\bk) = \phi^\dagger_{-\bq}(\bk + \bq)$ and $\Lambda_\bq(\bk) = \Lambda^\dagger_{-\bq}(\bk + \bq)$.
To simplify this expression further, we expand $\phi$ into a sum of generators
\beq
\phi_\bq(\bk) = \sum_{\gamma,\gamma' = \pm} \sum_{\mu=1}^{2 N_{\gamma,\gamma'}} \phi^\alpha_{\gamma,\gamma';\bq}(\bk) r^{\gamma,\gamma'}_\alpha, \qquad \tr [r^{\gamma_1,\gamma'_1}_\alpha]^\dagger r^{\gamma_2,\gamma'
_2}_\beta = 2 \delta_{\alpha, \beta} \delta_{\gamma_1,\gamma_2} \delta_{\gamma'_1, \gamma'_2}
\label{phit}
\eeq
where we have split the generators into those connecting the Chern sector $\gamma = \pm$ to $\gamma' = \pm$. In the Chern basis, they have the simple forms
\beq
r^{++} \propto \left(\begin{array}{cc}
    1 & 0 \\
    0 & 0
\end{array} \right), \qquad r^{+-} \propto \left(\begin{array}{cc}
    0 & 1 \\
    0 & 0
\end{array} \right), \qquad r^{-+} \propto \left(\begin{array}{cc}
    0 & 0 \\
    1 & 0
\end{array} \right), \qquad r^{--} \propto \left(\begin{array}{cc}
    0 & 0 \\
    0 & 1
\end{array} \right)
\eeq
For $\Lambda_\bq(\bk) = F_\bq(\bk) e^{i \Phi_\bq(\bk) \gamma_z}$, it is easy to see that 
\beq
\Lambda_\bq(\bk) r^{\gamma,\gamma'}_\alpha = F_\bq(\bk) e^{i \gamma \Phi_\bq(\bk)} r^{\gamma,\gamma'}_\alpha, \qquad r^{\gamma,\gamma'}_\alpha \Lambda_\bq(\bk)  = F_\bq(\bk) e^{i \gamma' \Phi_\bq(\bk)} r^{\gamma,\gamma'}_\alpha
\label{Lambdat}
\eeq
Substituting (\ref{phit}) in (\ref{E2k}) and using (\ref{Lambdat}), we get
\beq
    E_2 = \frac{1}{A} \sum_{\bq,\bq', \bk,\alpha,\gamma,\gamma'} V_{\bq'} F_{\bq'}(\bk) \phi^\alpha_{\gamma,\gamma',\bq}(\bk) \left\{ F_{\bq'}(\bk) [\phi^\alpha_{\gamma,\gamma',\bq}(\bk)]^*  - F_{\bq'}(\bk + \bq) [\phi^\alpha_{\gamma,\gamma',\bq}(\bk + \bq')]^*  e^{i [\gamma \Phi_{\bq'}(\bk + \bq) - \gamma'  \Phi_{\bq'}(\bk)]} \right\}
    \label{Hq}
\eeq
Comparing with the definition of $\H_\bq$ in Eq.~\eqref{EQphi}, this implies that, as anticipated, $\H_\bq$ has a block diagonal form in the $\gamma$ indices and is proportional to unity in the remaining internal indices with its explicit form given by
\beq
\H^{\mu,\nu}_{\gamma,\gamma';\bq}(\bk,\bk') = \frac{\delta_{\mu \nu}}{A} \sum_{\bq'} V_{\bq'} F_{\bq'}(\bk) \left\{ \delta_{\bk',\bk} F_{\bq'}(\bk)  - \delta_{\bk',[\bk + \bq']} F_{\bq'}(\bk + \bq) e^{i [\gamma \Phi_{\bq'}(\bk + \bq) - \gamma'  \Phi_{\bq'}(\bk)]} \right\}
\eeq
which reduces to Eq.~\eqref{Hgamma} in the main text for $\bq = 0$.

\section{Properties of $\rho$}
\label{app:RankRho}
In this appendix, we investigate the properties of the matrix $\rho$ defined as 
\beq
\rho_{\mu \nu} = \frac{1}{4} \tr Q [t_\mu, t_\nu]
\eeq
In the following, we will restrict ourselves to generators which anticommute with $Q$ for which $\tr Q [t_\mu, t_\nu] = 2 \tr Q t_\mu t_\nu$ leading to Eq.~\eqref{rhot} in the main text. In the following, we are going to show that when $t_\mu$ goes over all generators which anticommute with $Q$, $\rho$ is a full rank matrix with an equal number of $+1$ and $-1$ eigenvalues equal to $16 - \nu^2$. To this end, we start by choosing a basis where $Q$ is diagonal with $+1$ eigenvalues followed by $-1$ eigenvalues. Using the relation $\tr Q = 2\nu$, it is easy to see that the number of $\pm 1$ eigenvalues is $4 \pm \nu$. Now let us define the matrices $X_{ij}$ whose elements are given by
\beq
[X_{ij}]_{mn} = \delta_{i,m} \delta_{j,n}
\eeq
It is easy to show that $X_{ij}$ satisfy the following identities
\beq
X_{ij}^T = X_{ji}, \qquad \tr X^T_{ij} X_{nm} = 2\delta_{in} \delta_{jm}, \qquad Q X_{ij} = X_{ij} \begin{cases} 
+1 &: 1 \leq i \leq 4 + \nu \\
-1 &: 4 + \nu < i \leq 8 \end{cases}, \qquad X_{ij} Q = X_{ij}  \begin{cases} 
+1 &: 1 \leq j \leq 4 + \nu \\
-1 &: 4 + \nu < j \leq 8 \end{cases}
\eeq
We can then choose a basis of hermitian generators by restricting ourselves to $X_{ij}$ with $i = 1,\dots,4+\nu$ and $j = 4 + \nu + 1, \dots, 8$. Labelling such ordered pairs $(i_l,j_l)$ by an index $\nu$, we can define
\beq
t_{2\nu-1} = X_{i_\nu,j_\nu} + X^T_{i_\nu,j_\nu}, \qquad t_{2\nu} = i ( X_{i_\nu,j_\nu} - X^T_{i_\nu,j_\nu})
\eeq
which satisfy
\beq
\tr t_\mu t_\nu = 2\delta_{\mu,\nu}, \qquad Q t_{2\nu-1} = -i t_{2\nu}, \quad Q t_{2\nu} = i t_{2\nu - 1} 
\eeq
Thus, in this basis $\rho$ has a block diagonal form with $2 \times 2$ blocks given by $\sigma_y$. It follows that $\rho$ is a full rank matrix with an equal number of $+1$ and $-1$ eigenvalues equal to $16 - \nu^2$.

\section{Symmetry representations on the soft modes}
\label{app:SymRep}
In this appendix, we provide details for the procedure to explain the symmetry representations for the soft modes. We start by considering a general symmetry $\S$ with action, given by Eq.~\ref{Sck}. The soft modes are defined through the operator $\hat \phi_{n,\bq}$ given by Eq.~\ref{Wck}. The symmetry $\S$ acts on $\phi_{n,\bq}$ via the matrix representation $S_{mn}(\bq)$ (cf.~Eq.~\ref{SPhiHat}). This matrix can be obtained explicitly by first noting that the soft mode Hamiltonian $\H_\bq$ transforms under $\S$ as
\beq
[\S \H \S^{-1}]_\bq^{\mu \nu}(\bk, \bk') \equiv \sum_{\lambda, \rho} [\Gamma_\bq^{\lambda \mu}(\bk)]^* \H^{\lambda \rho}_{O\bq}(O \bk, O \bk') \Gamma^{\rho \nu}_\bq(\bk')
\eeq
where $\Gamma$ is defined through
\beq
U_\bk t_\mu U_{\bk + \bq}^\dagger = \sum_\nu \Gamma^{\nu \mu}_\bq(\bk) t_\nu
\eeq
For any $\S$ which leaves $Q$ invariant, i.e. $\S \in \G_Q$, $\H$ satisfies $\S \H \S^{-1} = \H$ which means that $\phi_{n,\bq}$ transforms as a representation under the action of $\S$:
\beq
    [S \phi]_{n,\bq}(\bk) \equiv \Gamma_\bq(\bk) \phi_{n,\bq}(\bk)   = \sum_m S_{nm}( \bq) \phi_{m,O \bq}(O \bk)
    \label{Sphi}
\eeq
The matrix $S_{mn}(\bq)$ can be obtained from the knowledge of the wavefunctions $\phi_{n,\bq}(\bk)$. These have a particularly simple form in the $\U(4) \times \U(4)$ limit as explained in the main text. In the following, we will employ this form to obtain the matrices $S$ explicitly.

We start by recalling the form of the symmetries when projected onto the flat bands which act in the Chern-pseudospin-spin basis as \cite{KIVCpaper, SkPaper}
\begin{gather}
    U_V c_\bk U_V^\dagger = e^{i \phi \eta_z} c_\bk, \quad S_{K/K'} c_\bk S^{-1}_{K/K'} = e^{i P_{K/K'} \bn \cdot \bs} c_\bk \nonumber \\
    C_2 c_\bk C_2^\dagger = \eta_x e^{i \theta_2(\bk)} c_{-\bk}, \quad \T c_\bk \T^{-1} = \eta_x \gamma_x c_{-\bk} \nonumber \\
     C_3 c_\bk C_3^\dagger = e^{i \theta_3(\bk) \gamma_z} c_{c_3 \bk}, \quad M_y c_\bk M_y^\dagger = \gamma_x e^{i \theta_y(\bk) \gamma_z} c_{m_y \bk}
\end{gather}
where we have defined $P_{K/K'} = \frac{1 \pm \eta_z}{2}$. Notice the $\bk$-dependent phase factors $\theta_{2,3,y}(\bk)$ which cannot be removed due to the band topology of the bands in the sublattice basis. In particular, the phase $\theta_3(\bk)$ is equal to 0 at $\Gamma$ and $\frac{2\pi}{3}$ at $K$ and $K'$ points.

In the $\U(4) \times \U(4)$ limit, there is a one-to-one correspondence between the lowest energy eigenstates of the soft modes Hamiltonian $\H_\bq$ and the generators $r^{\gamma,\gamma'}_\alpha$. To make correspondence manifest, we label the generator corresponding to the eigenstate $\phi_{n,\bq}$, $n = 1, \dots, 2 (16 - \nu^2)$, by the indices $(\gamma_n, \gamma'_n, \alpha_n)$. In this case, we can write the eigenstates as
\beq
\phi^{\alpha,\gamma,\gamma'}_{n,\bq}(\bk) = \phi^{\alpha,\gamma,\gamma'}_{\alpha_n,\gamma_n,\gamma'_n,\bq}(\bk) = \delta_{\alpha, \alpha_n} \delta_{\gamma, \gamma_n} \delta_{\gamma', \gamma'_n} \phi_{\gamma,\gamma';\bq}(\bk)
\eeq
where $\phi_{\gamma,\gamma';\bq}(\bk)$ is the lowest energy eigenfunction of the Hamiltonian $\H_{\gamma,\gamma';\bq}$ given in (\ref{Hq}). For $\bq = 0$, this has the simple form
\beq
\phi_{\gamma,\gamma';\bq = 0}(\bk) = \begin{cases}
\frac{1}{\sqrt{N}} &: \gamma = \gamma'\\
\Delta(\bk) &: \gamma = +, \gamma' = -\\
\Delta(\bk)^* &: \gamma = -, \gamma' = +
\end{cases}
\eeq
where $N$ is the number of points in the BZ and we assumed the normalization $\sum_\bk |\phi_{\gamma,\gamma';\bq}(\bk)|^2 = 1$. $\Delta(\bk)$ is inter-Chern soft mode wavefunction introduced in the main text. The form of $\Delta(\bk)$ depends on the gauge chosen for the flat band wave-functions. In the gauge chosen in Eq.~\ref{AGauge} in the main text, the phase of $\Delta(\bk)$ winds by $4\pi$ around the BZ due to a pair of vortices centered at $\Gamma$. On the other hand, in a periodic gauge, the phase of $\Delta(\bk)$ cannot wind around the BZ and extra vortices must appear elsewhere in the BZ. These extra vortices can be removed by a singular gauge transformation at the expense of making the gauge non-periodic. In numerics, it is usually a lot easier to construct the wavefunctions in the periodic gauge. One such gauge choice is shown in Fig.~\ref{fig:DeltaMagPh}. We can see that the phase of the wavefunctions winds by $6\pi$ rather than $4\pi$ around $\Gamma$ in addition to the presence of $2\pi$ vortices at the three $M$ points to cancel the total winding. By looking at the magnitude of the wavefunctions, we can see that the winding at the $M$ points is not associated with a vanishing of the magnitude of $\Delta(\bk)$ (except possibly at a single point) and thus can be removed by a singular gauge transformation. On the other hand, the vortices at $\Gamma$ are true vortices associated with the vanishing of the $|\Delta(\bk)|$. This can be more clearly illustrated by choosing a smaller value of $\kappa$ where the vanishing of the magnitude at $\Gamma$ becomes slower. 

\begin{figure}
    \centering
    \includegraphics[width = 0.8 \textwidth]{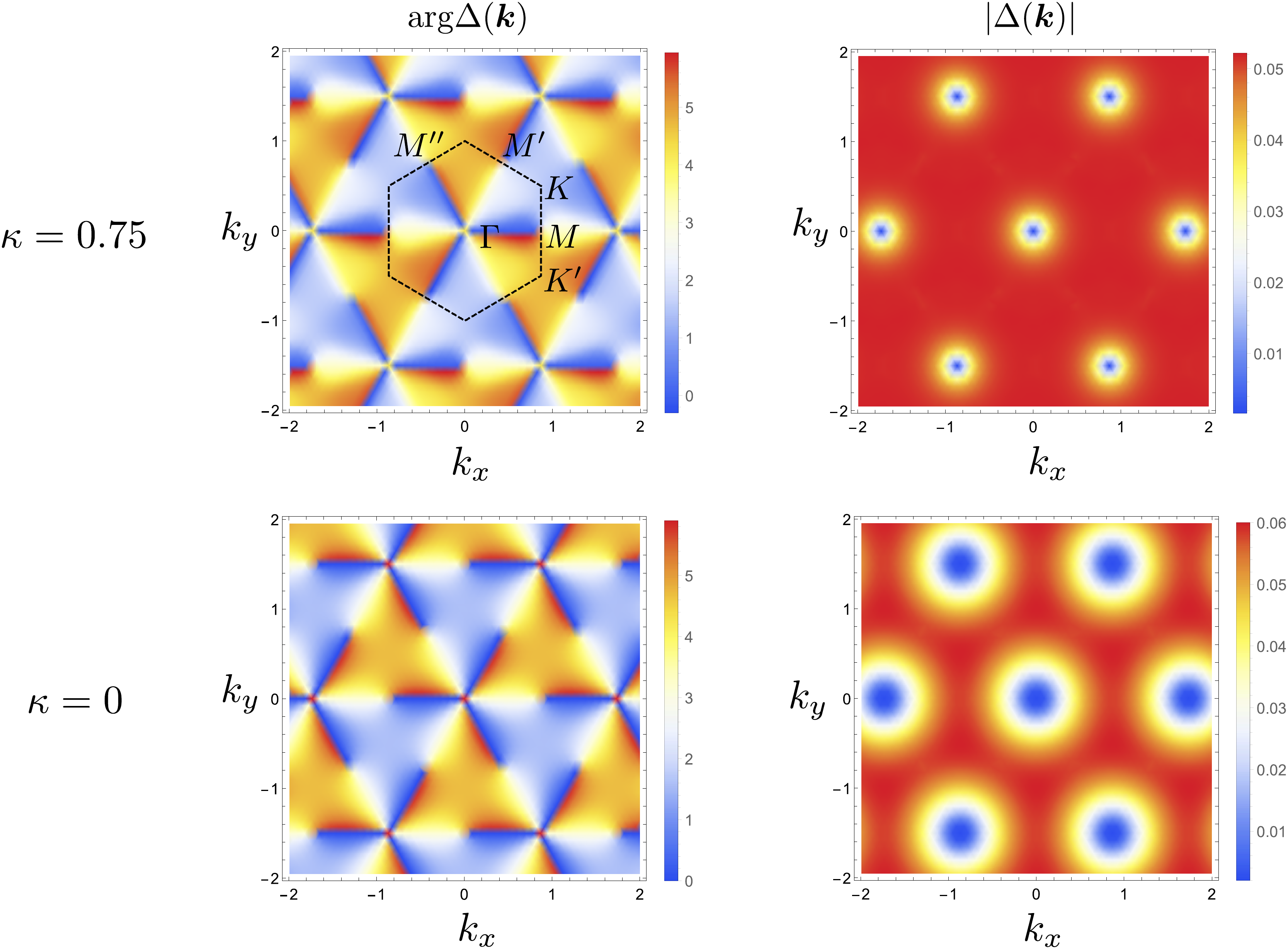}
    \caption{Magnitude and phase for the lowest eigenfunction of the inter-Chern soft mode Hamiltonian $\H^{+-}$ (at $\bq = 0$) denoted by $\Delta(\bk)$ in the periodic gauge for $\kappa = 0.75$ and $\kappa = 0$. The phase winds by $6 \pi$ around $\Gamma$ and by $-2\pi$ at $M$, $M'$ and $M''$. The phase winding at the $M$ points is not associated with vanishing of $|\Delta(\bk)|$ and thus can be removed by a singular gauge transformation.}
    \label{fig:DeltaMagPh}
\end{figure}

The symmetry representation matrix can now be expressed in the basis of the generators $r^{\gamma,\gamma'}_\mu$ using Eq.~\ref{Sphi} 
\beq
\Gamma^{\gamma_1, \gamma'_1, \alpha; \gamma_2, \gamma'_2, \beta}_\bq(\bk) \phi_{\gamma_2, \gamma'_2; \bq}(\bk) = S_{\gamma_2, \gamma'_2, \beta; \gamma_1, \gamma'_1, \alpha}(\bq) \phi_{\gamma_1, \gamma'_1; O\bq}(O \bk)
\label{GS}
\eeq
Note that there are no sums in the above equation. This equation can be solved for $S(\bq)$ as
\beq
S_{\gamma_2, \gamma'_2, \beta; \gamma_1, \gamma'_1, \alpha}(\bq) = \sum_\bk \phi_{\gamma_1, \gamma'_1; O\bq}(O \bk)^* \Gamma^{\gamma_1, \gamma'_1, \alpha; \gamma_2 \gamma'_2, \beta}_\bq(\bk) \phi_{\gamma_2 \gamma'_2; \bq}(\bk)
\label{Sq}
\eeq

Let us start with the global symmetries: $\U(1)$ valley charge conservation or $\SU(2)_K \times \SU(2)_{K'}$ spin rotation. These symmetries are characterized by a $\bk$-independent action $U$ and do not act spatially, i.e. $O = 1$. Furthermore, they act trivially on the Chern index $\gamma$. As a result, their action on the generators $r_\alpha^{\gamma, \gamma'}$ has the form
\beq
\Gamma^{\gamma_1, \gamma'_1, \alpha; \gamma_2, \gamma'_2, \beta}_\bq(\bk) = \frac{\delta_{\gamma_1, \gamma_2} \delta_{\gamma'_1, \gamma'_2}}{2} \tr U r_\beta^{\gamma_1 \gamma'_1} U^\dagger [r_\alpha^{\gamma_1 \gamma'_1}]^\dagger
\eeq
Substituting in Eq.~\ref{Sq}, we find
\beq
S_{\gamma_1, \gamma'_1, \alpha; \gamma_2, \gamma'_2, \beta}(\bq) = \frac{\delta_{\gamma_1, \gamma_2} \delta_{\gamma'_1, \gamma'_2}}{2} \tr U r_\alpha^{\gamma_1 \gamma'_1} U^\dagger [r_\beta^{\gamma_1 \gamma'_1}]^\dagger = \frac{1}{2} \tr U r_\alpha^{\gamma_1 \gamma'_1} U^\dagger [r_\beta^{\gamma_2 \gamma'_2}]^\dagger
\eeq
where we used the fact that $U$ leaves the Chern index invariant. This can be now expressed back in the hermitian basis $t_\mu$ (with $\tr t_\mu t_\nu = 2\delta_{\mu, \nu}$) as
\beq
S_{\mu \nu}(\bq) = \frac{1}{2} \tr U t_\mu U^\dagger t_\nu
\eeq

Next, let us consider $C_2$ symmetry with $U_\bk = \eta_x e^{i \theta_2(\bk)}$ and $O \bk = -\bk$. The symmetry action on the generators $r_\alpha^{\gamma, \gamma'}$ is given by
\beq
\Gamma^{\gamma_1, \gamma'_1, \alpha; \gamma_2, \gamma'_2, \beta}_\bq(\bk) = \frac{\delta_{\gamma_1, \gamma_2} \delta_{\gamma'_1, \gamma'_2}}{2} e^{i [\theta_2(\bk) - \theta_2(\bk + \bq)]} \tr \eta_x r_\beta^{\gamma_1 \gamma'_1} \eta_x [r_\alpha^{\gamma_1 \gamma'_1}]^\dagger
\eeq
We now restrict ourselves to the $\Gamma$ point $\bq = 0$ and substitute in (\ref{Sq}) to get
\beq
   S_{\gamma_1, \gamma'_1, \alpha; \gamma_2, \gamma'_2, \beta}(\Gamma) = \frac{\delta_{\gamma_1, \gamma_2} \delta_{\gamma'_1, \gamma'_2}}{2} \tr \eta_x r_\alpha^{\gamma_1 \gamma'_1} \eta_x [r_\beta^{\gamma_1 \gamma'_1}]^\dagger \sum_\bk \phi_{\gamma_1, \gamma'_1; \Gamma}(- \bk)^* \phi_{\gamma_1 \gamma'_1; \Gamma}(\bk)
\label{SqC2}
\eeq
For the periodic gauge choice of Fig.~\ref{fig:DeltaMagPh}, $\Delta(-\bk) = -\Delta(\bk)$ leading to $\sum_\bk \phi_{\gamma_1, \gamma'_1; \Gamma}(- \bk)^* \phi_{\gamma_2 \gamma'_2; \Gamma}(\bk) = \gamma_1 \gamma'_1$. Substituting in (\ref{SqC2}), we get
\beq
   S_{\gamma_1, \gamma'_1, \alpha; \gamma_2, \gamma'_2, \beta}(\Gamma) = \gamma_1 \gamma'_1 \frac{\delta_{\gamma_1, \gamma_2} \delta_{\gamma'_1, \gamma'_2}}{2} \tr \eta_x r_\alpha^{\gamma_1 \gamma'_1} \eta_x [r_\beta^{\gamma_1 \gamma'_1}]^\dagger = \tr \eta_x \gamma_z r_\alpha^{\gamma_1 \gamma'_1} \eta_x \gamma_z [r_\beta^{\gamma_2 \gamma'_2}]^\dagger
 \eeq
 Going back to the $t_\mu$ basis, we find
 \beq
 S^{C_2}_{\mu \nu}(\Gamma) = \frac{1}{2} \tr \eta_x \gamma_z t_\mu \eta_x \gamma_z t_\nu
 \eeq
 
 Next, we consider $C_3$ symmetry with $U_\bk = e^{i \theta_3(\bk) \gamma_z}$ and $O = c_3$. The symmetry action on the generators is
\beq
 \Gamma^{\gamma_1, \gamma'_1, \alpha; \gamma_2, \gamma'_2, \beta}_\bq(\bk) = \frac{\delta_{\gamma_1, \gamma_2} \delta_{\gamma'_1, \gamma'_2}}{2} e^{i [\gamma \theta_3(\bk) - \gamma' \theta_3(\bk + \bq)]} \delta_{\alpha \beta}
\eeq
Resticting to the $\Gamma$ point, we get
\beq
   S_{\gamma_1, \gamma'_1, \alpha; \gamma_2, \gamma'_2, \beta}(\Gamma) = \frac{\delta_{\gamma_1, \gamma_2} \delta_{\gamma'_1, \gamma'_2} \delta_{\alpha \beta}}{2}  \sum_\bk e^{i (\gamma_1 - \gamma'_1) \theta_3(\bk)} \phi_{\gamma_1, \gamma'_1; \Gamma}(c_3 \bk)^* \phi_{\gamma_1 \gamma'_1; \Gamma}(\bk)
\label{SqC2}
\eeq
For $\gamma_1 = \gamma'_1$, it is easy to see that the $\bk$ integral yields 1. For $\gamma_1 = -\gamma'_1$, we can verify by a direct calculation that the $\bk$ integral yields $e^{-\frac{2\pi i}{3} \gamma_1}$ which gives
\beq
 S_{\gamma_1, \gamma'_1, \alpha; \gamma_2, \gamma'_2, \beta}(\Gamma) = \frac{\delta_{\gamma_1, \gamma_2} \delta_{\gamma'_1, \gamma'_2} \delta_{\alpha \beta}}{2} e^{\frac{2\pi i}{3} (\gamma_1 - \gamma'_1)}
 \label{S3}
\eeq
Any easier way is to note that Eq.~\ref{GS} holds for every $\bk$ and thus can be used to determine $S(\bq)$ by choosing a $C_3$ invariant momentum $\bk$ for which $\phi_\Gamma(\bk)$ does not vanish. Choosing $\bk = K$ and noting that there is no phase winding at this point for the gauge of Fig.~\ref{fig:DeltaMagPh}, we can immediately read off $S(\Gamma) = \frac{\delta_{\gamma_1, \gamma_2} \delta_{\gamma'_1, \gamma'_2} \delta_{\alpha \beta}}{2} e^{\theta_3(K) (\gamma_1 - \gamma'_1)}$ which is the same as (\ref{S3}) since $\theta_3(K) = \frac{2\pi}{3}$. If we now go back to the $t_\mu$ basis, we find
 \beq
 S^{C_3}_{\mu \nu}(\Gamma) = \frac{1}{2} \tr e^{\frac{2\pi}{3} \gamma_z} t_\mu e^{-\frac{2\pi}{3} \gamma_z} t_\nu
 \eeq
 
 We next consider $M_y$ which flips the Chern sector so its action on the generators is
\beq
 \Gamma^{\gamma_1, \gamma'_1, \alpha; \gamma_2, \gamma'_2, \beta}_\bq(\bk) =  \frac{\delta_{\gamma_1, -\gamma_2} \delta_{\gamma_2, -\gamma_2} }{2} e^{-i [\gamma_1 \theta_y(\bk) - \gamma'_1 \theta_y(\bk + \bq)]} \tr \gamma_x r_\beta^{\gamma_1, \gamma'_1} \gamma_x [r_\beta^{-\gamma_1 ,-\gamma'_1}]^\dagger
\eeq
Substituting in (\ref{Sq}), we get
\beq
    S_{\gamma_1, \gamma'_1, \alpha; \gamma_2, \gamma'_2, \beta}(\Gamma) = \frac{\delta_{\gamma_1, \gamma_2} \delta_{\gamma'_1, \gamma'_2}}{2} \tr \gamma_x r_\alpha^{\gamma_1 \gamma'_1} \gamma_x [r_\beta^{\gamma_1 \gamma'_1}]^\dagger \sum_\bk e^{-i (\gamma_1 - \gamma'_1) \theta_y(\bk)} \phi_{-\gamma_1, -\gamma'_1; \Gamma}(m_y \bk)^* \phi_{\gamma_1 \gamma'_1; \Gamma}(\bk)
\eeq
We can verify by direct evaluation that the $\bk$ summation yields 1 regardless of $\gamma_1$ and $\gamma'_1$ which yields
\beq
    S_{\gamma_1, \gamma'_1, \alpha; \gamma_2, \gamma'_2, \beta}(\Gamma) = \frac{\delta_{\gamma_1, \gamma_2} \delta_{\gamma'_1, \gamma'_2}}{2} \tr \gamma_x r_\alpha^{\gamma_1 \gamma'_1} \gamma_x [r_\beta^{\gamma_1 \gamma'_1}]^\dagger = \frac{1}{2} \tr \gamma_x r_\alpha^{\gamma_1 \gamma'_1} \gamma_x [r_\beta^{\gamma_2 \gamma'_2}]^\dagger
\eeq
In the hermitian basis $t_\mu$, this becomes
\beq
S^{M_y}_{\mu,\nu}(\Gamma) = \frac{1}{2} \tr \gamma_x t_\mu \gamma_x t_nu
\eeq

Finally, we consider time-reversal symmetry whose action on the generators is
\beq
 \Gamma^{\gamma_1, \gamma'_1, \alpha; \gamma_2, \gamma'_2, \beta}_\bq(\bk) =  \frac{\delta_{\gamma_1, -\gamma_2} \delta_{\gamma_2, -\gamma_2} }{2} \tr \gamma_x \eta_x [r_\alpha^{\gamma_1, \gamma'_1}]^* \gamma_x \eta_x [r_\beta^{-\gamma_1 ,-\gamma'_1}]^\dagger
\eeq
Substituting in (\ref{Sq}), we get
\beq
    S_{\gamma_1, \gamma'_1, \alpha; \gamma_2, \gamma'_2, \beta}(\Gamma) = \frac{\delta_{\gamma_1, \gamma_2} \delta_{\gamma'_1, \gamma'_2}}{2} \tr \gamma_x \eta_x r_\alpha^{\gamma_1 \gamma'_1} \gamma_x \eta_x [r_\beta^{\gamma_1 \gamma'_1}]^\dagger \sum_\bk \phi_{-\gamma_1, -\gamma'_1; \Gamma}(-\bk) \phi_{\gamma_1 \gamma'_1; \Gamma}(\bk) = \frac{1}{2} \tr \gamma_x \eta_x r_\alpha^{\gamma_1 \gamma'_1} \gamma_x \eta_x [r_\beta^{\gamma_2 \gamma'_2}]^\dagger
\eeq
This becomes in the $t_\mu$ basis
\beq
S^\T_{\mu,\nu} = \frac{1}{2} \tr \gamma_x \eta_x t_\mu \gamma_x \eta_x t_\nu^*
\eeq

\end{widetext}

\end{document}